\documentclass{aa}
\usepackage[varg]{txfonts}
\usepackage{xcolor}
\usepackage{natbib}
\usepackage{soul}
\usepackage{amssymb}
\usepackage{amsmath}
\usepackage{graphicx}
\usepackage{hyperref}
\usepackage{soul}
\usepackage{array}

\hypersetup{
    colorlinks=true,
    linkcolor=blue,
    filecolor=magenta,
    urlcolor=blue,
    citecolor=blue,
}

\bibpunct{(}{)}{;}{a}{}{,}

\newcommand{\ero}{eROSITA\xspace}
\newcommand{\rosat}{\textit{ROSAT}\xspace}
\newcommand{\xmm}{\textit{XMM-Newton}\xspace}
\newcommand{\nustar}{\textit{NuSTAR}\xspace}
\newcommand{\integral}{\textit{INTEGRAL}\xspace}
\newcommand{\rxte}{\textit{RXTE}\xspace}
\newcommand{\swiftbat}{\textit{Swift}/BAT\xspace}
\newcommand{\swift}{\textit{Swift}\xspace}
\newcommand{\maxi}{\textit{MAXI}\xspace}

\newcommand{\logns}{$\log N$-$\log S$\xspace}
\newcommand{\lognl}{$\log N$-$\log L$\xspace}
\newcommand{\gamcas}{$\gamma\,\mathrm{Cas}$\xspace}

\definecolor{myBlue}{rgb}{0,0.5,1}

\usepackage[normalem]{ulem}

\title{The low luminosity end of Galactic HMXBs with \ero}
\subtitle{Evidence for sustained accretion in BeXRBs}

\author{A.~Zainab\inst{\ref{inst:fau}\corrauth{aafiazainab.ansar@fau.de}}, P.~Thalhammer\inst{\ref{inst:fau}}, N.~Zalot\inst{\ref{inst:fau}} \email{nicolas.zalot@fau.de},  A.~Avakyan\inst{\ref{inst:tue}}, J.~Stierhof\inst{\ref{inst:fau}} \email{jakob.stierhof@fau.de}, 
E.~Sokolova-Lapa\inst{\ref{inst:fau}} \email{ekaterina.sokolova-lapa@fau.de}, V.~Doroshenko\inst{\ref{inst:omega}}, V.~Grinberg\inst{\ref{esa-estec}}, 
P.~Kretschmar\inst{\ref{esa-esac}} \email{Peter.Kretschmar@esa.int}, G.~Lipunova\inst{\ref{inst:fau}} \email{galina.lipunova@fau.de}, N.~Islam\inst{\ref{inst:manipal}}, 
M.~R.~Schreiber\inst{\ref{inst:utfsm}},
C.~Kirsch\inst{\ref{inst:fau}} \email{christian.ck.kirsch@fau.de}, S.~H\"ammerich\inst{\ref{inst:fau}}\email{steven.haemmerich@fau.de}, P.~Weber\inst{\ref{inst:fau}} \email{philipp.ph.weber@fau.de}, R.~Ballhausen\inst{\ref{inst:godd},\ref{inst:umcp}} \email{ballhaus@umd.edu}, 
A.~Rouco~Escorial\inst{\ref{inst:star}}, 
J.~Coley\inst{\ref{inst:godd},\ref{inst:how}} \email{joel.coley@Howard.edu}, R.~Rothschild\inst{\ref{inst:ucsd}} \email{rrothschild@ucsd.edu}, 
K.~Pottschmidt\inst{\ref{inst:godd},\ref{inst:umcp}}\thanks{deceased 17 June 2025}, 
J.~Wilms\inst{\ref{inst:fau}} \email{joern.wilms@sternwarte.uni-erlangen.de}}

\authorrunning{author}

\institute{Dr.\ Karl-Remeis Sternwarte and Erlangen Centre for Astroparticle Physics, Friedrich-Alexander Universit\"at Erlangen-N\"urnberg, Sternwartstr.~7, 96049 Bamberg, Germany \label{inst:fau}
\and
Universit\"at T\"ubingen, Institut f\"ur Astronomie und Astrophysik T\"ubingen, Sand 1, 72076 T\"ubingen, Germany \label{inst:tue}
\and
Omega Lambda Tech GmbH, Germany \label{inst:omega}
\and 
European Space Agency (ESA), European Space Research and Technology Centre (ESTEC), Keplerlaan 1, 2201 AZ Noordwijk, The Netherlands\label{esa-estec}
\and
European Space Agency (ESA), European Space Astronomy Centre (ESAC), Camino Bajo del Castillo s/n, 28692 Villanueva de
la Ca{\~n}ada, Madrid, Spain \label{esa-esac}
\and
Manipal Centre for Natural Sciences, Manipal Academy of Higher Education, Manipal, Karnataka 576104, India
\label{inst:manipal}
\and
Departamento de F{\'i}sica, Universidad T{\'e}cnica Federico Santa Mar{\'i}a, Av. Espa{\~n}a 1680, Valpara{\'i}so, Chile
\label{inst:utfsm}
\and
NASA Goddard Space Flight Center, Astrophysics Science Division, Greenbelt, MD 20771, USA
\label{inst:godd}
\and
University of Maryland College Park, Department of Astronomy, College Park, MD 20742, USA
\label{inst:umcp} 
\and
Starion Espa\~{n}a S.L.U, calle Chile 10, oficina 247, 28290 Las Rozas, Madrid, Spain \label{inst:star}  
\and
Department of Physics and Astronomy, Howard University, Washington, DC 20059, USA
\label{inst:how}
\and
Department of Astronomy and Astrophysics, University of California, San Diego, 9500 Gilman Dr., La Jolla, CA 92093-0424, USA
\label{inst:ucsd} 
}

\begin{document} 
\abstract{
We present a first look at the Galactic population of heretofore known HMXBs as observed by \textit{SRG}/eROSITA during its first four surveys. \ero's sensitivity of $\sim10^{-13}\mathrm{erg}\,\mathrm{s}^{-1}\,\mathrm{cm}^{-2}$ ,
translating to $10^{32}$--$10^{34}\,\mathrm{erg}\,\mathrm{s}^{-1}$ in luminosity
for most known HMXBs in the Milky Way, has thus far never been reached by any wide-area survey instrument. We present the extended \lognl distribution of known HMXBs reaching down to $10^{32}\,\mathrm{erg}\,\mathrm{s}^{-1}$ using \ero, and show the large scatter that can be induced by source intrinsic variability. We present sub-type resolved luminosity distributions, showing that the Supergiant X-ray binaries (SgXBs) and Be X-ray binaries (BeXRBs) occupy different parts of the overall distribution, and reanalyse \rxte/ASM and \maxi data for comparison to \ero. The luminosity regime uncovered by \ero allows a systematic study of the ``transient'' BeXRBs, which are typically below the detection threshold of monitors outside of outburst, and whose low luminosity behavior has been a longstanding question. Signatures of stable accretion at low luminosities have been observed with pointed instruments for a fraction of the overall sample, so far. With the eROSITA results, we posit that accretion outside of outburst is likely the norm, since a vast majority (${>}80$\%) of BeXRBs are detected at luminosities at least an order of magnitude higher than expected for the most X-ray luminous Be stars. We discuss the observed luminosity in the context of cold disk accretion and the ``propeller'' mechanism, with considerations of cooling emission from the neutron star. We highlight a small subpopulation of ``isolated'' Be-stars that reach X-ray luminosities comparable to the least luminous BeXRBs, hinting at the presence of companions.}

\maketitle
\nolinenumbers

\section{Introduction}

High Mass X-ray Binaries (HMXBs) have been a subject of extensive research since the advent of X-ray astronomy. As individual sources, they serve as laboratories for accretion onto compact objects, mainly but not exclusively, highly magnetised neutron stars, since the vast majority of HMXBs have been found to host neutron stars \citep{fortin2023,neumann2023}. As populations, they serve as an important tracer of Galactic star formation rate \citep{grimm2002,persic2004,mineo2011,mineo2014}, and studying their (sub-)populations is crucial for constraints on stellar and binary evolution scenarios.

HMXBs are divided into two main subclasses, based on the spectral and luminosity types of the companion to the compact object. In the predominantly persistent Supergiant X-ray Binaries (SgXB),  the donor is a massive OB-type supergiant star. In the predominantly transient Be X-ray Binaries (BeXRB), the donor is a main-sequence Be star with a circumstellar decretion disk giving rise to emission lines in its optical spectrum  \citep[hence the `e'; see e.g.,][]{reig2011}. SgXBs are further subdivided in two main categories: ``Classical'' SgXBs are persistent sources detected at stable luminosities of a few ${\sim}10^{35}$--$10^{36}\mathrm{erg}\,\mathrm{s}^{-1}$, while Supergiant Fast X-ray Transients (SFXTs) are markedly different in their variability behavior \citep{sidoli2018}, spending most of their time in quiescence and increasing in luminosity by up to five orders of magnitude during short flaring states. BeXRBs, apart from a handful displaying atypically low dynamic range \citep[e.g.,][]{reig1999}, have been studied as outbursting transients \citep{reig2011}. In addition, there is a small number of HMXBs characterised by extremely strong obscuration, called obscured HMXBs, typically associated with an sgB[e] companion star \citep[see e.g.,][]{bartlett2019}. 

The overall HMXB population can be described using X-ray luminosity functions (XLF).  While XLFs have been obtained in both soft and hard energy bands down to ${\sim}10^{35}\mathrm{erg}\,\mathrm{s}^{-1}$ based on \rxte/ASM and \integral data \citep{grimm2002,lutovinov2013}, the lower luminosity end in these studies was largely unconstrained due to the limited sensitivity of the available data  \citep{doroshenko2014}. The sensitivity has improved with later all-sky monitors like \swiftbat \citep{cusumano2010,voss2010a} and MAXI \citep{matsuoka2009}, which due to the nature of their measurement strategies, primarily capture outbursting sources but can extend down to a few $10^{34}\,\mathrm{erg}\,\mathrm{s}^{-1}$ for nearby systems. Given that HMXBs are highly variable sources \citep{sidoli2018}, extending the distribution to lower luminosities is necessary to characterize the true population. 

This naturally begs the question: which systems populate the lower luminosity end of the HMXB distribution? Typical supergiant systems can be observed for most of their orbit at a mostly stable luminosity \citep{sidoli2018}. Thus, reduced accretion scenarios are deemed especially important in the case of the ``transient'' BeXRBs and SFXTs. As a predominant subclass, the behavior of BeXRBs outside of outbursts is particularly relevant to the lower luminosity end of the XLF. Outside of their luminous outbursts, they were often considered to be in ``off'' states, where accretion onto the neutron star ceases. This state was attributed to either a lack of matter or to the ``propeller regime'', where centrifugal forces eject all or most of the accretion flow from the system \citep{illarionov1975}. 

Over the last decade more and more evidence has accrued for individual BeXRBs showing sustained accretion at low mass accretion rates and luminosities \citep[e.g.,][]{tsygankov2017,ballhausen2017,doroshenko2022}. This behaviour is only known for ${\sim}$20\% of the known Milky Way sample of HXMBs, and its universality is yet to be tested systematically using an unbiased sample \citep[see][for previous studies of BeXRBs at low luminosity]{tsygankov2017a, zalot2026}. On the other hand, some sources show behaviour more in agreement with a ``propeller'' effect \citep{roucoescorial2017,roucoescorial2020}, albeit much fewer, in comparison to the 20\% that accrete outside of outbursts. SFXTs also benefit from sensitive unbiased observations since they are typically observed during flares, which last about 100--1000\,s each, exhibiting low duty cycles and high dynamic ranges. \citet{bozzo2015} have previously characterised their low luminosity behaviour using \integral data and showed that they inhabit a lower luminosity space in comparison to SgXBs. Evaluating the degree to which accretion in these systems is inhibited at low luminosity and testing its predicted effect on the XLF are crucial to this work. 

The extended ROentgen Survey Imaging Telescope Array \citep[eROSITA,][]{Merloni12,predehl2021} on board the Spectrum-Roentgen-Gamma \citep[\textit{Spektr-RG}, \textit{SRG};][]{sunyaev2021a}, whose all-sky survey capability and improved sensitivity compared to its predecessor \citep[\rosat,][launched in 1990]{boller2016}, is particularly suited for the task. Especially for sources with very rare outbursts or flares, the deep sensitivity of \ero allows detection down to much lower luminosities, enabling for the first time a study of transient HMXB populations in individual snapshots, in contrast to averaged data. Finally, as an all-sky survey, \ero offers an unbiased view of the sample, and allows for observations of sources at varying luminosity and spectral states. \ero has been predicted to increase the number of known HMXBs in the Galaxy by a factor two \citep{doroshenko2014}. Confirmation of the candidate HMXBs is pending detailed follow-up studies and is part of ongoing work. Here, we focus only on the hitherto known sample of HMXBs, whose characterisation at low luminosity is imperative to identify faint HMXBs in the survey. 

 \begin{figure}
    \centering
    \includegraphics[width=0.45\textwidth]{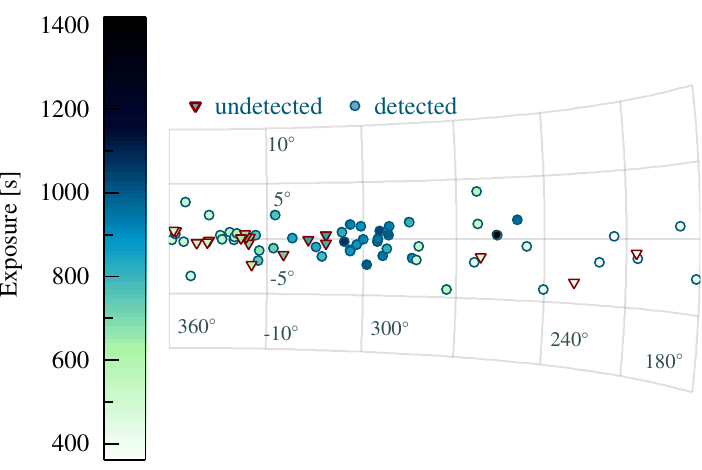}
    \caption{Known HMXBs in the Western hemisphere depicted with the effective exposure time at their corresponding sky position, shown in a Hammer projection of the Galactic plane. All sources in our sample are within $5^\circ$ latitude of the galactic plane. They each get effective exposure of ${\sim}400$--$1400\,\mathrm{s}$ in the combined eRASS1:4 data.}
    \label{fig:HMXBs_plane}
\end{figure}

The paper is structured as follows. First, we present in Sect.~\ref{sec:sample} the compilation of the source sample from existing catalogs to produce the corresponding \ero catalog and the extraction of \ero data products for each source. We then analyze the resulting distribution of fluxes and derived luminosities in Sect.~\ref{sec:fluxes}, followed by a comparison to previous instruments with \rxte/ASM and \maxi in Sect.~\ref{sec:previnst}. For the rest of the manuscript, we focus primarily on the BeXRBs, since they dominate the luminosity regime uncovered by \ero, and briefly investigate the isolated Be star population for comparison to the BeXRBs in Sect.~\ref{sec:bestar}. The implications of our results for BeXRB behaviour at low luminosity are discussed in Sect.~\ref{sec:lowlum}, along with a brief comparison to SgXB and SFXT studies. We offer an outlook of this work on inferences for the XLF of HMXBs, in Sect.~\ref{sec:outlook-xlf}, as well as on implications for isolated Be-stars in Sect.~\ref{sec:BeStars}, before ending the article with our conclusions in Sect.~\ref{sec:conc}.

\section{Sample, Cross-matching, and Data} \label{sec:sample}
To study the population of HMXBs as observed by \ero, we need an updated HMXB catalog for the Milky Way, which we can then use to cross-match with the \ero catalog \citep{merloni2024}. As part of the German (\emph{eROSITA\_DE}) consortium, within the data rights agreement we have access only to the Western Galactic hemisphere (Western hemisphere, hereafter), defined to the west of Sgr A$^{*}$, $l_{\mathrm{II}}\,{\simeq0}^{\circ}$. We therefore present results solely from this region. 

\subsection{HMXBs: source sample and catalog creation}
We first created our catalog based on \textit{XRBcats}, published and kept up to date by \citet{neumann2023}, taking into account the results of \citet{fortin2023}, and making further changes and updates where appropriate. The changes mainly pertain to accurate classification of the individual sources into the subcategories of HMXBs, and are further elucidated upon in Appendix~\ref{app:catalog}. Filtering \textit{XRBcats} HMXB catalog for the Western Galactic hemisphere resulted in 83 sources. Of those, our revised literature-based categorisation leads to a catalog of 72 sources, with subtypes as listed in Table~\ref{tab:ero_numbers}.

\subsection{Cross-matching with \ero}
The scanning strategy of \ero entails covering the entire sky in six months, and is achieved by a shifting rotation axis in 1\degr\,$\mathrm{day}^{-1}$ increments \citep{predehl2021,merloni2024}. Each point in the sky is observed ${\sim}$6\,times for ${\lesssim}$40\,s at a time, over the course of the day. At the ecliptic poles, each point is scanned more than six times, as these are where the great circles of each individual scan intersect. For the galactic plane, where HMXBs primarily reside, the effective exposure in a single sky survey varies between ${\sim}$100--350\,s (cumulatively depicted in Fig.~\ref{fig:HMXBs_plane}). Within the German consortium, each eRASS\footnote{eRASS refers to a single sky survey. eRASS1--4 refers to each eRASS analysed individually, while eRASS:4 refers to combined results from stacking all four eRASSes.} has a corresponding catalog of detected sources. 

We performed a simple geometric matching between the cleaned HMXB catalog and the eRASS1 main catalog \citep{merloni2024}, with an angular separation cutoff of ${\sim}$15'', which we converged on after conducting cross-matching with various values of angular separation up to a maximum of 15'' \citep[see, e.g.,][for works applying  the same criterion]{boller2025}. Further, we applied a cut on the detection likelihood assigned by eSASS, \texttt{DET\_LIKE}, and excluded all sources with $\texttt{DET\_LIKE}<10$ as spurious detections. Using SIMBAD queries and visual inspection, we cross checked each resulting match for duplicate detections (in case the sources were situated close by) and contaminating neighboring sources. 

The sources which did not fulfill the above criteria were considered not detected, and were flagged as such, to facilitate computation of flux upper limits. The detected and non-detected sources were separated for ease of data reduction. We repeated this procedure with the subsequent eRASS catalogs. eRASS1--3 catalogs are now publicly available \citep{ramos-ceja2026}, while eRASS4 is only accessible internally\footnote{For eRASS4, we used the catalog processed using \texttt{020} on 2023 December 12 and includes positional and photometric corrections.}. After cross-matching of our compilation of HMXBs with the eRASS catalogs, we find \ero detections of 41--47 sources of the total 72 sources in the Western hemisphere, in eRASS1--4. The distribution of the detections between subclasses was as listed in Table~\ref{tab:ero_numbers}.

\begin{table}
\caption{HMXBs in the updated catalog for the Western hemisphere, and those detected by \ero in each sky survey, divided by subclass.}
\renewcommand{\arraystretch}{1.32} 
    \renewcommand{\tabcolsep}{2mm}
    \centering
    \begin{tabular}{cccccccc}
    \hline
    \hline
    & N & BE & SG & SF & RSG & MQ & Candidate/ \\
    & & & & & & & Confused \\
    \hline
    Total & 72 & 33 & 24 & 7 & 1 & 1 & 4/3 \\
    \hline
    eRASS1 & 44 & 21 & 14 & 5 & 1 & 1 & 0/2  \\
    eRASS2 & 46 & 24 & 14 & 5 & 1 & 0 & 0/2  \\
    eRASS3 & 47 & 25 & 13 & 4 & 1 & 1 & 1/2  \\
    eRASS4 & 41 & 20 & 10 & 6 & 1 & 1 & 1/2  \\
    \hline

    \end{tabular}
    \label{tab:ero_numbers}
\end{table}

\subsection{\ero Data Reduction} \label{sec:ero}

To measure fluxes we extracted data products using eSASS version 211214 \citep{brunner2022}, processing version 020. We used different methods of extraction depending on whether or not the source was detected and part of the eRASS1B catalog \citep{merloni2024}. 

For each detected source, we extracted data products using eSASS' \texttt{CATPREP} option, which allows catalogs formatted for eSASS to be input as source information. The input catalog must contain information on the coordinates ($\alpha$,$\delta$; ICRS) of the sources, their corresponding source and background count rates as detected by \ero in the relevant energy band (\texttt{ML\_CTS\_0}, and \texttt{ML\_BKG\_0} for band 0, in this case) and the specific identifier assigned to each \ero detected source (\texttt{DETUID}). For non-detections, we used source regions centered on the source positions obtained from the HMXB catalog, with fixed radii of ${\sim}10$'' and background regions defined as annuli around the source, with fixed outer radii of ${\sim}50$''. 

\section{Flux and Luminosity Distributions} \label{sec:fluxes}
\subsection{Spectral fitting and  flux determination}
\label{subsec:fluxes}
For sources detected with $\gtrsim$50\,counts, we used the extracted \ero spectra for spectral analysis. We described the spectra using the two most basic models typically used for HMXBs in the soft X-ray band \citep[see e.g.,][]{delia2013}, an absorbed power law (\texttt{TBabs*powerlaw} in \texttt{ISIS} notation) and an absorbed black body (\texttt{TBabs*bbody} in \texttt{ISIS} notation). For the fit, we allowed the absorption to vary around $N_{\mathrm{H}}$ obtained from XRBcats \citep{neumann2023},\footnote{The catalog lists absorption from fits to \swift/XRT data of Galactic sources \citep{evans2020}.} if available, or around the Galactic column from the HI4PI survey \citep[][applying the \citet{willingale2013} correction for X-rays]{bekhti2016}, if not. As intrinsic absorption can be highly variable, often luminosity-dependent, for instance as a consequence of ionization \citep[like in highly obscured systems;][]{sidoli2022}, and hard to constrain for many sources, we did not include a further partial covering absorption component beyond the line of sight absorption for a consistent treatment of the sources. We visually inspected the fits for any peculiarities with diagnostics such as that shown in Fig.~\ref{fig:eg_spec} and compared C-stat values for the different models. We then computed corresponding fluxes for the best fit values using the convolution model \texttt{cflux} in \texttt{ISIS\footnote{\url{https://space.mit.edu/cxc/software/isis/manual.pdf}}}. 

\begin{figure*}
    \centering
    \includegraphics[width=0.425\textwidth]{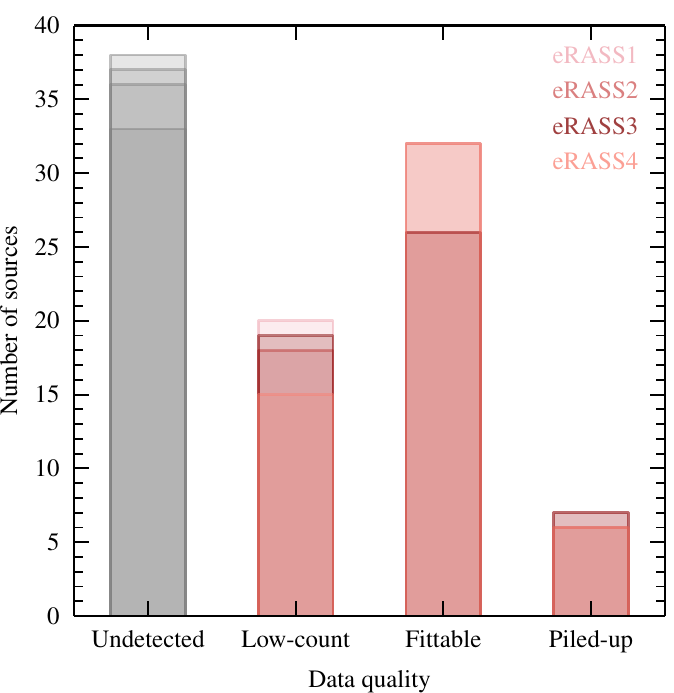}
    \includegraphics[width=0.45\textwidth]{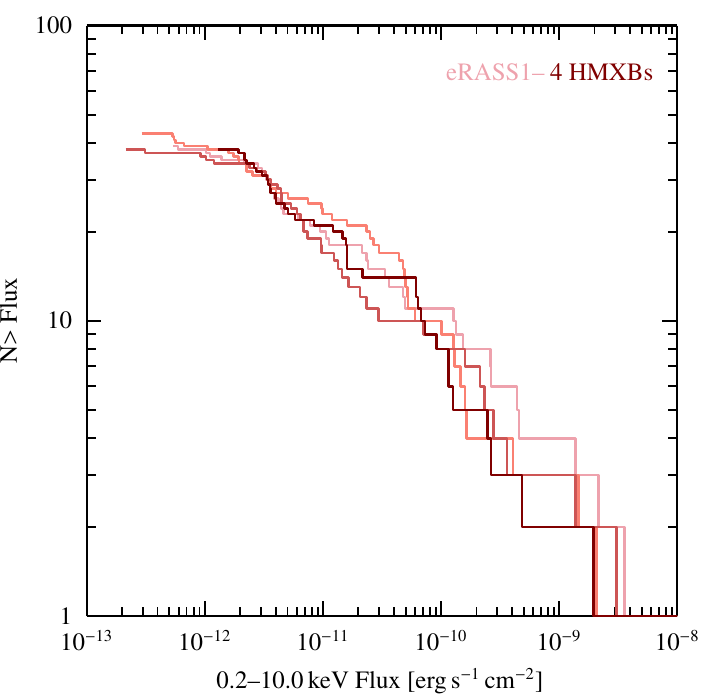} 
    \caption{\textbf{Left:} The distribution of spectra available for each eRASS and their corresponding treatment. Spectra with $\leq1000$\,counts and $\geq50$\,counts were used for spectral fitting, those with $\gtrsim1000$\,counts were treated for pile-up, and for the remainder, the parameters were fixed (at $\Gamma=1$ or $kT=1\,\mathrm{keV}$) for flux estimation. \textbf{Right:} The \logns distribution computed using \ero fluxes obtained according to Sect.~\ref{subsec:fluxes} in eRASS1:4. The \ero \logns distribution reaches down to fluxes of $10^{-13}\,\mathrm{erg}\,\mathrm{cm}^{-2}\,\mathrm{s}^{-1}$, which is where \ero is complete up to 99\% \citep{weber2026}.}  
    \label{fig:ero_logns}
\end{figure*}

Especially bright sources with spectra containing $\gtrsim$1000 counts needed to be checked for pile-up. We used the Simulation of X-ray Telescopes \citep[SIXTE;][]{dauser2019} software package, which has been used successfully to model pile-up in eROSITA detectors \citep[e.g.,][]{konig2022}, and found that 1\% pile-up appears at count rates of 2.5\,$\mathrm{cts}\,\mathrm{s}^{-1}$ so that some especially bright sources are affected by up to 5\% pile-up. We re-extracted the 6--7 sources in each eRASS with very high count-rates by excising the core and generating annuli regions with an inner radius scaled according to the observed count-rate and subsequent pile-up fraction. We modelled the resulting spectra as above. The difference between the uncorrected and corrected distributions for eRASS1 is shown in Fig.~\ref{fig:pileup}, 

To estimate fluxes for sources with spectra ${\lesssim}$50 counts, we fixed parameters at $\Gamma=1$ for the absorbed power law and $kT=1\,\mathrm{keV}$ for the black body model. For highly absorbed sources we used the absorption value available from the literature as summarized by \citet{neumann2023} to avoid unrealistically low flux upper limits. We estimated upper flux limits for non-detected sources by fixing the parameters at the above values and taking the 90\% upper confidence bound of the flux measurement. These are documented in the catalog. The distribution of the spectra over the afore-mentioned categories is depicted in the left panel of Fig.~\ref{fig:ero_logns}. 

\begin{figure}
    \centering
    \includegraphics[width=\linewidth]{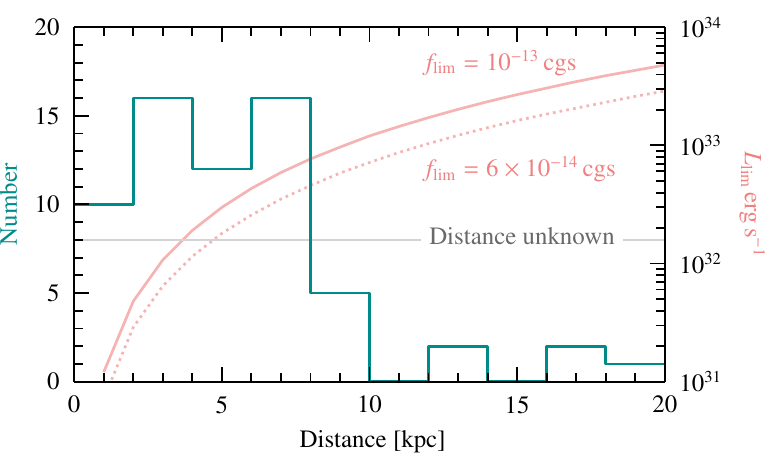}
    \caption{Left hand axis and histogram: Number of sources in a given distance range (teal) and number of sources with unknown distances (grey). Right hand axis: Limiting luminosity to which \ero is complete within a sensitivity limit of $10^{-13}\mathrm{erg}\,\mathrm{s}\,\mathrm{cm}^{-2}$ marked for the 0.2--10.0\,keV (solid) and the 0.2--2.3\,keV (dotted) bands. \ero is complete down to a 0.2--10.0\,keV luminosity of $10^{33}\,\mathrm{erg}\,\mathrm{s}^{-1}$ within ${\sim}10\,\mathrm{kpc}$.}
    \label{fig:dist}
\end{figure}

The fluxes are shown in Fig.~\ref{fig:ero_logns} in a \logns distribution, restricted to sources with detection likelihood $\texttt{DET\_LIKE}\gtrsim10$. In a single survey, \ero's sensitivity at places with the least effective exposure is ${\sim}10^{-13}\,\mathrm{erg}\,\mathrm{cm}^{-2}\,\mathrm{s}^{-1}$ \citep{merloni2024}, such that we are effectively complete with respect to the initial sample above this limit. This implies that any non-detections above this limit are more likely source-intrinsic. The luminosity and distance covered with this sensitivity are shown in Fig.~\ref{fig:dist} along with the number of sources detected at each distance, with the sensitivity limits for energy ranges 0.2--10.0\,keV and 0.2--2.3\,keV marked. 

\subsection{The HMXB $\log N$-$\log S$ as measured with \ero}
\begin{figure*}
    \centering
    \includegraphics[width=0.48\textwidth]{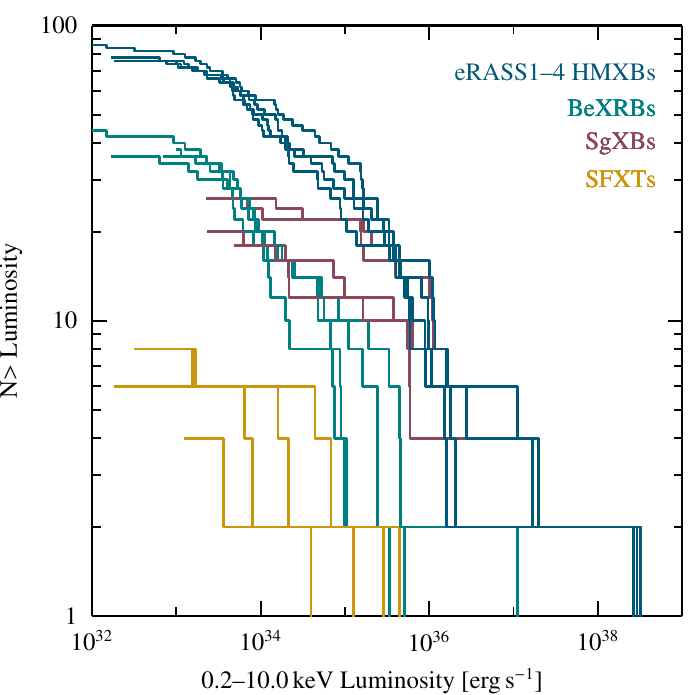}
    \hfill
    \includegraphics[width=0.48\textwidth]{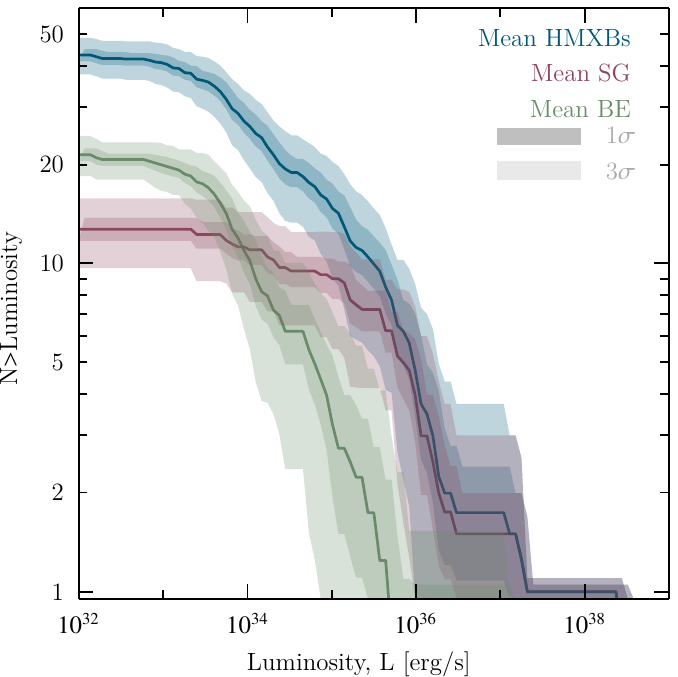}
    \caption{\textbf{Left:} The \lognl distributions of HMXBs and their subclasses as detected by \ero in eRASS1 to eRASS4, multiplied by 2 to project to the whole sky. The subclasses show marked variability in eRASS1--4, depending on the flaring states of individual sources. \textbf{Right:} The fits to eRASS1--4 data of the HMXB sample, as well as the BeXRB and SgXB subclass samples, shown along with 1000 iterations of \lognl distributions computed by sampling randomly from an eRASS observation for each source. The intrinsic source variability results in significant scatter of the overall HMXB distribution, while there is an evident distinction in the luminosity range occupied by BeXRBs and SgXBs.}  
    \label{fig:ero_lum_types}
\end{figure*}
We computed luminosity distributions in the 0.2--10.0\,keV band for eRASS1--4 based on distances from Gaia Data Release 3 \citep[][following \citealt{bailer-jones2021b}]{collaboration2023a}. For systems without reliable Gaia distances we used distances as listed in \textit{XRBcat}. 

We present the luminosity distributions 
of HMXBs and their subtypes measured in
each eRASS in Fig.~\ref{fig:ero_lum_types}. 
The variability in the subclass luminosity distributions contributes to the variance observed in the overall HMXB distributions. The higher end of the luminosity function is dominated by the classical SgXBs, while the lower luminosity range is dominated by BeXRBs. We did not find any BeXRBs at luminosities above $10^{37}\mathrm{erg}\mathrm{s}^{-1}$, the typical luminosity reached during Type~II outbursts \citep{okazaki2001}, while four systems were caught during Type I outbursts recorded by \maxi or \swiftbat. SFXTs on the other hand have been sampled both during flares and at low states.

As the eROSITA HMXB \lognl distributions show a clear turn-over (Fig.~\ref{fig:ero_lum_types}), based on previous studies \citep[see e.g.,][]{lutovinov2013,voss2010a} we empirically describe the eRASS1--4 distributions for all HMXBs, and for the SgXBs and BeXRBs, with a broken power law of the form 
\begin{equation}
\label{eq:bknplaw}
A(L_\mathrm{X}) =  \left \{ \begin{array}{lll}
          K L_\mathrm{X,34}^{-\Gamma_1} & \mbox{for $L_\mathrm{X} \leq L_\mathrm{break}$} \\
        \\
          K \left(L_\mathrm{break}/10^{34}\,\mathrm{erg}\,\mathrm{s}^{-1}\right)^{\Gamma_2-\Gamma_1} L_\mathrm{X,34}^{-\Gamma_2} &
          \mbox{otherwise} \\
                \end{array}
        \right.
\end{equation}
where $L_\mathrm{X}$ is the X-ray luminosity, $L_{\mathrm{X},34}=L_\mathrm{X}/10^{34}\,\mathrm{erg}\,\mathrm{s}^{-1}$, $\Gamma_1$ and $\Gamma_2$ are the two slopes of the distribution, and $L_{\mathrm{break}}$ is the break luminosity. The normalisation constant, $K$,  corresponds to the number of sources above $10^{34}\,\mathrm{erg\,s}^{-1}$. Best-fit parameters are listed in Table~\ref{tab:fit_results}, while the Kolmogorov-Smirnoﬀ test (KS test) probabilities computed by determining the maximum difference between the distribution and the applied model are given in Table~\ref{tab:kstest}. To estimate uncertainties, we created 1000 luminosity distributions by randomly assigning a flux value from a Gaussian distribution around the confidence bounds of the original flux measurement from each eRASS. We report the $2\sigma$ percentiles computed from the distributions of each of the fit parameters. 

\begin{table}
\caption{Fit results for the phenomenological description of the \lognl distributions.}
\renewcommand{\arraystretch}{1.32} 
    \renewcommand{\tabcolsep}{1mm}
    \centering
    \begin{tabular}{ccccc}
Sample & Norm ($K$)  & $\Gamma_{1}$ & $\Gamma_{2}$ & $L_{\mathrm{break}}$ \\ 
& & & & $\times 10^{34}\mathrm{erg}\,\mathrm{s}^{-1}$ \\
         \hline
HMXBs  & & & \\
\hline
\rxte/ASM  & 5 & 0.61 & -- & -- \\
eRASS1 & $23.6^{+2.1}_{-0.8}$ & $0.21\pm0.04$ & $0.55^{+0.45}_{-0.01}$ & $9.24^{+23.15}_{-0.07}$ \\
eRASS2 & $27.2^{+1.1}_{-1.0}$ & $0.15^{+0.04}_{-0.02}$ & $1.04\pm0.11$ & $13.7^{+7.1}_{-0.8}$ \\
eRASS3  & $30^{+7}_{-6}$ & $0.07^{+0.21}_{-0.04}$ & $0.44^{+0.14}_{-0.04}$ & $0.48^{+6.37}_{-0.07}$ \\
eRASS4 & $23.6^{+1.1}_{-1.5}$ & $0.24\pm0.04$ & $1.54^{+0.15}_{-0.28}$ & $40.2^{+13.7}_{-1.2}$ \\
 \hline
SgXBs & & & \\
\hline
eRASS1 & $11.7^{+0.4}_{-1.3}$ & $0.09^{+0.05}_{-0.05}$ & $0.57^{+1.05}_{-0.03}$ & $29^{+58}_{-5}$ \\
 eRASS2  & $12.5^{+1.0}_{-1.1}$ & $0.15^{+0.04}_{-0.06}$ & $2.6^{+0.7}_{-1.9}$ & $78^{+13}_{-63}$ \\
 eRASS3  & $9.8^{+1.0}_{-1.4}$ & $0.17\pm0.08$ & $0.41^{+8.60}_{-0.09}$ & $9^{+48}_{-8}$ \\
eRASS4 & $7.62^{+1.68}_{-0.12}$ & $0.13^{+0.09}_{-0.01}$ & $0.52^{+4.64}_{-0.01}$ & $8.7^{+156.2}_{-0.5}$ \\
 \hline
BeXRBs & & & \\
\hline
eRASS1 & $12.1^{+4.0}_{-2.1}$ & $0.09^{+0.13}_{-0.04}$ & $0.95^{+0.10}_{-0.23}$ & $0.37^{+0.59}_{-0.03}$ \\
eRASS2 & $13^{+5}_{-4}$ & $0.12^{+0.19}_{-0.07}$ & $0.60^{+2.98}_{-0.15}$ & $0.5^{+15.9}_{-0.4}$ \\
eRASS3 & $10.8^{+7.6}_{-1.7}$ & $0.27^{+0.19}_{-0.23}$ & $1.0^{+8.0}_{-0.4}$ & $1.6^{+8.0}_{-1.3}$ \\
eRASS4  & $9.8^{+6.1}_{-0.9}$ & $0.29^{+0.09}_{-0.25}$ & $6.2^{+2.8}_{-5.8}$ & $43^{+7}_{-43}$ \\
 \hline
    \end{tabular}
    \label{tab:fit_results}
\end{table}

The parameters of the fits vary significantly among the four eRASS, especially $\Gamma_2$ and $L_{\mathrm{break}}$. Evidently, the overall shape depends on how many sources are flaring or in brighter states. This is further discussed in Appendix~\ref{app:fits}. To estimate the overall variation in the \lognl distribution, we created 5000 equally likely distributions. The detected fluxes of different sources can be considered independent, so in each distribution every source is randomly assigned a luminosity $L$ corresponding to its detection in one of the eRASSes. The result is shown in Fig.~\ref{fig:ero_lum_types} (right panel), and indicates a significant difference in shape of the overall luminosity distribution of HMXBs based on intrinsic variability and statistical uncertainties. The distinct luminosity regimes occupied by the subclasses are also evident. The fit parameters corresponding to the simulated distributions are shown and discussed in Appendix~\ref{app:fits}. The highest, lone luminosity bin in the SgXB sample (and the overall HMXBs) corresponds to Circinus X-1, which is unique in its young age \citep{heinz2013} and regular bursting nature \citep{linares2010}\footnote{Circinus X-1 was initially classed as a low mass X-ray binary but has since been identified to have a B5-A0 type supergiant \citep{jonker2007}, and is considered a peculiar HXMB with a weakly magnetised neutron star.}.  

\subsection{Comparison to previous $\log N$-$\log S$-measurements} \label{sec:previnst}

\begin{figure*}
    \centering
     \includegraphics[width=0.48\textwidth]{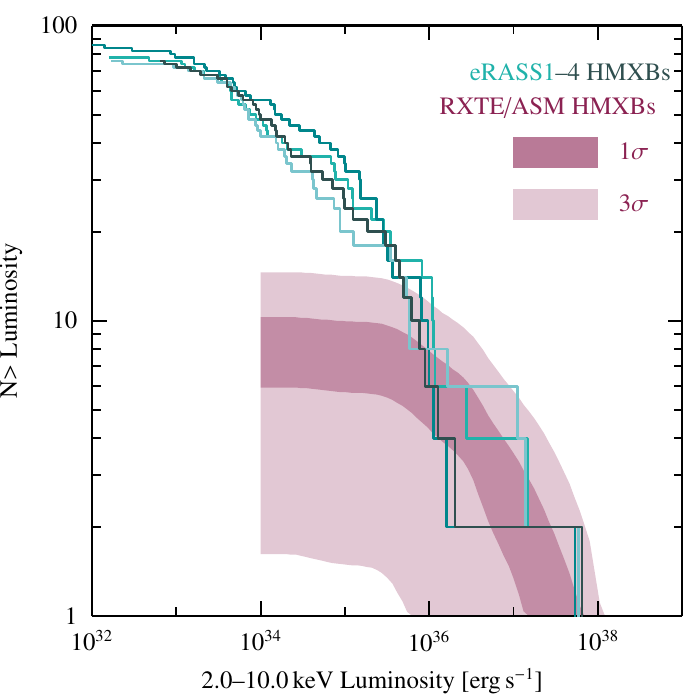}
     \hfill
      \includegraphics[width=0.48\textwidth]{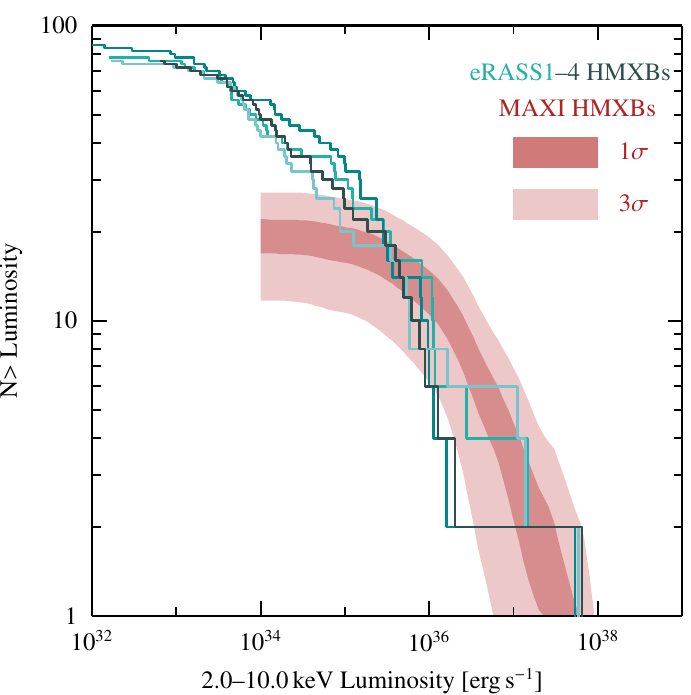}
    \caption{\rxte/ASM (left) and \maxi (right) \lognl taking variability into account by random sampling from the monitoring lightcurves of each source, along with the eRASS1--4 distributions. \rxte/ASM is sensitive up to $10^{36}\,\mathrm{erg}\,\mathrm{s}^{-1}$, while \maxi reaches a few $10^{35}\,\mathrm{erg}\,\mathrm{s}^{-1}$. Both are consistent with the \ero distribution above this threshold. The large variance of the HMXB \lognl distribution is evident in both cases.}  
    \label{fig:rxte_maxi_logns}
\end{figure*}

An important point at this juncture is that BeXRBs, which thus far have predominantly been studied as transients during bright outbursts, have been detected at low luminosities for a majority of eROSITA observations, and that the \lognl is dominated by BeXRB at lower luminosities. We discuss this in more detail in later sections of the paper. Before we do so, however, it is important to confirm that the \ero results agree with those previous studies, especially given that the methodologies of different studies of the Galactic \logns were not consistent with each other.

\citet{grimm2002} constructed the \logns of Galactic X-ray binaries in a comparable energy range to that of \ero by averaging the 1\,d binned \rxte/ASM lightcurves available at that time (see left panel of Fig.~\ref{fig:grimm_logns}). Since \citeauthor{grimm2002} average the data, however, their \logns is not representative of the \logns based on short exposures, such as those employed here (or in typical observations of other galaxies). To see whether \rxte/ASM agrees with \ero, we therefore generated 50000 iterations of \rxte/ASM \logns distributions based on selecting single random rate data points from the 1\,d binned ASM lightcurves of all known HMXBs. In this way, each source is sampled at all stages of activity, with the overall distribution displaying significant variability between individual realizations. We display the variation in terms of $3\sigma$ wide shaded regions measured from these 50000 realisations in Fig.~\ref{fig:rxte_maxi_logns} (Left). Further details of the procedure are given in Appendix~\ref{app:rxte}. The \logns measured with \rxte/ASM in this way. The shape of the \lognl distribution is consistent for the luminosity range in common between the two instruments, down to $10^{36}\,\mathrm{erg}\,\mathrm{s}^{-1}$. Beyond this point, the \rxte/ASM distributions plateau, and are severely affected by incompleteness effects due to both sensitivity and the existing source sample at the time.

In a follow up to the \rxte/ASM analysis of \citet{grimm2002},  \citet{islam2016} also took variability into account, assuming that the true count rate corresponding to each light curve bin of a source was normally distributed with the detected count rate as the mean and the ASM uncertainty as standard deviation. Our choice to sample randomly directly from the lightcurves without using a Gaussian probability distribution is based on the assumption that the \rxte/ASM lightcurves contain a sufficient number of bins to provide representative sampling of each source's luminosity function when sampling directly from their expectation values. We also found this method to be more suitable comparison with the unbiased detections by \ero, since the latter corresponds to a random snapshot of the sky, as opposed to an averaged distribution consisting of all \rxte/ASM lightcurves, regardless of activity at a specific time. Despite these differences, the power-law index describing the HMXB XLF reported by \citet{islam2016}, 0.48 with a variance of 0.19, is in agreement with our fit to \rxte/ASM with an index of 0.61 and falls between $\Gamma_1$ and $\Gamma_2$ found in the \ero fits (see Table~\ref{tab:fit_results}). The latter is not surprising, given their model is a single power law. The difference to \ero is most apparent at lower luminosities where the precise treatment of marginally detected \rxte/ASM sources introduces systematic uncertainties.

We repeated the above exercise using light curves from the currently operating monitor \maxi. Figure~\ref{fig:rxte_maxi_logns} (right) compares the \maxi luminosity distribution, compared to \ero. Once again, the \ero detections are consistent down to the luminosity below which \maxi begins to plateau. \rxte/ASM and \maxi data also consistent with each other. They both observe outbursts $\gtrsim10^{36}\,\mathrm{erg}\,\mathrm{s}^{-1}$, but do not extend significantly below $10^{35}\,\mathrm{erg}\,\mathrm{s}^{-1}$. The two instruments were active over two different time periods, with different outburst patterns allowing \rxte/ASM to observe more luminous outbursts in its time than \maxi. 
Similar to the \rxte/ASM results, Fig.~\ref{fig:rxte_maxi_logns} shows that the largest discrepancies are for low luminosities.

Finally, on resolving the \rxte/ASM and \maxi distributions by subclass (see Appendix~\ref{app:rxte}) it becomes evident that the luminosity functions measured so far have been dominated by SgXBs despite them making up a smaller fraction of the Galactic HMXBs. We thus suppose that the work based on previous instruments with sensitivity such that only luminosities $\gtrsim10^{36}\,\mathrm{erg}\,\mathrm{s}^{-1}$ are probed, misrepresents the overall HMXB slope, and has an inherent bias towards persistent SgXBs, since BeXRBs, if represented, are only observed in outbursts, which are exceedingly rare (see Fig.~\ref{fig:rxte_lum_types}). 

\section{Comparison to Be stars} \label{sec:bestar}

 The comparatively high fraction of BeXRBs detected at lower luminosity is of particular interest, since it may indicate ongoing low luminosity accretion. However, Be stars, the donors in BeXRBs, are also known to be X-ray emitters \citep{naze2014, naze2018}. To check whether the eROSITA detections of BeXRBs at luminosities $\lesssim10^{34}\,\mathrm{erg}\,\mathrm{s}^{-1}$ are due to accretion onto the neutron star, or partially originating from its donor, we need to compare the BeXRB \lognl with that of (presumably) isolated Be-stars. 

To obtain a catalog of Be stars with \ero detections in eRASS1:4, we retraced the steps undertaken by \citet{naze2023}. Specifically, we used the Be Star Spectra (BeSS) catalog \citep{neiner2011} and applied quality cuts using Gaia DR3 data \citep{collaboration2023a}, to arrive at 832 stars in the Western hemisphere. We then cross-matched this list with the eRASS1B catalog to obtain the eROSITA detected sample, leaving 170 stars (${\sim}20\%$) \citep[in agreement with][]{naze2023}. We fitted the \ero spectra of these stars with absorbed black body spectra, using eRASS:4 data to optimise for signal-to-noise ratio. We also analysed eRASS1 data, as a representative case. 

\begin{figure}
    \centering
    \includegraphics[width=0.45\textwidth]{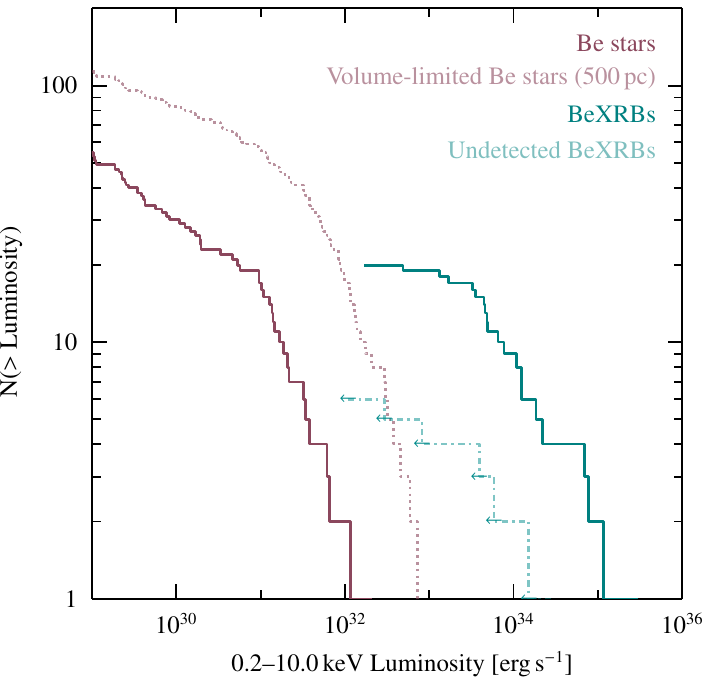}
    \caption{\ero \lognl distributions of BeXRBs (teal) shown alongside that of Be stars (violet), using eRASS1 as a representative case. A volume-limited sample of Be stars within 500\,pc (violet, dashed) is also shown in lieu of a volume-corrected distribution. BeXRBs are evidently detected at higher luminosity by at least an order of magnitude compared to the volume-limited isolated Be star sample. The luminosity upper limits on the BeXRBs not detected in any eRASS (teal, dashed) are also in excess of the volume-limited sample. There is however, a region of overlap between the least luminous BeXRBs and the most luminous Be stars.}  
    \label{fig:bex_be_comp}
\end{figure}

Similar to \citet{naze2023}, we construct a \lognl distribution from the \ero detections of all of these sources. In addition, again following \citet{naze2023}, to avoid a bias towards anomalously luminous X-ray-bright Be stars, we also construct a less biased sample by limiting the Be-\lognl to a volume limited to within 500\,pc. The \lognl of the detected BeXRBs shows that a vast majority of the BeXRBs detected by \ero have significant excess luminosities compared to those of all Galactic Be stars with  X-ray detections, hinting at accretion being the norm at low luminosity (Fig.~\ref{fig:bex_be_comp}). Compared to the volume limited Be-star \lognl, which ends at ${\sim}10^{32}\,\mathrm{erg}\,\mathrm{s}^{-1}$, we additionally find that the BeXRBs detected by \ero are more X-ray luminous than expected for their Be donors, and the undetected systems also have upper limits above this threshold. However, there is a region of overlap between the total Be-star sample and the BeXRBs in the luminosity range $10^{32}$--$10^{33}\,\mathrm{erg}\,\mathrm{s}^{-1}$ where disentangling the most luminous Be stars from low luminosity BeXRBs becomes important. 

 \begin{figure}
    \centering
    \includegraphics[width=0.45\textwidth]{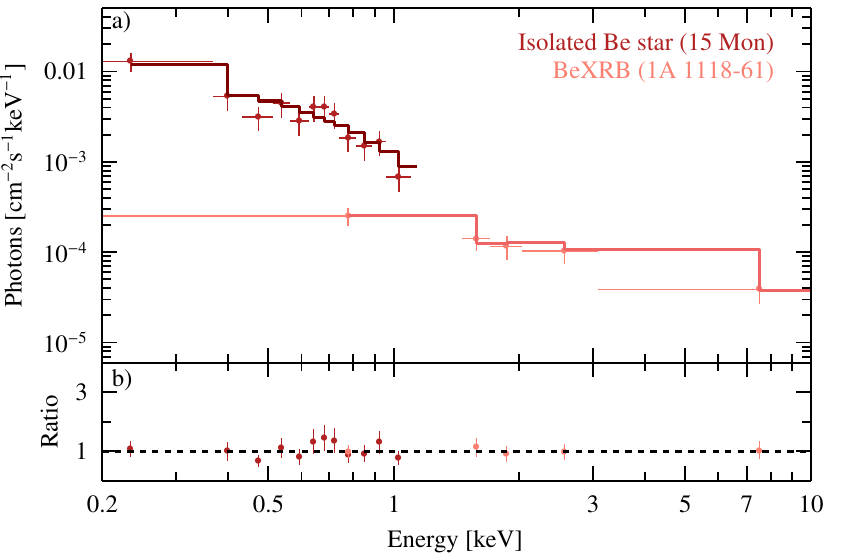}
    \caption{Comparison of a typical spectrum of an isolated Be star, and a BeXRB at low luminosity. The Be star spectrum is purely thermal and drops off after 1\,keV. The BeXRB (1A~1118$-$61) is found here in the stable low luminosity state, and its spectrum is significantly harder, and prominent above 1\,keV. }    
    \label{fig:comp_spec}
\end{figure}

 In the event of sufficient data, the two scenarios -- an isolated Be star and a ``quiescent'' BeXRB -- can be reasonably discerned based on their X-ray spectral shapes (Fig.~\ref{fig:comp_spec}). Since a majority of the Be star sample spectra are unsuitable for spectral modelling, however, we focus on their luminosity in soft and hard energy ranges \citep[see also][for a more detailed look including hardness ratios]{naze2023}. We find that a handful of systems known to be hitherto ``isolated'' are brighter than the majority of them, especially in the hard band above 2\,keV, and at comparable luminosity to the least luminous BeXRBs. In their overall 0.2--10.0\,keV flux -- corresponding to the luminosity regime $10^{32}$--$10^{33}\,\mathrm{erg}\,\mathrm{s}^{-1}$ --  Be stars and BeXRBs show significant overlap (shown in Fig.~\ref{fig:flux_lum}, left). However,  in the hard band, the Be stars have significantly less flux compared to BeXRBs as evidenced by the shifting of the Be data points in Fig.~\ref{fig:flux_lum} (right). 
 Further, the regime straddled by low luminosity BeXRBs and these unusually bright Be stars is also populated by a handful of \gamcas systems, as well as two Be+sdOB systems, which can also display X-ray emission up to $10^{32}\,\mathrm{erg}\,\mathrm{s}^{-1}$ \citep{naze2022}. Colliding wind binaries can also show up at these luminosities, but can be easily discernible at other wavelengths \citep[e.g.,][]{blomme2024, h.e.s.s.collaboration2020}. 

 To probe their hard X-ray emission and ascertain the type of system, a selected sample of the Be stars showing hard X-ray emission are part of an ongoing campaign with \nustar initiated by us (Fig.~\ref{fig:flux_lum}, triangles). So far, LV Mus has already been observed by \nustar and shows a power law tail,  not expected for isolated Be stars, and potentially indicative of a neutron star companion. Another, Cl Pismis 17 3, has also been observed by \nustar. The observation revealed a lower flux than expected, and is further plagued by stray light, such that its behaviour cannot currently be ascertained without future work. The flux upper limit is still consistent with \gamcas behaviour. Further details will be provided in a future work (Zainab et al. in prep). 

\section{What are BeXRBs doing at low luminosities?}\label{sec:lowlum}

Given that \ero has detected a high fraction of BeXRBs (${\sim}$60\%) at luminosities below $10^{35}\,\mathrm{erg}\,\mathrm{s}^{-1}$, and that they are significantly more luminous than the average Be-star population, we now discuss the implications of their low luminosity detections. We start by describing the detections of BeXRBs observationally, primarily focusing on their luminosity, before summarising the possible physical scenarios that could support the detections. We then offer a brief comparison to studies of SgXBs and SFXTs, and comment on the commonality in the behaviour of HMXBs at low mass accretion rates. 

\subsection{eROSITA view of BeXRBs}

\begin{figure}
    \centering
    \includegraphics[width=0.45\textwidth]{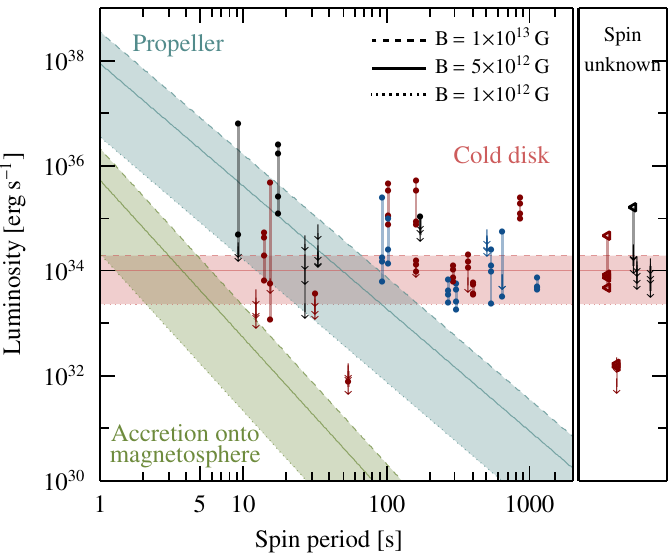}
    \caption{Luminosities and NS spins of Galactic BeXRBs. Points from each eRASS scan are connected for readability. A separate panel shows luminosities of systems without a known spin period. The transition luminosities towards cold disk accretion according to \citet[][salmon]{tsygankov2017} and into propeller regime \citep[][teal]{illarionov1975} are shown for a range of $B$-fields, as well as the minimum luminosity that can be expected from accretion onto the magnetosphere \citep[][olive green]{corbet1997}. Sources without known spin periods are indicated as triangles or arrows on the right, depending on whether they are detected or not detected in eRASS1--4. The maroon data points correspond to eRASS1--4 detections of sources with no confirmed hard X-ray emission, while the dark blue data points are confirmed to show hard X-rays by \nustar \citep[][and references therein]{zalot2026}. Systems that have no known distance or an uncertain large distance are depicted in black, with luminosities computed for 15\,kpc. We observe most BeXRBs straddling the cold disk regime, several that are often detected around the propeller limit and above the estimates for emission due to accretion onto the magnetosphere, and a few that go mostly undetected and have upper limits below $10^{33}\mathrm{erg}\,\mathrm{s}^{-1}$.}
    \label{fig:prop-colddisk}
\end{figure}

The vast majority of BeXRBs detected by \ero are detected at 0.2--10\,keV luminosities ${\gtrsim}10^{33}\,\mathrm{erg}\,\mathrm{s}^{-1}$, which translates to approximately $3\times10^{33}\,\mathrm{erg}\,\mathrm{s}^{-1}$ in bolometric luminosity. This implies that in a single eRASS $\gtrsim$60\% of the known BeXRBs are detected at luminosities in excess of what is expected from Be stars. When combining the eRASS1:4 data (and excluding snapshots caught during Type I outburst), the detection fraction increases to 80\%. For many of the non-detected sources, the \ero scans add up to $\lesssim$1\,ks of effective exposure time and, on average, only about ${\sim}500$\,s (Fig.~\ref{fig:HMXBs_plane}). It is therefore not unlikely that some of these non-detections are simply due to the sensitivity limit. However, they can still provide useful information based on their upper limits. 

Figure~\ref{fig:prop-colddisk} shows the luminosity at which each BeXRB in the Western hemisphere was observed by \ero (or upper limits where the source was not detected), as a function of the known spin period of the accreting neutron star\footnote{The eROSITA exposure is not long enough to estimate spin periods. These are instead taken from \textit{XRBcats} \citep{neumann2023}.}. Systems where the distance is not known are represented in black and the luminosity is computed for a large distance of 15\,kpc. 

Three kinds of populations emerge in the figure -- the largest of them having spin periods ${\gtrsim}100\,\mathrm{s}$ clustered around luminosities of ${\sim}10^{33}$--$10^{35}\,\mathrm{erg}\,\mathrm{s}^{-1}$. Another set, with spin periods $\lesssim100\,\mathrm{s}$ is found having a luminosity $\gtrsim10^{36}\,\mathrm{erg}\,\mathrm{s}^{-1}$, compatible with outbursting states when detected, and dropping to a non-detection with an upper limit at lower luminosity. The final few sources, also with short spin periods, are not detected the majority of the time and have stringent upper limits, some $<10^{32}\,\mathrm{erg}\,\mathrm{s}^{-1}$. These three populations, based on their positions in Fig.~\ref{fig:prop-colddisk}, may hint at different physical mechanisms. As such, the interpretation of detection as an indication of ongoing accretion may require different degrees of weighting for each group. 

The first group of sources is the most convincing case for ongoing stable accretion outside of outburst. These sources are detected a vast majority of the time at luminosities of $10^{33}$--$10^{34}\,\mathrm{erg}\,\mathrm{s}^{-1}$. Moreover, the sources depicted in blue in Fig.~\ref{fig:prop-colddisk} have confirmed \textsl{NuSTAR} detections of hard X-ray emission at low luminosity \citep[see][and references therein]{zalot2026}. 

Among the second set of sources, there are three systems that were found in luminous states multiple times. These seem to be undergoing regular outbursts owing to shorter orbital periods (20--50\,d, see Table~\ref{tab:bexrbs_info}). They are each detected only once at low luminosity and only have two non-detections among them. These non-detections have a corresponding high upper limit, such that accretion at low luminosity cannot be ruled out. 

The third set is populated by sources that are predominantly not detected, with stringent upper limits at lower luminosities in comparison: Ginga~0834$-$430, LS~992, AX~J1700.2$-$4220, and IGR J11305$-$6256 (whose spin period is unknown), the latter two having the lowest upper limits $\lesssim10^{32}\,\mathrm{erg}\,\mathrm{s}^{-1}$. These are among only six sources that are not detected in any of the eRASS scans, the rest of which do not have either distance or spin estimates. LS~992 and AX~J1700.2$-$4220 cross the threshold of detection upon combining the eRASS data, but the latter has a very low luminosity ${\sim}10^{32}\,\mathrm{erg}\,\mathrm{s}^{-1}$ in the 2.0--10.0\,keV band.

\subsection{The physical accretion mechanism of BeXRB at low luminosities} \label{sec:pmech}

We now discuss the potential physical scenarios at work in the BeXRB observed with \ero, with the caveat that there are many factors of uncertainty such as that the source luminosity that may be off by up to a factor 10, due to neglecting geometrical effects and other uncertainties \citep[see][]{falkner2026, kuhnel2017, markozov2024}, the upper limits being sensitive to the assumed absorption value, as well as the magnetic field being unknown for most of these systems. 
 
\citet{illarionov1975} proposed that the low number of X-ray binaries detected in the early 1970s, was likely due to centrifugal inhibition preventing matter from being accreted onto the surface of neutron stars by their rotating magnetospheres. 

The transition to this behavior from accretion is typically described using the relative extents of characteristic radii from the neutron star. A necessary condition for accretion onto the neutron star is $R_\mathrm{m} < \min(R_{\mathrm{capt}}, R_{\mathrm{co}})$ \citep[see, e.g.,][for more details]{bozzo2008,stierhof2025}. Here, $R_\mathrm{m}$ is the magnetospheric radius, where the pressure of the neutron star's magnetic field is balanced by the ram pressure of the incoming plasma (for disk accretion, the inner truncation radius of the disk); $R_{\mathrm{co}}$ is the corotation radius, where the neutron star's angular velocity equals the Keplerian angular velocity; and $R_{\mathrm{capt}}$ is the radius within which matter is gravitationally captured by the neutron star, equal to the accretion radius $R_{\mathrm{acc}}$ for wind accretion. The ``propeller'' in the propeller effect refers to the scenario where, during a significant drop in mass accretion rate, $R_\mathrm{m}$ is pushed outwards and exceeds $R_{\mathrm{co}}$, so that the rotating magnetosphere centrifugally inhibits accretion. For disk accretion with $R_\mathrm{m}$ moderately above $R_{\mathrm{co}}$, matter may instead be trapped and accreted at a reduced rate; \citealt{perna2006}); if $R_\mathrm{m}$ exceeds even $R_{\mathrm{acc}}$, accretion is inhibited magnetically \citep[see, e.g.,][for a detailed summary of various accretion regimes]{bozzo2008,martinez-nunez2017}. In both cases, accretion is halted or strongly reduced, which leads to an absence of sources at luminosities where either barrier is at work, and no pulsations are expected to be seen \citep[e.g.,][]{corbet1997}.

Over the last decade, new observations have added further detail to this picture. Observations of GRO~J1008$-$57 \citep{tsygankov2017} and GX~304$-$1 \citep{escorial2018a}, showed signatures of stable accretion at luminosities of ${\sim}10^{34}$--$10^{35}\,\mathrm{erg}\,\mathrm{s}^{-1}$. In the latter case, pulsations were also detected, clearly indicating accretion down to the surface of the neutron star. In the SMC, AX~J0049.4$-$7323 \citep{ducci2018} showed signs of accretion outside of outburst. Then, \citet{tsygankov2019a} observed a transition in spectral states in A~0535$+$26 outside of outburst. In each of the above cases, the broadband X-ray spectrum switches from an absorbed power law with an exponential cutoff shape to displaying two distinct components separated by a dip \citep[see][for the proposed physical mechanisms explaining the components]{sokolova-lapa2021, mushtukov2021}. This spectral shape was also observed for decades in X~Persei, which is always detected at luminosities of $10^{33}$--$10^{34}\,\mathrm{erg}\,\mathrm{s}^{-1}$, with many physical scenarios proposed to explain its origin \citep{doroshenko2012}. 
See \citet{zalot2026} for a 
systematic analysis of archival NuSTAR observations of 11 BeXRBs at low luminosity (${\sim}$20\% of the overall sample), showing that the shape of the double peaked spectra appears to be  related to the magnetic field strength. 

One mechanism invoked to explain stable accretion at low luminosities is the ``cold disk accretion'', where the accretion disk around the neutron star is maintained at a low enough temperature that it is not sufficiently ionised for efficient accretion \citep{tsygankov2017}\footnote{This model is based on work conducted on disks in LMXBs and accreting white dwarfs by \citet{lasota2001}.}. The corresponding accretion rates give rise to  a luminosity of $10^{33}$--$10^{34}\,\mathrm{erg}\,\mathrm{s}^{-1}$. \citet{tsygankov2017} estimated that, among other factors, the onset of this mechanism depends on the magnetic field of the neutron star, and obtained a threshold luminosity where this could occur. For the propeller regime, on the other hand, the limiting luminosity is expected to additionally depend on the spin period of the neutron star \citep{illarionov1975}. 
 
 In Fig.~\ref{fig:prop-colddisk}, we depict the different luminosity thresholds corresponding to these scenarios. Following \citet{tsygankov2017}, the disk becomes cold and only weakly ionised, below
\begin{equation}
\label{eq:lcold}
L_\mathrm{cold} \sim 9\times10^{33}\, k^{1.5}\, B_{12}^{0.86}\, M_{1.4}^{0.28}\, R_6^{1.57}\,\mathrm{erg}\,\mathrm{s}^{-1} \quad ,
\end{equation}
where $B_{12}$, $M_{1.4}$, and $R_6$ are the magnetic field, the mass, and the radius $R$ of the neutron star in units of $10^{12}\,\mathrm{G}$, $1.4\,M_\odot$, and $10^6\,\mathrm{cm}$, respectively. The factor $k=R_\mathrm{m}/R_\mathrm{A}$, relates the size of the magnetosphere to the Alfvén radius \citep[see][for details]{stierhof2025} and is assumed to be 0.5 for disk accretion.
The  limiting luminosity for the onset of the propeller regime \citep{stella1986, tsygankov2016} is,
\begin{equation}
\label{eq:lprop}
L_\mathrm{prop} \sim 4\times10^{37}\, k^{7/2}\, B_{12}^{2}\, P^{-7/3}\, M_{1.4}^{-2/3}\, R_6^{5}\,\mathrm{erg}\,\mathrm{s}^{-1} \quad .
\end{equation}
\citet{corbet1997} suggested that low-level emission could still occur due to matter being accreted onto and subsequently stalled at the magnetospheric surface rather than being accreted onto the neutron star itself such that  only the potential energy released down to $R_\mathrm{m}$ is released, and the observed luminosity is reduced by a factor $R/R_\mathrm{m}\simeq R/R_\mathrm{c}$ with respect to $L_\mathrm{prop}$.

The teal area in  Fig.~\ref{fig:prop-colddisk} represents the limiting luminosity below  which the neutron star would be expected to propel, computed for a range of typical magnetic fields between $10^{12}$--$10^{13}\,\mathrm{G}$ \citep[see][]{staubert2019a}. The horizontal reddish area indicates the threshold at which cold disk accretion sets in. The propeller phenomenon is predicted to take precedence if a system drops below the limiting luminosity $L_\mathrm{prop}$ before it reaches its corresponding cold disk luminosity. According to propeller theory, the luminosity should drop by several orders of magnitude once the mass accretion rate drops to the threshold corresponding to $L_\mathrm{cold}$, precluding its detection, at least in hard X-rays.

The sources found at luminosities ${>}10^{36}\,\mathrm{erg}\,\mathrm{s}^{-1}$ in a single eRASS are detected at lower luminosities in other eRASSes, and are sometimes even detected at a luminosity corresponding to the lowest propeller threshold (${\sim}10^{34}\,\mathrm{erg}\,\mathrm{s}^{-1}$ for a magnetic field of ${\sim}10^{12}\,\mathrm{G}$). This is the case for 2S~1553$-$542, Swift~J1626.6$-$5156, and MAXI~J0903$-$531. A scenario has been suggested in which leakage from the magnetospheric boundary onto the neutron star allows a small amount of matter to be accreted if the neutron star’s period of rotation still allows for some gravitational capture, with expected luminosities ${\lesssim}10^{33}\,\mathrm{erg}\,\mathrm{s}^{-1}$. Pulsations are expected in this scenario \citep[][and references therein]{roucoescorial2017}. An accretion disk in the ``trapped'' regime, when the matter piles up in the inner regions of the disk near the corotation radius, can also provide low accretion rates and pulsations \citep{kluzniak2007, dangelo2010}. In this context, the magnetospheric radius $R_\mathrm{m}$ can also be significantly smaller than the  Alfv\'en radius $R_\mathrm{A}$, and remain close to the corotation radius down to low luminosities \citep[see, e.g.,][on disc-magnetosphere interaction]{bozzo2018a}. \citet{wijnands2017} suggested that during particularly bright outbursts the neutron star crust can be heated to such high temperatures that it can sustain emission at a brighter luminosity than the expected ``quiescent'' state, with measured luminosity above $10^{33}\,\mathrm{erg}\,\mathrm{s}^{-1}$. 

It is unclear if the detection of the short spin systems close to the luminosity gap corresponds to active accretion within a cold disk scenario or leakage through the magnetospheric boundary even while the propeller mechanism is at hand, or a brief intermediate plateau stage during the system's transition to a lower luminosity state \citep[see][also for 4U~0115$+$63]{roucoescorial2017,xiao2025}.
It is intriguing that these sources are still predominantly detected at higher luminosities and very rarely forego detection. On the other hand, the predominantly non-detected sources are more in line with those BeXRBs that are reported in literature to show a drop in flux of several orders of magnitude and no hard X-ray emission or pulsations \citep{elshamouty2016a,tsygankov2016}. We highlight two of these with particularly low luminosity upper limits. GS~0834$-$430, whose distance is contested\footnote{\citet{israel2000} identified its counterpart to be a Be-type star at a distance of 4.5\,kpc. However, the nearest Gaia counterpart with the same V-band magnitude has a parallax placing the star within 1\,kpc of us. We adopt the larger distance for consistency with the measurement by \citep{tsygankov2017a}.}, has not shown any major outbursts since 2012 \citep{miyasaka2013} despite showing regular outburst behaviour in the 1990s \citep{wilson1997}. \citet{tsygankov2017a} suggested that its low X-ray luminosity pointed to thermal emission from the neutron star, long after an outburst. AX~J1700.2$-$4220 was discovered as a faint source in an ASCA survey \citep{anderson2014}, and is sparsely studied in literature apart from the determination of its orbital period \citep{corbet2017}. 

Although \ero does not provide enough statistics for detailed spectral or timing studies of the sample, these detections are indicative of source behaviour at low luminosity \citep[see also][for a \xmm and \textit{Chandra} sample]{tsygankov2017a}, as also evidenced by the \ero detections of sources with confirmed hard X-ray emission (blue in Fig.~\ref{fig:prop-colddisk}). The detection of short spin-period systems which are detected in the same luminosity regime (${\sim}10^{34}\,\mathrm{erg}\,\mathrm{s}^{-1}$) is intriguing and necessitates X-ray follow-up to confirm their long term behaviour outside outburst -- specifically at hard energies to look for an emerging double-hump spectrum. The possibility of accretion below $10^{32}\,\mathrm{erg}\,\mathrm{s}^{-1}$ should also be ascertained with hard X-ray follow-up for the sources that seem to consistently drop below such a low level of luminosity. Overall, however, \ero further cements the clear trend that has emerged in the BeXRB population. A vast majority of them are found at luminosities consistent with ongoing accretion at an intermediate luminosity in a random snapshot of the X-ray sky.  

\subsection{Connection to the SgXB and SFXT studies}
\label{sec:sgxb}
In the case of SgXBs, of the 24 in the Western hemisphere, 50--60\% are detected in each eRASS scan and 66\% are detected in the eRASS:4 data. We attribute the relatively high fraction of non-detection for persistent systems to the high intrinsic absorption typical for these systems, which is especially a problem for the soft X-ray range \ero is sensitive to, and is especially relevant for the SgXBs farther away detected by \integral \citep{walter2007,martinez-nunez2017, kretschmar2019}. Some, albeit not many, are also detected at lower luminosity, in line with \citet{lutovinov2013}. 

For SFXTs, which are generally characterised by low duty cycles \citep{sidoli2018}, we detect 3--5 of the 7 known in the Western hemisphere, in each of the eRASS scans. The \ero non-detections agree with expected luminosities for the ``quiescent state'' \citep[e.g.,][]{intzand2005,sidoli2008,bozzo2010}.
\citet{romano2015} offers a clear picture that SFXTs are active for a significant fraction of time above a luminosity of $10^{33}\,\mathrm{erg}\,\mathrm{s}^{-1}$, in accordance with the \ero detections. The low accretion rate in conjunction with flaring states has been explained using magnetic gating \citep{grebenev2007, bozzo2008} or the formation of a hot quasi-static shell \citep{shakura2012, shakura_2018}. The latter mechanism has also been put forth for SgXBs which show ``off states'' in their lightcurves, albeit for significantly shorter periods compared to SFXTs, or even more starkly different, BeXRBs \citep{shakura2013}. 

We reiterate that the \ero non-detections cannot be self-sufficiently interpreted without additional information on intrinsic absorption of these sources. However, it bears mentioning that a vast majority of these systems contain neutron stars, such that some commonality between them and the BeXRBs, is expected, especially pertaining to what happens close to the neutron star. The main difference arises from both SgXBs and SFXTs expected to be primarily wind-fed systems, accreting spherically without a persistent disk in close, low-eccentricity orbits \citep{davidson1973}. However, in the scenario where stable accretion at low luminosity is sustained for BeXRBs, long term orbital monitoring is necessary to understand signatures of cold disk accretion better, specifically the contribution of stellar wind from the Be donor. An example where stellar wind contribution may be relevant is GX~304$-$1, a source that has shown no detectible outbursts in more than a decade\footnote{\url{https://swift.gsfc.nasa.gov/results/transients/weak/GX304-1/}} but has been detected at a stable intermediate luminosity several times since its last outburst \citep{sokolova-lapa2021,zalot2026}. \citet{roucoescorial2019} discussed that long-term \swift/XRT monitoring showed slight increase in flux even away from periastron, which is at odds with the idea of a cold disk as a stable reservoir that is progressively depleted when away from the donor star on a long eccentric orbit. The wider implications of this remain to be seen.

\section{Take-aways for the XLF from eROSITA}
\label{sec:outlook-xlf}

We now turn to the implications of our results on the overall population studies and offer takeaways from the expanded luminosity distributions shown in Sect.~\ref{sec:fluxes} and Sect.~\ref{sec:previnst}. 
In their work on the HMXB XLF in soft X-rays, \citet{grimm2003} attempted to extend the luminosity function down to $10^{34}\,\mathrm{erg}\,\mathrm{s}^{-1}$, but suggested that a cutoff would not be necessary down to such luminosity. This is in disagreement with the \ero data, which clearly requires a break. This is consistent with reports of hard X-ray luminosity distributions by \citet{lutovinov2013}, and \citet{voss2010a}, and predictions by \citet{doroshenko2014}. The break luminosity, $\Gamma_1$, and $\Gamma_2$ that are measured for eRASS1--4, are broadly consistent with those predicted by \citep{doroshenko2014}, within uncertainties. The comparison of the measured normalisation requires analysis of the total sample expected to be uncovered by \ero, instead of only the known Galactic HMXBs. 

A more complete XLF would require accurate characterisation of candidate HMXBs found in the \ero survey, and modelling of HMXB population density in the Galaxy. This is part of future work, in addition to combining these soft X-ray results with hard X-ray information, where the underlying accretion processes are better accessible. The \ero information should be combined with sensitive measurements obtained with hard X-ray surveys, or by conducting unbiased survey-like follow-up. Such an undertaking is both time-intensive and observationally expensive, but is currently underway as part of a \textit{NuSTAR} campaign for a volume limited sample (at a distance of 7.5\,kpc) for the BeXRBs contained within the Western hemisphere. Data from the \textit{Einstein Probe} mission \citep{yuan2022}, which is currently conducting a highly sensitive survey of the X-ray sky, would be especially useful for expanding on these studies to the Eastern hemisphere, as would be a release of the \ero data from the Eastern hemisphere.

The changes in the shape of the XLF introduced by the outbursting and flaring behaviour of individual HMXBs occur regardless of the energy range. As \citet{islam2016} already suggested, variability effects must be taken into account while studying luminosity functions of X-ray binaries for the Milky Way. This is especially important in order for the inferences made for our Galaxy to be comparable to results obtained for external galaxies, where often only a few snapshots are available \citep[e.g.,][]{lehmer2019, mineo2014}. 

We also show that even qualitatively splitting the luminosity distribution of HMXBs in the Milky Way between the subclasses, BeXRBs, SgXBs and the much smaller sample of SFXTs, is a useful exercise because it hints at which sources are likely to be detected in a random snapshot of a distant galaxy. In the decade of \rxte/ASM observations, only one or two BeXRBs were consistently detected (see Fig.~\ref{fig:rxte_lum_types}, left panel), since they spend the vast majority of their lifetime at luminosities below detection thresholds of current monitors. BeXRBs were only included in the luminosity distribution in outburst. Since, \rxte/ASM and \maxi sensitivity for the Milky Way is analogous to what has been achieved in other galaxies so far, these results imply that in the case of similar populations of extragalactic BeXRBs and SgXBs to Galactic populations, the average HMXB detection in a distant galaxy is much more likely to have a supergiant companion. This only begins to change at luminosities below $10^{35}\,\mathrm{erg}\,\mathrm{s}^{-1}$, where the BeXRBs are detected most of the time. The shape of the \lognl also depends on the subclasses, and can be parameterised differently depending on the composition of the overall HMXB sample, as discussed in Appendix~\ref{app:fits}, using Fig.~\ref{fig:fit_pars}, with the caveat that these results are not normalised for the entire Galaxy.  These results should be corroborated with HMXB studies of the Magellanic Clouds \citep{shtykovskiy2005,haberl2016, kaltenbrunner2026}, to test their applicability. Although resolving sub-populations for galaxies farther away is unfeasible, the overall variability information could be compared with variability attributed to stochasticity \citep{kyritsis2026}. 

While the low luminosity end is largely inaccessible in other galaxies, it holds important information for binary evolution scenarios. In view of this, we attempt to put the comparison to isolated Be stars in the context of binary population studies in Sect.~\ref{sec:BeStars} before concluding this article in Sect.~\ref{sec:conc}. 

\section{Are some Be stars harboring quiet BeXRBs?}
\label{sec:BeStars}

As discussed in Sect.~\ref{sec:bestar}, the emission from a Be star can in a minority of cases be comparable to the least luminous BeXRBs. While the excess luminosity of known BeXRBs compared to what is typically observed from isolated Be stars is taken as evidence for ongoing accretion, we now address the origin of excess X-ray emission in the case of the brightest and most luminous Be stars. Previously, \citet{naze2018} showed that there is a subsample of Be stars that have higher X-ray luminosities, and harder X-ray spectra. Some of these are expected to be \gamcas objects, while colliding winds and magnetic activity are proposed as alternative causes. They do not rule out the presence of compact object companions. These results were further supported by the survey data from \ero \citep{naze2023}. To explain anomalously hard X-ray luminosities, as \citet{naze2023} pointed out, binary interaction is the most favoured scenario. There is also extensive theoretical support from the supposition of binarity. Be stars make up 20\% of the population of B-type main sequence stars. Their characteristic circumstellar disks are thought to be formed by rapid rotation, which is a consistent property for the entire population of Be stars \citep{zorec2016,zorec2017}. Many hypotheses have been offered for the origin of this rapid rotation. \citet{bodenheimer1995} suggested that it is due to the angular momentum of the star's parent molecular cloud, \citet{granada2013} likened it to spin-up events from contraction of the stellar core, and \citet{shao2014} and \citet{bodensteiner2020} posit that it is a result of mass transfer during binary interaction. However, the binarity fraction of Be and B stars were found to be similar. A number of studies thereafter considered the possibility that the Be star's companion star is stripped during binary interaction resulting in accumulation of circumstellar material. \citet{dodd2024} show, based on a much more comprehensive sample of both B-type and Be-type stars, that there is a significant lack of Be stars in binary systems seen by Gaia compared to B stars. They suggest that this might be due to the companions being undetected by the methods used, suggesting stripped companions as a possible explanation. This is supported both by theoretical work \citep{el-badry2021} and observational detection of such systems \citep{bodensteiner2020a,frost2022}. In addition, \citet{bodensteiner2020} do not find any close main sequence companions to Be stars, also supporting the stripped star hypothesis. However, intriguingly \citet{dodd2024} also note that the Be stars do not show up in binaries at separations suitable for mass transfer, and attribute this to the presence of higher order companions. The model proposes that the third companion induces migration of the Be star and the stripped star, allowing for them to be close enough for mass transfer to occur, stripping one star and giving the Be star a circumstellar disk. \citet{hastings2021} show that although observationally the binary fraction of Be stars appears low, with valid theoretical estimates, one can arrive at initial properties for the population that are consistent with most Be stars being binary interaction products, with the post-merger results skewing the currently observed binary fraction. They also show that even in the case where one can explain Be star formation along a single star pathway, many Be stars do not fit this criterion. 

Previously, the most luminous Be stars which may be considered to have binary companions were not assumed to be potential HMXBs due to their comparatively low luminosity for neutron star binaries. However, now that it has become clear that many BeXRBs can exist for several years without going into outburst, it is worth revisiting the bright, hard and luminous Be stars that do not show obvious evidence of the \gamcas phenomenon as potential BeXRB candidates. 

\section{Summary and Conclusion}

\label{sec:conc}
We studied the \ero data of the known sample of HMXBs in the Milky Way and found that the HMXB XLF can be extended by ${\gtrsim3}$ orders of magnitude. The survey is complete down to ${\sim}10^{33}\,\mathrm{erg}\,\mathrm{s}^{-1}$ for a distance of 10\,kpc with a sensitivity limit of ${\sim}10^{-13}\,\mathrm{erg}\,\mathrm{s}^{-1}\mathrm{cm}^{-2}$. We also split the HMXB XLF into the two main subclasses, the predominantly persistent SgXBs and the predominantly transient BeXRBs, and found that there is a clear separation in the luminosity space occupied by each subclass. While SgXBs dominate the higher luminosity regime, BeXRBs are a significant contributor to the low luminosity end of the HMXB XLF. The \ero distribution necessitates a broken power law, in contrast to the \rxte/ASM results \citep[][which sample luminosities $>10^{36}\,\mathrm{erg}\,\mathrm{s}^{-1}$, where a break is not required]{grimm2002,islam2016}, similar to what was proposed by \citet{voss2010a} and \citet{lutovinov2013}. We reported on the variability of the luminosity distributions using Monte Carlo modeling based on the four eRASS detections. On modelling the corresponding X-ray luminosity distributions with a broken power law, we also find that the break luminosity varies significantly between the subclasses, while both indices show higher variance for the BeXRB sample (see Fig.~\ref{fig:lognl_fit_pars}). We then compared the \ero results with previous work conducted using \rxte/ASM in the same energy range, and conducted a variability analysis by sampling directly from the long-term \rxte/ASM monitoring lightcurves of each source. We repeated this process using data from the currently operating \maxi monitor. We find that above $10^{35}\,\mathrm{erg}\,\mathrm{s}^{-1}$ the \ero, \rxte/ASM, and \maxi distributions are consistent. 

We emphasise the high detection fraction of BeXRBs (${\gtrsim60\%}$ in a single eRASS and ${\gtrsim80\%}$ on combining the eRASS1:4 data) at low luminosities $10^{33}$--$10^{35}\,\mathrm{erg}\,\mathrm{s}^{-1}$, which lends further credence to their stable accretion outside of outburst. We present the BeXRBs in luminosity and spin period space, to study the effect of the propeller regime predicted at low luminosity. Most BeXRBs in the Western hemisphere are found persistently around luminosities of $10^{33}$--$10^{35}\,\mathrm{erg}\,\mathrm{s}^{-1}$, consistent with the threshold where accretion from a cold disk is predicted to occur and, for long spin periods, above the propeller limit. A handful of short-spin sources are detected below the propeller line, which suggests that matter still reaches the neutron star. While these detections do not exclude the possibility of a propeller phase, it is intriguing to find them detected most of the time. Importantly, four BeXRBs have stringent upper limits of ${\lesssim}10^{32}$--$10^{33}\,\mathrm{erg}\,\mathrm{s}^{-1}$: three lie below the propeller limit, and the fourth, with unknown spin period, lies below the cold-disk regime. We propose them as candidates for the propeller regime or for a halted mass supply.

The accretion models for low luminosity activity require further theoretical work, and cannot currently be discerned from the available data. However, the \ero detections do point to two populations, those BeXRBs that have stringent upper limits below the luminosity where stable accretion has been observed to occur in other systems and those that are consistently detected below outburst luminosities. Nevertheless, for the known sample there seems to only be a marginal effect of a potential propeller phase on the luminosity distribution down to a few $\times10^{32}\,\mathrm{erg}\,{s}^{-1}$. We can now set an upper limit of fewer than 20\% for systems that could be consistently hindered by the propeller effect, with the caveat that this is based only on the Western hemisphere. In addition, the prevalence of low luminosity long-period BeXRBs also suggests that there may be quiescent BeXRBs among hard and luminous isolated Be stars. 

\begin{acknowledgements}
We acknowledge funding from Deutsche Forschungsgemeinschaft under the umbrella of DFG Forschungsgruppe WI 1860/17-2 (eRO STEP 2, P5). We regret the untimely passing of our co-author, Katja~Pottschmidt. Her sharp insight and incisive questions helped propel many a project forward, and her kindness and mentorship remain ingrained in many of us. She is sorely missed by all those who knew her, and the research of highly magnetized accreting neutron stars is not the same in her absence. This work also extensively cites the work of Jan Robrade, who was a pivotal part of the \textit{eROSITA} consortium, and passed away recently. He will be missed. This work is based on data from eROSITA, the soft X-ray instrument aboard SRG, a joint Russian-German science mission supported by the Russian Space Agency (Roskosmos), in the interests of the Russian Academy of Sciences represented by its Space Research Institute (IKI), and the Deutsches Zentrum für Luft- und Raumfahrt (DLR). The SRG spacecraft was built by Lavochkin Association (NPOL) and its subcontractors, and is operated by NPOL with support from the Max Planck Institute for Extraterrestrial Physics (MPE). The development and construction of the eROSITA X-ray instrument was led by MPE, with contributions from the Dr. Karl Remeis Observatory Bamberg \& ECAP (FAU Erlangen-Nuernberg), the University of Hamburg Observatory, the Leibniz Institute for Astrophysics Potsdam (AIP), and the Institute for Astronomy and Astrophysics of the University of Tübingen, with the support of DLR and the Max Planck Society. The Argelander Institute for Astronomy of the University of Bonn and the Ludwig Maximilians Universität Munich also participated in the science preparation for eROSITA. The eROSITA data shown here were processed using the eSASS/NRTA software system developed by the German eROSITA consortium. This research has made use of a collection of ISIS functions (ISISscripts) provided by ECAP/Remeis observatory and MIT (\url{https://www.sternwarte.uni-erlangen.de/isis/}), data and software provided by the High Energy Astrophysics Science Archive Research Center (HEASARC), which is a service of the Astrophysics Science Division at NASA/GSFC and data from the European Space Agency (ESA) mission \textit{Gaia} (\url{https://www.cosmos.esa.int/gaia}), processed by the \textit{Gaia} Data Processing and Analysis Consortium (DPAC, \url{https://www.cosmos.esa.int/web/gaia/dpac/consortium}).
Funding for the DPAC has been provided by national institutions, in particular the institutions participating in the \textit{Gaia} Multilateral Agreement. The material was additionally supported by NASA under award number 80GSFC24M0006.
\end{acknowledgements}

\bibliographystyle{aa} 
\bibliography{Low-lum-paper}

@article{anderson2014,
  title = {{{CHASING THE IDENTIFICATION OF}} {{{\emph{ASCA}}}} {{GALACTIC OBJECTS}} ({{ChIcAGO}}): {{AN X-RAY SURVEY OF UNIDENTIFIED SOURCES IN THE GALACTIC PLANE}}. {{I}}. {{SOURCE SAMPLE AND INITIAL RESULTS}}},
  author = {Anderson, Gemma E. and Gaensler, B. M. and Kaplan, David L. and Slane, Patrick O. and Muno, Michael P. and Posselt, Bettina and Hong, Jaesub and Murray, Stephen S. and Steeghs, Danny T. H. and Brogan, Crystal L. and Drake, Jeremy J. and Farrell, Sean A. and Benjamin, Robert A. and Chakrabarty, Deepto and Drew, Janet E. and Finley, John P. and Grindlay, Jonathan E. and Lazio, T. Joseph W. and Lee, Julia C. and Mauerhan, Jon C. and Van Kerkwijk, Marten H.},
  year = 2014,
  month = apr,
  volume = {212},
  number = {1},
  pages = {13},
  doi = {10.1088/0067-0049/212/1/13},
  journal = {\apjs}
}

@article{bailer-jones2021b,
  title = {Estimating {{Distances}} from {{Parallaxes}}. {{V}}. {{Geometric}} and {{Photogeometric Distances}} to 1.47 {{Billion Stars}} in {{Gaia Early Data Release}} 3},
  author = {{Bailer-Jones}, C. A. L. and Rybizki, J. and Fouesneau, M. and Demleitner, M. and Andrae, R.},
  year = 2021,
  month = mar,
  volume = {161},
  pages = {147},
  publisher = {IOP},
  doi = {10.3847/1538-3881/abd806},
  journal = {\aj}
}

@article{ballhausen2017,
  title = {Looking at {{A}} 0535+26 at Low Luminosities with {{NuSTAR}}},
  author = {Ballhausen, Ralf and Pottschmidt, Katja and F{\"u}rst, Felix and Wilms, J{\"o}rn and Tomsick, John A. and Schwarm, Fritz-Walter and Stern, Daniel and Kretschmar, Peter and Caballero, Isabel and Harrison, Fiona A. and Boggs, Steven E. and Christensen, Finn E. and Craig, William W. and Hailey, Charles J. and Zhang, William W.},
  year = 2017,
  month = dec,
  volume = {608},
  pages = {A105},
  publisher = {EDP Sciences},
  doi = {10.1051/0004-6361/201730845},
  langid = {english},
  journal = {\aap}
}

@article{bartlett2019,
  title = {{{CI Camelopardalis}}: {{The}} First {{sgB}}[e]-High Mass {{X-ray}} Binary Twenty Years on: {{A}} Supernova Imposter in Our Own {{Galaxy}}?},
  author = {Bartlett, E. S. and Clark, J. S. and Negueruela, I.},
  year = 2019,
  month = feb,
  volume = {622},
  pages = {A93},
  publisher = {EDP Sciences},
  doi = {10.1051/0004-6361/201834315},
  langid = {english},
  journal = {\aap}
}

@article{bekhti2016,
  title = {{{HI4PI}}: A Full-Sky {{H}} i Survey Based on {{EBHIS}} and {{GASS}}},
  author = { {Ben Bekhti}, N. and Fl{\"o}er, L. and Keller, R. and Kerp, J. and Lenz, D. and Winkel, B. and Bailin, J. and Calabretta, M. R. and Dedes, L. and Ford, H. A. and Gibson, B. K. and Haud, U. and Janowiecki, S. and Kalberla, P. M. W. and Lockman, F. J. and {McClure-Griffiths}, N. M. and Murphy, T. and Nakanishi, H. and Pisano, D. J. and {Staveley-Smith}, L.},
  year = 2016,
  month = oct,
  volume = {594},
  pages = {A116},
  publisher = {EDP Sciences},
  doi = {10.1051/0004-6361/201629178},
  langid = {english},
  journal = {\aap}
}

@article{blomme2024,
  title = {The Colliding-Wind Binary {{HD}} 168112},
  author = {Blomme, R. and Rauw, G. and Volpi, D. and Naz{\'e}, Y. and {Abdul-Masih}, M.},
  year = 2024,
  month = jul,
  volume = {687},
  pages = {A106},
  publisher = {EDP Sciences},
  doi = {10.1051/0004-6361/202450227},
  langid = {english},
  journal = {\aap}
}

@article{bodenheimer1995,
  title = {Angular {{Momentum Evolution}} of {{Young Stars}} and {{Disks}}},
  author = {Bodenheimer, Peter},
  year = 1995,
  month = sep,
  volume = {33},
  number = {Volume 33, 1995},
  pages = {199--238},
  publisher = {ARA&A},
  doi = {10.1146/annurev.aa.33.090195.001215},
  langid = {english},
  journal = {\araa}
}

@article{bodensteiner2020,
  title = {Investigating the Lack of Main-Sequence Companions to Massive {{Be}} Stars},
  author = {Bodensteiner, J. and Shenar, T. and Sana, H.},
  year = 2020,
  month = sep,
  volume = {641},
  pages = {A42},
  doi = {10.1051/0004-6361/202037640},
  journal = {\aap}
}

@article{bodensteiner2020a,
  title = {Is {{HR}} 6819 a Triple System Containing a Black Hole? - {{An}} Alternative Explanation},
  author = {Bodensteiner, J. and Shenar, T. and Mahy, L. and Fabry, M. and Marchant, P. and {Abdul-Masih}, M. and Banyard, G. and Bowman, D. M. and Dsilva, K. and Frost, A. J. and Hawcroft, C. and Reggiani, M. and Sana, H.},
  year = 2020,
  month = sep,
  volume = {641},
  pages = {A43},
  publisher = {EDP Sciences},
  doi = {10.1051/0004-6361/202038682},
  langid = {english},
  journal = {\aap}
}

@article{boller2016,
  title = {Second {{ROSAT}} All-Sky Survey ({{2RXS}}) Source Catalogue},
  author = {Boller, {T}. and Freyberg, M. J. and Tr{\"u}mper, J. and Haberl, F. and Voges, W. and Nandra, K.},
  year = 2016,
  month = apr,
  volume = {588},
  pages = {A103},
  doi = {10.1051/0004-6361/201525648},
  langid = {english},
  journal = {\aap}
}

@article{boller2025,
  title = {The {{eROSITA DR1}} Variability Catalogue},
  author = {Boller, {T}. and Salvato, M. and Buchner, J. and Freyberg, M. and Haberl, F. and Maitra, C. and Schwope, A. and Robrade, J. and Rukdee, S. and Rau, A. and Grotova, I. and Waddell, S. and Ni, Q. and Krumpe, M. and Georgakakis, A. and Merloni, A. and Nandra, K.},
  year = 2025,
  month = aug,
  volume = {700},
  pages = {A61},
  publisher = {EDP},
  doi = {10.1051/0004-6361/202449355},
  journal = {\aap}
}

@article{bozzo2008,
  title = {Are {{There Magnetars}} in {{High-Mass X-Ray Binaries}}? {{The Case}} of {{Supergiant Fast X-Ray Transients}}},
  author = {Bozzo, E. and Falanga, M. and Stella, L.},
  year = 2008,
  month = aug,
  volume = {683},
  number = {2},
  pages = {1031},
  publisher = {IOP Publishing},
  doi = {10.1086/589990},
  langid = {english},
  journal = {\apj}
}

@article{bozzo2010,
  title = {The Supergiant Fast {{X-ray}} Transients {{XTE J1739-302}} and {{IGR J08408-4503}} in Quiescence with {{XMM-Newton}}},
  author = {Bozzo, E. and Stella, L. and Ferrigno, C. and Giunta, A. and Falanga, M. and Campana, S. and Israel, G. and Leyder, J. C.},
  year = 2010,
  month = sep,
  volume = {519},
  pages = {A6},
  doi = {10.1051/0004-6361/201014095},
  langid = {english},
  journal = {\aap}
}

@article{bozzo2015,
  title = {Supergiant Fast {{X-ray}} Transients as an under-Luminous Class of Supergiant {{X-ray}} Binaries},
  author = {Bozzo, E. and Romano, P. and Ducci, L. and Bernardini, F. and Falanga, M.},
  year = 2015,
  month = feb,
  volume = {55},
  number = {4},
  eprint = {1411.4470},
  primaryclass = {astro-ph},
  pages = {1255--1263},
  doi = {10.1016/j.asr.2014.11.012},
  archiveprefix = {arXiv},
  journal = {Adv. Space Res.}
}

@article{bozzo2018a,
  title = {Magnetospheric Radius of an Inclined Rotator in the Magnetically Threaded Disk Model},
  author = {Bozzo, E. and Ascenzi, S. and Ducci, L. and Papitto, A. and Burderi, L. and Stella, L.},
  year = 2018,
  month = sep,
  volume = {617},
  pages = {A126},
  doi = {10.1051/0004-6361/201732004},
  langid = {english},
  journal = {\aap}
}

@article{brunner2022,
  title = {The {{eROSITA Final Equatorial Depth Survey}} ({{eFEDS}}) - {{X-ray}} Catalogue},
  author = {Brunner, H. and Liu, T. and Lamer, G. and Georgakakis, A. and Merloni, A. and Brusa, M. and Bulbul, E. and Dennerl, K. and Friedrich, S. and Liu, A. and Maitra, C. and Nandra, K. and {Ramos-Ceja}, M. E. and Sanders, J. S. and Stewart, I. M. and Boller, T. and Buchner, J. and Clerc, N. and Comparat, J. and Dwelly, T. and Eckert, D. and Finoguenov, A. and Freyberg, M. and Ghirardini, V. and Gueguen, A. and Haberl, F. and Kreykenbohm, I. and Krumpe, M. and Osterhage, S. and Pacaud, F. and Predehl, P. and Reiprich, T. H. and Robrade, J. and Salvato, M. and Santangelo, A. and Schrabback, T. and Schwope, A. and Wilms, J.},
  year = 2022,
  month = may,
  volume = {661},
  pages = {A1},
  publisher = {EDP Sciences},
  doi = {10.1051/0004-6361/202141266},
  langid = {english},
  journal = {\aap}
}

@article{cody2025,
  title = {The {{Optical Photometric Variability}} of {{Herbig Ae}}/{{Be Stars}} from {{TESS}}},
  author = {Cody, Ann Marie and Hillenbrand, Lynne A. and Chandragiri, Shreya and Morgan, Marvin},
  year = 2025,
  month = dec,
  volume = {994},
  pages = {253},
  publisher = {IOP},
  doi = {10.3847/1538-4357/ae119a},
  journal = {\apj}
}

@article{collaboration2023a,
  title = {Gaia {{Data Release}} 3. {{Summary}} of the Content and Survey Properties},
  author = {{Gaia Collaboration} and Vallenari, A. and Brown, A. G. A. and Prusti, T. and {de Bruijne}, J. H. J. and Arenou, F. and Babusiaux, C. and Biermann, M. and Creevey, O. L. and Ducourant, C. and Evans, D. W. and Eyer, L. and Guerra, R. and Hutton, A. and Jordi, C. and Klioner, S. A. and Lammers, U. L. and Lindegren, L. and Luri, X. and Mignard, F. and Panem, C. and Pourbaix, D. and Randich, S. and Sartoretti, P. and Soubiran, C. and Tanga, P. and Walton, N. A. and {Bailer-Jones}, C. a. L. and Bastian, U. and Drimmel, R. and Jansen, F. and Katz, D. and Lattanzi, M. G. and {van Leeuwen}, F. and Bakker, J. and Cacciari, C. and Casta{\~n}eda, J. and De Angeli, F. and Fabricius, C. and Fouesneau, M. and Fr{\'e}mat, Y. and Galluccio, L. and Guerrier, A. and Heiter, U. and Masana, E. and Messineo, R. and Mowlavi, N. and Nicolas, C. and Nienartowicz, K. and Pailler, F. and Panuzzo, P. and Riclet, F. and Roux, W. and Seabroke, G. M. and Sordo, R. and Th{\'e}venin, F. and {Gracia-Abril}, G. and Portell, J. and Teyssier, D. and Altmann, M. and Andrae, R. and Audard, M. and {Bellas-Velidis}, I. and Benson, K. and Berthier, J. and Blomme, R. and Burgess, P. W. and Busonero, D. and Busso, G. and C{\'a}novas, H. and Carry, B. and Cellino, A. and Cheek, N. and Clementini, G. and Damerdji, Y. and Davidson, M. and {de Teodoro}, P. and Nu{\~n}ez Campos, M. and Delchambre, L. and Dell'Oro, A. and Esquej, P. and {Fern{\'a}ndez-Hern{\'a}ndez}, J. and Fraile, E. and Garabato, D. and {Garc{\'i}a-Lario}, P. and Gosset, E. and Haigron, R. and Halbwachs, J.-L. and Hambly, N. C. and Harrison, D. L. and Hern{\'a}ndez, J. and Hestroffer, D. and Hodgkin, S. T. and Holl, B. and Jan{\ss}en, K. and {Jevardat de Fombelle}, G. and Jordan, S. and {Krone-Martins}, A. and Lanzafame, A. C. and L{\"o}ffler, W. and Marchal, O. and Marrese, P. M. and Moitinho, A. and Muinonen, K. and Osborne, P. and Pancino, E. and Pauwels, T. and {Recio-Blanco}, A. and Reyl{\'e}, C. and Riello, M. and Rimoldini, L. and Roegiers, T. and Rybizki, J. and Sarro, L. M. and Siopis, C. and Smith, M. and Sozzetti, A. and Utrilla, E. and {van Leeuwen}, M. and Abbas, U. and {\'A}brah{\'a}m, P. and Abreu Aramburu, A. and Aerts, C. and Aguado, J. J. and Ajaj, M. and {Aldea-Montero}, F. and Altavilla, G. and {\'A}lvarez, M. A. and Alves, J. and Anders, F. and Anderson, R. I. and Anglada Varela, E. and Antoja, T. and Baines, D. and Baker, S. G. and {Balaguer-N{\'u}{\~n}ez}, L. and Balbinot, E. and Balog, Z. and Barache, C. and Barbato, D. and Barros, M. and Barstow, M. A. and Bartolom{\'e}, S. and Bassilana, J.-L. and Bauchet, N. and Becciani, U. and Bellazzini, M. and Berihuete, A. and Bernet, M. and Bertone, S. and Bianchi, L. and Binnenfeld, A. and {Blanco-Cuaresma}, S. and Blazere, A. and Boch, T. and Bombrun, A. and Bossini, D. and Bouquillon, S. and Bragaglia, A. and Bramante, L. and Breedt, E. and Bressan, A. and Brouillet, N. and Brugaletta, E. and Bucciarelli, B. and Burlacu, A. and Butkevich, A. G. and Buzzi, R. and Caffau, E. and Cancelliere, R. and {Cantat-Gaudin}, T. and Carballo, R. and Carlucci, T. and Carnerero, M. I. and Carrasco, J. M. and Casamiquela, L. and Castellani, M. and {Castro-Ginard}, A. and Chaoul, L. and Charlot, P. and Chemin, L. and Chiaramida, V. and Chiavassa, A. and Chornay, N. and Comoretto, G. and Contursi, G. and Cooper, W. J. and Cornez, T. and Cowell, S. and Crifo, F. and Cropper, M. and Crosta, M. and Crowley, C. and Dafonte, C. and Dapergolas, A. and David, M. and David, P. and {de Laverny}, P. and De Luise, F. and De March, R. and De Ridder, J. and {de Souza}, R. and {de Torres}, A. and {del Peloso}, E. F. and {del Pozo}, E. and Delbo, M. and Delgado, A. and Delisle, J.-B. and Demouchy, C. and Dharmawardena, T. E. and Di Matteo, P. and Diakite, S. and Diener, C. and Distefano, E. and Dolding, C. and Edvardsson, B. and Enke, H. and Fabre, C. and Fabrizio, M. and Faigler, S. and Fedorets, G. and Fernique, P. and Fienga, A. and Figueras, F. and Fournier, Y. and Fouron, C. and Fragkoudi, F. and Gai, M. and {Garcia-Gutierrez}, A. and {Garcia-Reinaldos}, M. and {Garc{\'i}a-Torres}, M. and Garofalo, A. and Gavel, A. and Gavras, P. and Gerlach, E. and Geyer, R. and Giacobbe, P. and Gilmore, G. and Girona, S. and Giuffrida, G. and Gomel, R. and Gomez, A. and {Gonz{\'a}lez-N{\'u}{\~n}ez}, J. and {Gonz{\'a}lez-Santamar{\'i}a}, I. and {Gonz{\'a}lez-Vidal}, J. J. and Granvik, M. and Guillout, P. and Guiraud, J. and {Guti{\'e}rrez-S{\'a}nchez}, R. and Guy, L. P. and Hatzidimitriou, D. and Hauser, M. and Haywood, M. and Helmer, A. and Helmi, A. and Sarmiento, M. H. and Hidalgo, S. L. and Hilger, T. and H{\l}adczuk, N. and Hobbs, D. and Holland, G. and Huckle, H. E. and Jardine, K. and Jasniewicz, G. and {Jean-Antoine Piccolo}, A. and {Jim{\'e}nez-Arranz}, {\'O} and Jorissen, A. and Juaristi Campillo, J. and Julbe, F. and Karbevska, L. and Kervella, P. and Khanna, S. and Kontizas, M. and Kordopatis, G. and Korn, A. J. and K{\'o}sp{\'a}l, {\'A} and {Kostrzewa-Rutkowska}, Z. and Kruszy{\'n}ska, K. and Kun, M. and Laizeau, P. and Lambert, S. and Lanza, A. F. and Lasne, Y. and Le Campion, J.-F. and Lebreton, Y. and Lebzelter, T. and Leccia, S. and Leclerc, N. and {Lecoeur-Taibi}, I. and Liao, S. and Licata, E. L. and Lindstr{\o}m, H. E. P. and Lister, T. A. and Livanou, E. and Lobel, A. and Lorca, A. and Loup, C. and Madrero Pardo, P. and Magdaleno Romeo, A. and Managau, S. and Mann, R. G. and Manteiga, M. and Marchant, J. M. and Marconi, M. and Marcos, J. and Marcos Santos, M. M. S. and Mar{\'i}n Pina, D. and Marinoni, S. and Marocco, F. and Marshall, D. J. and Martin Polo, L. and {Mart{\'i}n-Fleitas}, J. M. and Marton, G. and Mary, N. and Masip, A. and Massari, D. and {Mastrobuono-Battisti}, A. and Mazeh, T. and McMillan, P. J. and Messina, S. and Michalik, D. and Millar, N. R. and Mints, A. and Molina, D. and Molinaro, R. and Moln{\'a}r, L. and Monari, G. and Mongui{\'o}, M. and Montegriffo, P. and Montero, A. and Mor, R. and Mora, A. and Morbidelli, R. and Morel, T. and Morris, D. and Muraveva, T. and Murphy, C. P. and Musella, I. and Nagy, Z. and Noval, L. and Oca{\~n}a, F. and Ogden, A. and Ordenovic, C. and Osinde, J. O. and Pagani, C. and Pagano, I. and Palaversa, L. and Palicio, P. A. and {Pallas-Quintela}, L. and Panahi, A. and {Payne-Wardenaar}, S. and Pe{\~n}alosa Esteller, X. and Penttil{\"a}, A. and Pichon, B. and Piersimoni, A. M. and Pineau, F.-X. and Plachy, E. and Plum, G. and Poggio, E. and Pr{\v s}a, A. and Pulone, L. and Racero, E. and Ragaini, S. and Rainer, M. and Raiteri, C. M. and Rambaux, N. and Ramos, P. and {Ramos-Lerate}, M. and Re Fiorentin, P. and Regibo, S. and Richards, P. J. and Rios Diaz, C. and Ripepi, V. and Riva, A. and Rix, H.-W. and Rixon, G. and Robichon, N. and Robin, A. C. and Robin, C. and Roelens, M. and Rogues, H. R. O. and Rohrbasser, L. and {Romero-G{\'o}mez}, M. and Rowell, N. and Royer, F. and Ruz Mieres, D. and Rybicki, K. A. and Sadowski, G. and S{\'a}ez N{\'u}{\~n}ez, A. and Sagrist{\`a} Sell{\'e}s, A. and Sahlmann, J. and Salguero, E. and Samaras, N. and Sanchez Gimenez, V. and Sanna, N. and Santove{\~n}a, R. and Sarasso, M. and Schultheis, M. and Sciacca, E. and Segol, M. and Segovia, J. C. and S{\'e}gransan, D. and Semeux, D. and Shahaf, S. and Siddiqui, H. I. and Siebert, A. and Siltala, L. and Silvelo, A. and Slezak, E. and Slezak, I. and Smart, R. L. and Snaith, O. N. and Solano, E. and Solitro, F. and Souami, D. and Souchay, J. and Spagna, A. and Spina, L. and Spoto, F. and Steele, I. A. and Steidelm{\"u}ller, H. and Stephenson, C. A. and S{\"u}veges, M. and Surdej, J. and Szabados, L. and {Szegedi-Elek}, E. and Taris, F. and Taylor, M. B. and Teixeira, R. and Tolomei, L. and Tonello, N. and Torra, F. and Torra, J. and Torralba Elipe, G. and Trabucchi, M. and Tsounis, A. T. and Turon, C. and Ulla, A. and Unger, N. and Vaillant, M. V. and {van Dillen}, E. and {van Reeven}, W. and Vanel, O. and Vecchiato, A. and Viala, Y. and Vicente, D. and Voutsinas, S. and Weiler, M. and Wevers, T. and Wyrzykowski, {\L} and Yoldas, A. and Yvard, P. and Zhao, H. and Zorec, J. and Zucker, S. and Zwitter, T.},
  year = 2023,
  month = jun,
  volume = {674},
  pages = {A1},
  doi = {10.1051/0004-6361/202243940},
  langid = {english},
  journal = {\aap}
}

@article{corbet1997,
  title = {A {{Low-Amplitude X-Ray}} and {{Optical Outburst}} from the {{Periodic Transient A0538}}--66: {{Accretion}} onto a {{Magnetosphere}}?},
  author = {Corbet, R. H. D. and Charles, P. A. and Southwell, K. A. and Smale, A. P.},
  year = 1997,
  month = feb,
  volume = {476},
  number = {2},
  pages = {833},
  publisher = {IOP Publishing},
  doi = {10.1086/303644},
  langid = {english},
  journal = {\apj}
}

@article{corbet2017,
  title = {Diverse {{Long-term Variability}} of {{Five Candidate High-mass X-Ray Binaries}} from {{Swift Burst Alert Telescope Observations}}},
  author = {Corbet, Robin H. D. and Coley, Joel B. and Krimm, Hans A.},
  year = 2017,
  month = sep,
  volume = {846},
  number = {2},
  pages = {161},
  publisher = {The American Astronomical Society},
  doi = {10.3847/1538-4357/aa8638},
  langid = {english},
  journal = {\apj}
}

@article{cusumano2010,
  title = {The {{Palermo Swift-BAT}} Hard {{X-ray}} Catalogue - {{II}}. {{Results}} after 39 Months of Sky Survey},
  author = {Cusumano, G. and {La Parola}, V. and Segreto, A. and Mangano, V. and Ferrigno, C. and Maselli, A. and Romano, P. and Mineo, T. and Sbarufatti, B. and Campana, S. and Chincarini, G. and Giommi, P. and Masetti, N. and Moretti, A. and Tagliaferri, G.},
  year = 2010,
  month = feb,
  volume = {510},
  pages = {A48},
  publisher = {EDP Sciences},
  doi = {10.1051/0004-6361/200811184},
  langid = {english},
  journal = {\aap}
}

@article{dangelo2010,
  title = {Episodic Accretion on to Strongly Magnetic Stars},
  author = {D'Angelo, Caroline R. and Spruit, Hendrik C.},
  year = 2010,
  month = aug,
  journal = {\mnras},
  volume = {406},
  number = {2},
  pages = {1208},
  doi = {10.1111/j.1365-2966.2010.16749.x},
  langid = {english}
}

@article{dauser2019,
  title = {{{SIXTE}}: A Generic {{X-ray}} Instrument Simulation Toolkit},
  author = {Dauser, Thomas and Falkner, Sebastian and Lorenz, Maximilian and Kirsch, Christian and Peille, Philippe and Cucchetti, Edoardo and Schmid, Christian and Brand, Thorsten and Oertel, Mirjam and Smith, Randall and Wilms, J{\"o}rn},
  year = 2019,
  month = sep,
  volume = {630},
  pages = {A66},
  doi = {10.1051/0004-6361/201935978},
  langid = {english},
  journal = {\aap}
}

@article{davidson1973,
  title = {Neutron-{{Star Accretion}} in a {{Stellar Wind}}: {{Model}} for a {{Pulsed X-Ray Source}}},
  author = {Davidson, Kris and Ostriker, Jeremiah P.},
  year = 1973,
  month = jan,
  volume = {179},
  pages = {585--598},
  doi = {10.1086/151897},
  langid = {english},
  journal = {\apj}
}

@article{delia2013,
  title = {The Seven Year {{Swift-XRT}} Point Source Catalog ({{1SWXRT}})},
  author = {D'Elia, V. and Perri, M. and Puccetti, S. and Capalbi, M. and Giommi, P. and Burrows, D. N. and Campana, S. and Tagliaferri, G. and Cusumano, G. and Evans, P. and Gehrels, N. and Kennea, J. and Moretti, A. and Nousek, J. A. and Osborne, J. P. and Romano, P. and Stratta, G.},
  year = 2013,
  month = mar,
  volume = {551},
  pages = {A142},
  doi = {10.1051/0004-6361/201220863},
  langid = {english},
  journal = {\aap}
}

@article{dodd2024,
  title = {Gaia Uncovers Difference in {{B}} and {{Be}} Star Binarity at Small Scales: Evidence for Mass Transfer Causing the {{Be}} Phenomenon},
  author = {Dodd, Jonathan M and Oudmaijer, Ren{\'e} D and Radley, Isaac C and Vioque, Miguel and Frost, Abigail J},
  year = 2024,
  month = jan,
  volume = {527},
  number = {2},
  pages = {3076--3086},
  doi = {10.1093/mnras/stad3105},
  journal = {\mnras}
}

@article{doroshenko2012,
  title = {The Hard {{X-ray}} Emission of {{X Persei}}},
  author = {Doroshenko, V. and Santangelo, A. and Kreykenbohm, I. and Doroshenko, R.},
  year = 2012,
  month = apr,
  volume = {540},
  pages = {L1},
  doi = {10.1051/0004-6361/201218878},
  journal = {\aap}
}

@article{doroshenko2014,
  title = {Population of the {{Galactic X-ray}} Binaries and {{eRosita}}},
  author = {Doroshenko, V. and Ducci, L. and Santangelo, A. and Sasaki, M.},
  year = 2014,
  month = jul,
  volume = {567},
  pages = {A7},
  publisher = {EDP Sciences},
  doi = {10.1051/0004-6361/201423766},
  langid = {english},
  journal = {\aap}
}

@article{doroshenko2022,
  title = {{{SRGA J124404}}.1-632232/{{SRGU J124403}}.8-632231: A New {{X-ray}} Pulsar Discovered in the All-Sky Survey by {{SRG}}},
  author = {Doroshenko, V. and Staubert, R. and Maitra, C. and Rau, A. and Haberl, F. and Santangelo, A. and Schwope, A. and Wilms, J. and Buckley, D. A. H. and Semena, A. and Mereminskiy, I. and Lutovinov, A. and Gromadzki, M. and Townsend, L. J. and Monageng, I. M.},
  year = 2022,
  month = may,
  volume = {661},
  eprint = {2106.14539},
  primaryclass = {astro-ph},
  pages = {A21},
  doi = {10.1051/0004-6361/202141147},
  archiveprefix = {arXiv},
  journal = {\aap}
}

@article{ducci2018,
  title = {In-Depth Study of Long-Term Variability in the {{X-ray}} Emission of the {{Be}}/{{X-ray}} Binary System {{AX J0049}}.4-7323},
  author = {Ducci, L. and Romano, P. and Malacaria, C. and Ji, L. and Bozzo, E. and Santangelo, A.},
  year = 2018,
  month = jun,
  volume = {614},
  pages = {A34},
  publisher = {EDP Sciences},
  doi = {10.1051/0004-6361/201731922},
  langid = {english},
  journal = {\aap}
}

@article{ducci2023,
  title = {Modelling the Expected Very High Energy {$\gamma$}-Ray Emission from Accreting Neutron Stars in {{X-ray}} Binaries},
  author = {Ducci, L and Romano, P and Vercellone, S and Santangelo, A},
  year = 2023,
  month = nov,
  volume = {525},
  number = {3},
  pages = {3923--3945},
  doi = {10.1093/mnras/stad2440},
  journal = {\mnras}
}

@article{el-badry2021,
  title = {A Stripped-Companion Origin for {{Be}} Stars: Clues from the Putative Black Holes {{HR}} 6819 and {{LB-1}}},
  author = {{El-Badry}, Kareem and Quataert, Eliot},
  year = 2021,
  month = apr,
  volume = {502},
  number = {3},
  pages = {3436--3455},
  doi = {10.1093/mnras/stab285},
  journal = {\mnras}
}

@article{elshamouty2016a,
  title = {The Soft {{X-ray}} Spectrum of the High-Mass {{X-ray}} Binary {{V0332}}+53 in Quiescence},
  author = {Elshamouty, Khaled G. and Heinke, Craig O. and Chouinard, Rhys},
  year = 2016,
  month = nov,
  journal = {\mnras},
  volume = {463},
  number = {1},
  pages = {78},
  doi = {10.1093/mnras/stw1940},
  langid = {english}
}

@article{escorial2018a,
  title = {Discovery of Accretion-Driven Pulsations in the Prolonged Low {{X-ray}} Luminosity State of the {{Be}}/{{X-ray}} Transient {{GX}} 304-1},
  author = {{Rouco Escorial}, A. and van den Eijnden, J. and Wijnands, R.},
  year = 2018,
  month = dec,
  volume = {620},
  eprint = {1811.01453},
  primaryclass = {astro-ph},
  pages = {L13},
  doi = {10.1051/0004-6361/201834572},
  archiveprefix = {arXiv},
  journal = {\aap}
}

@article{evans2020,
  title = {{{2SXPS}}: {{An Improved}} and {{Expanded Swift X-Ray Telescope Point-source Catalog}}},
  author = {Evans, P. A. and Page, K. L. and Osborne, J. P. and Beardmore, A. P. and Willingale, R. and Burrows, D. N. and Kennea, J. A. and Perri, M. and Capalbi, M. and Tagliaferri, G. and Cenko, S. B.},
  year = 2020,
  month = apr,
  volume = {247},
  number = {2},
  pages = {54},
  doi = {10.3847/1538-4365/ab7db9},
  langid = {english},
  journal = {\apjs}
}

@article{falkner2026,
  title = {Modeling Accretion Columns in Accretion-Powered Pulsars - {{II}}. {{Directly}} Observable Column Emission},
  author = {Falkner, S. and {Sokolova-Lapa}, E. and Schwarm, F.-W. and Bissinger, M. and Ballhausen, R. and Postnov, K. A. and Dauser, T. and Hemphill, P. B. and Pottschmidt, K. and F{\"u}rst, F. and Kretschmar, P. and Sch{\"o}nherr, G. and Wolff, M. T. and Becker, P. A. and Falanga, M. and Ferrigno, C. and Rothschild, R. E. and Staubert, R. and Kreykenbohm, I. and Wilms, J.},
  year = 2026,
  month = jul,
  volume = {711},
  pages = {A225},
  publisher = {EDP Sciences},
  doi = {10.1051/0004-6361/202453643},
  langid = {english},
  journal = {\aap}
}

@article{faltova2026,
  title = {High-Mass {{X-ray}} Binaries as Members of Open Clusters},
  author = {Faltov{\'a}, N. and Paunzen, E. and Pri{\v s}egen, M.},
  year = 2026,
  month = apr,
  volume = {708},
  pages = {A234},
  publisher = {EDP Sciences},
  doi = {10.1051/0004-6361/202558406},
  langid = {english},
  journal = {\aap}
}

@article{fortin2022,
  title = {Constraints to Neutron-Star Kicks in High-Mass {{X-ray}} Binaries with {{Gaia EDR3}}},
  author = {Fortin, Francis and Garc{\'i}a, Federico and Chaty, Sylvain and {Chassande-Mottin}, Eric and Bunzel, Adolfo Simaz},
  year = 2022,
  month = sep,
  volume = {665},
  pages = {A31},
  publisher = {EDP Sciences},
  doi = {10.1051/0004-6361/202140853},
  langid = {english},
  journal = {\aap}
}

@article{fortin2023,
  title = {A Catalogue of High-Mass {{X-ray}} Binaries in the {{Galaxy}}: From the {{INTEGRAL}} to the {{Gaia}} Era},
  author = {Fortin, Francis and Garc{\'i}a, Federico and Bunzel, Adolfo Simaz and Chaty, Sylvain},
  year = 2023,
  month = mar,
  volume = {671},
  pages = {A149},
  publisher = {EDP Sciences},
  doi = {10.1051/0004-6361/202245236},
  langid = {english},
  journal = {\aap}
}

@article{frost2022,
  title = {{{HR}} 6819 Is a Binary System with No Black Hole - {{Revisiting}} the Source with Infrared Interferometry and Optical Integral Field Spectroscopy},
  author = {Frost, A. J. and Bodensteiner, J. and Rivinius, Th and Baade, D. and Merand, A. and Selman, F. and {Abdul-Masih}, M. and Banyard, G. and Bordier, E. and Dsilva, K. and Hawcroft, C. and Mahy, L. and Reggiani, M. and Shenar, T. and Cabezas, M. and Hadrava, P. and Heida, M. and Klement, R. and Sana, H.},
  year = 2022,
  month = mar,
  volume = {659},
  pages = {L3},
  publisher = {EDP Sciences},
  doi = {10.1051/0004-6361/202143004},
  langid = {english},
  journal = {\aap}
}

@article{granada2013,
  title = {Populations of Rotating Stars - {{II}}. {{Rapid}} Rotators and Their Link to {{Be-type}} Stars},
  author = {Granada, A. and Ekstr{\"o}m, S. and Georgy, C. and Krti{\v c}ka, J. and Owocki, S. and Meynet, G. and Maeder, A.},
  year = 2013,
  month = may,
  volume = {553},
  pages = {A25},
  publisher = {EDP Sciences},
  doi = {10.1051/0004-6361/201220559},
  langid = {english},
  journal = {\aap}
}

@article{grebenev2007,
  title = {The First Observation of {{AX J1749}}.1-2733 in a Bright {{X-ray}} State---{{Another}} Fast Transient Revealed by {{INTEGRAL}}},
  author = {Grebenev, S. A. and Sunyaev, R. A.},
  year = 2007,
  month = mar,
  journal = {Astron. Lett.},
  volume = {33},
  number = {3},
  pages = {149--158},
  doi = {10.1134/S1063773707030024},
  langid = {english}
}

@article{grimm2002,
  title = {The {{Milky Way}} in {{X-rays}} for an Outside Observer: {{Log}}( {{{\emph{N}}}} )-{{Log}}( {{{\emph{S}}}} ) and Luminosity Function of {{X-ray}} Binaries from {{RXTE}}/{{ASM}} Data},
  author = {Grimm, H.-J. and Gilfanov, M. and Sunyaev, R.},
  year = 2002,
  month = sep,
  volume = {391},
  number = {3},
  pages = {923--944},
  doi = {10.1051/0004-6361:20020826},
  langid = {english},
  journal = {\aap}
}

@article{grimm2003,
  title = {X-Ray {{Binaries}} in the {{Milky Way}} and {{Other Galaxies}}},
  author = {Grimm, Hans-Jakob and Gilfanov, Marat and Sunyaev, Rashid},
  year = 2003,
  month = dec,
  volume = {3},
  number = {S1},
  pages = {257--269},
  doi = {10.1088/1009-9271/3/S1/257},
  langid = {english},
  journal = {\cjaa}
}

@article{haberl2016,
  title = {High-Mass {{X-ray}} Binaries in the {{Small Magellanic Cloud}}},
  author = {Haberl, F. and Sturm, R.},
  year = 2016,
  month = feb,
  volume = {586},
  pages = {A81},
  publisher = {EDP Sciences},
  doi = {10.1051/0004-6361/201527326},
  langid = {english},
  journal = {\aap}
}

@article{hastings2021,
  title = {Stringent Upper Limit on {{Be}} Star Fractions Produced by Binary Interaction},
  author = {Hastings, B. and Langer, N. and Wang, C. and Schootemeijer, A. and Milone, A. P.},
  year = 2021,
  month = sep,
  volume = {653},
  pages = {A144},
  publisher = {EDP Sciences},
  doi = {10.1051/0004-6361/202141269},
  langid = {english},
  journal = {\aap}
}

@article{heinz2013,
  title = {The {{Youngest Known X}}-Ray {{Binary}}: {{Circinus X-1}} and Its {{Natal Supernova Remnant}}},
  author = {Heinz, S. and others},
  year = 2013,
  journal = {Astrophys. J.},
  volume = {779},
  pages = {171},
  doi = {10.1088/0004-637X/779/2/171}
}

@article{h.e.s.s.collaboration2020,
  title = {Detection of Very-High-Energy {$\gamma$}-Ray Emission from the Colliding Wind Binary {$\eta$} {{Car}} with {{H}}.{{E}}.{{S}}.{{S}}.},
  author = {{H. E. S. S. Collaboration} and Abdalla, H. and Adam, R. and Aharonian, F. and Ait Benkhali, F. and Ang{\"u}ner, E. O. and Arakawa, M. and Arcaro, C. and Armand, C. and Armstrong, T. and Ashkar, H. and Backes, M. and Barbosa Martins, V. and Barnard, M. and Becherini, Y. and Berge, D. and Bernl{\"o}hr, K. and Blackwell, R. and B{\"o}ttcher, M. and Boisson, C. and Bolmont, J. and Bonnefoy, S. and Bregeon, J. and Breuhaus, M. and Brun, F. and Brun, P. and Bryan, M. and B{\"u}chele, M. and Bulik, T. and Bylund, T. and Caroff, S. and Carosi, A. and Casanova, S. and Cerruti, M. and Chand, T. and Chandra, S. and Chen, A. and Colafrancesco, S. and Cotter, G. and Cury{\l}o, M. and Davids, I. D. and Davies, J. and Deil, C. and Devin, J. and {deWilt}, P. and Dirson, L. and {Djannati-Ata{\"i}}, A. and Dmytriiev, A. and Donath, A. and Doroshenko, V. and Dyks, J. and Egberts, K. and Eichhorn, F. and Emery, G. and Ernenwein, J.-P. and Eschbach, S. and Feijen, K. and Fegan, S. and Fiasson, A. and Fontaine, G. and Funk, S. and F{\"u}{\ss}ling, M. and Gabici, S. and Gallant, Y. A. and Gat{\'e}, F. and Giavitto, G. and Giunti, L. and Glawion, D. and Glicenstein, J. F. and Gottschall, D. and Grondin, M.-H. and Hahn, J. and Haupt, M. and Heinzelmann, G. and Henri, G. and Hermann, G. and Hinton, J. A. and Hofmann, W. and Hoischen, C. and Holch, T. L. and Holler, M. and H{\"o}rbe, M. and Horns, D. and Huber, D. and Iwasaki, H. and Jamrozy, M. and Jankowsky, D. and Jankowsky, F. and {Jardin-Blicq}, A. and Joshi, V. and {Jung-Richardt}, I. and Kastendieck, M. A. and Katarzy{\'n}ski, K. and Katsuragawa, M. and Katz, U. and Khangulyan, D. and Kh{\'e}lifi, B. and King, J. and Klepser, S. and Klu{\'z}niak, W. and Komin, {\relax Nu}. and Kosack, K. and Kostunin, D. and Kreter, M. and Lamanna, G. and Lemi{\`e}re, A. and {Lemoine-Goumard}, M. and Lenain, J.-P. and Leser, E. and Levy, C. and Lohse, T. and Lypova, I. and Mackey, J. and Majumdar, J. and Malyshev, D. and Malyshev, D. and Marandon, V. and Marchegiani, P. and Marcowith, A. and Mares, A. and {Mart{\'i}-Devesa}, G. and Marx, R. and Maurin, G. and Meintjes, P. J. and Moderski, R. and Mohamed, M. and Mohrmann, L. and Moore, C. and Morris, P. and Moulin, E. and Muller, J. and Murach, T. and Nakashima, S. and Nakashima, K. and {de Naurois}, M. and Ndiyavala, H. and Niederwanger, F. and Niemiec, J. and Oakes, L. and O'Brien, P. and Odaka, H. and Ohm, S. and {de Ona Wilhelmi}, E. and Ostrowski, M. and Panter, M. and Parsons, R. D. and Perennes, C. and Petrucci, P.-O. and Peyaud, B. and Piel, Q. and Pita, S. and Poireau, V. and Priyana Noel, A. and Prokhorov, D. A. and Prokoph, H. and P{\"u}hlhofer, G. and Punch, M. and Quirrenbach, A. and Raab, S. and Rauth, R. and Reimer, A. and Reimer, O. and Remy, Q. and Renaud, M. and Rieger, F. and Rinchiuso, L. and Romoli, C. and Rowell, G. and Rudak, B. and {Ruiz-Velasco}, E. and Sahakian, V. and Sailer, S. and Saito, S. and Sanchez, D. A. and Santangelo, A. and Sasaki, M. and Scalici, M. and Schlickeiser, R. and Sch{\"u}ssler, F. and Schulz, A. and Schutte, H. M. and Schwanke, U. and Schwemmer, S. and {Seglar-Arroyo}, M. and Senniappan, M. and Seyffert, A. S. and Shafi, N. and Shiningayamwe, K. and Simoni, R. and Sinha, A. and Sol, H. and Specovius, A. and Spencer, S. and {Spir-Jacob}, M. and Stawarz, {\L}. and Steenkamp, R. and Stegmann, C. and Steppa, C. and Takahashi, T. and Tavernier, T.},
  year = 2020,
  month = mar,
  volume = {635},
  pages = {A167},
  publisher = {EDP},
  doi = {10.1051/0004-6361/201936761},
  journal = {\aap}
}

@article{illarionov1975,
  title = {Why the {{Number}} of {{Galactic X-ray Stars Is}} so {{Small}}?},
  author = {Illarionov, A. F. and Sunyaev, R. A.},
  year = 1975,
  month = feb,
  journal = {\aap},
  volume = {39},
  pages = {185},
  langid = {english}
}

@article{intzand2005,
  title = {Chandra Observation of the Fast {{X-ray}} Transient {{IGR J17544-2619}}: Evidence for a Neutron Star?},
  author = {In 'T Zand, J. J. M.},
  year = 2005,
  month = oct,
  volume = {441},
  number = {1},
  pages = {L1-L4},
  doi = {10.1051/0004-6361:200500162},
  langid = {english},
  journal = {\aap}
}

@article{islam2016,
  title = {Effects of Variability of {{X-ray}} Binaries on the {{X-ray}} Luminosity Functions of {{Milky Way}}},
  author = {Islam, Nazma and Paul, Biswajit},
  year = 2016,
  month = aug,
  volume = {47},
  eprint = {1602.08287},
  primaryclass = {astro-ph},
  pages = {81--87},
  doi = {10.1016/j.newast.2016.02.008},
  archiveprefix = {arXiv},
  journal = {\na}
}

@article{israel2000,
  title = {The Discovery of the Optical/{{IR}} Counterpart of the 12-s Transient {{X-ray}} Pulsar {{GS}} 0834-43},
  author = {Israel, G. L. and Covino, S. and Campana, S. and Polcaro, V. F. and Roche, P. and Stella, L. and Di Paola, A. and Lazzati, D. and Mereghetti, S. and Giallongo, E. and Fontana, A. and Verrecchia, F.},
  year = 2000,
  month = may,
  volume = {314},
  number = {1},
  pages = {87--91},
  publisher = {Oxford University Press (OUP)},
  doi = {10.1046/j.1365-8711.2000.03404.x},
  langid = {english},
  journal = {\mnras}
}

@article{jonker2007,
  title = {Detection of the Radial Velocity Curve of the {{B5-A0}} Supergiant Companion Star of {{Cir X-1}}?},
  author = {Jonker, P. G. and Nelemans, G. and Bassa, C. G.},
  year = 2007,
  month = jan,
  volume = {374},
  pages = {999--1005},
  publisher = {OUP},
  doi = {10.1111/j.1365-2966.2006.11210.x},
  journal = {\mnras}
}

@article{kaltenbrunner2026,
  title = {A Comprehensive Catalogue of High-Mass {{X-ray}} Binaries in the {{Large Magellanic Cloud}} Detected during the First {{eROSITA}} All-Sky Survey},
  author = {Kaltenbrunner, D. and Maitra, C. and Haberl, F. and Bodensteiner, J. and Bogensberger, D. and Buckley, D. A. H. and Cioni, M. R. L. and Greiner, J. and Monageng, I. and Udalski, A. and Vasilopoulos, G. and Willer, R.},
  year = 2026,
  month = mar,
  volume = {707},
  eprint = {2602.08152},
  primaryclass = {astro-ph},
  pages = {A225},
  doi = {10.1051/0004-6361/202555693},
  archiveprefix = {arXiv},
  journal = {\aap}
}

@article{kluzniak2007,
  title = {Magnetically {{Torqued Thin Accretion Disks}}},
  author = {Klu{\'z}niak, W. and Rappaport, S.},
  year = 2007,
  month = dec,
  volume = {671},
  number = {2},
  pages = {1990},
  publisher = {IOP Publishing},
  doi = {10.1086/522954},
  langid = {english},
  journal = {\apj}
}

@article{konig2022,
  title = {X-Ray Detection of a Nova in the Fireball Phase},
  author = {K{\"o}nig, Ole and Wilms, J{\"o}rn and Arcodia, Riccardo and Dauser, Thomas and Dennerl, Konrad and Doroshenko, Victor and Haberl, Frank and H{\"a}mmerich, Steven and Kirsch, Christian and Kreykenbohm, Ingo and Lorenz, Maximilian and Malyali, Adam and Merloni, Andrea and Rau, Arne and Rauch, Thomas and Sala, Gloria and Schwope, Axel and Suleimanov, Valery and Weber, Philipp and Werner, Klaus},
  year = 2022,
  month = may,
  journal = {Nature},
  volume = {605},
  number = {7909},
  pages = {248--250},
  doi = {10.1038/s41586-022-04635-y},
  langid = {english}
}

@article{kretschmar2019,
  title = {Advances in {{Understanding High-Mass X-ray Binaries}} with {{INTEGRALand Future Directions}}},
  author = {Kretschmar, Peter and F{\"u}rst, Felix and Sidoli, Lara and Bozzo, Enrico and {Alfonso-Garz{\'o}n}, Julia and Bodaghee, Arash and Chaty, Sylvain and Chernyakova, Masha and Ferrigno, Carlo and Manousakis, Antonios and Negueruela, Ignacio and Postnov, Konstantin and Paizis, Adamantia and Reig, Pablo and {Rodes-Roca}, Jos{\'e} Joaqu{\'i}n and Tsygankov, Sergey and Bird, Antony J. and {Bissinger n{\'e} K{\"u}hnel}, Matthias and Blay, Pere and Caballero, Isabel and Coe, Malcolm J. and Domingo, Albert and Doroshenko, Victor and Ducci, Lorenzo and Falanga, Maurizio and Grebenev, Sergei A. and Grinberg, Victoria and Hemphill, Paul and Kreykenbohm, Ingo and {Kreykenbohm n{\'e} Fritz}, Sonja and Li, Jian and Lutovinov, Alexander A. and {Mart{\'i}nez-N{\'u}{\~n}ez}, Silvia and {Mas-Hesse}, J. Miguel and Masetti, Nicola and McBride, Vanessa A. and Neronov, Andrii and Pottschmidt, Katja and Rodriguez, J{\'e}r{\^o}me and Romano, Patrizia and Rothschild, Richard E. and Santangelo, Andrea and Sguera, Vito and Staubert, R{\"u}diger and Tomsick, John A. and Torrej{\'o}n, Jos{\'e} Miguel and Torres, Diego F. and Walter, Roland and Wilms, J{\"o}rn and {Wilson-Hodge}, Colleen A. and Zhang, Shu},
  year = 2019,
  month = dec,
  journal = {New Astron. Rev.},
  volume = {86},
  pages = {101546},
  doi = {10.1016/j.newar.2020.101546},
  langid = {english}
}

@article{kuhnel2017,
  title = {Evidence for Different Accretion Regimes in {{GRO J1008-57}}},
  author = {K{\"u}hnel, Matthias and F{\"u}rst, Felix and Pottschmidt, Katja and Kreykenbohm, Ingo and Ballhausen, Ralf and Falkner, Sebastian and Rothschild, Richard E. and Klochkov, Dmitry and Wilms, J{\"o}rn},
  year = 2017,
  month = nov,
  volume = {607},
  pages = {A88},
  doi = {10.1051/0004-6361/201629131},
  langid = {english},
  journal = {\aap}
}

@article{kyritsis2026,
  title = {Prescriptions for the Stochasticity Effect on the Integrated {{X-ray}} Luminosity of Star-Forming Galaxies:{{Implications}} for Selecting Star-Forming Galaxies and {{AGN}} in {{X-ray}} Surveys},
  author = {Kyritsis, Elias and Zezas, Andreas and Kovlakas, Konstantinos},
  year = 2026,
  month = jun,
  publisher = {arXiv},
  eprint = {arXiv:2606.30624},
  doi = {10.48550/arXiv.2606.30624},
  langid = {english},
  journal = {ArXiv e-prints}
}

@article{lasota2001,
  title = {The Disc Instability Model of Dwarf Novae and Low-Mass {{X-ray}} Binary Transients},
  author = {Lasota, Jean-Pierre},
  year = 2001,
  month = jun,
  journal = {New Astron. Rev.},
  volume = {45},
  number = {7},
  pages = {449--508},
  doi = {10.1016/S1387-6473(01)00112-9},
  langid = {english}
}

@article{lehmer2019,
  title = {X-{{Ray Binary Luminosity Function Scaling Relations}} for {{Local Galaxies Based}} on {{Subgalactic Modeling}}},
  author = {Lehmer, Bret D. and Eufrasio, Rafael T. and Tzanavaris, Panayiotis and {Basu-Zych}, Antara and Fragos, Tassos and Prestwich, Andrea and Yukita, Mihoko and Zezas, Andreas and Hornschemeier, Ann E. and Ptak, Andrew},
  year = 2019,
  month = jun,
  volume = {243},
  number = {1},
  pages = {3},
  doi = {10.3847/1538-4365/ab22a8},
  langid = {english},
  journal = {\apjs}
}

@article{linares2010,
  title = {The {{Return}} of the {{Bursts}}: {{Thermonuclear Flashes}} from {{Circinus X-1}}},
  author = {Linares, M. and Watts, A. and Altamirano, D. and Soleri, P. and Degenaar, N. and Yang, Y. and Wijnands, R. and Casella, P. and Homan, J. and Chakrabarty, D. and Rea, N. and {Armas-Padilla}, M. and Cavecchi, Y. and Kalamkar, M. and Kaur, R. and Patruno, A. and {van der Klis}, M.},
  year = 2010,
  month = aug,
  volume = {719},
  pages = {L84-L89},
  publisher = {IOP},
  doi = {10.1088/2041-8205/719/1/L84},
  journal = {\apj}
}

@article{lutovinov2013,
  title = {Population of Persistent High-Mass {{X-ray}} Binaries in the {{Milky Way}}},
  author = {Lutovinov, A. A. and Revnivtsev, M. G. and Tsygankov, S. S. and Krivonos, R. A.},
  year = 2013,
  month = may,
  volume = {431},
  pages = {327--341},
  publisher = {OUP},
  doi = {10.1093/mnras/stt168},
  journal = {\mnras}
}

@article{markozov2024,
  title = {Apparent Luminosity and Pulsed Fraction Affected by Gravitational Lensing of Accretion Columns in Bright {{X-ray}} Pulsars},
  author = {Markozov, Ivan D and Mushtukov, Alexander A},
  year = 2024,
  month = jan,
  volume = {527},
  number = {3},
  pages = {5374--5384},
  doi = {10.1093/mnras/stad3248},
  journal = {\mnras}
}

@article{martinez-nunez2017,
  title = {Towards a Unified View of Inhomogeneous Stellar Winds in Isolated Supergiant Stars and Supergiant High Mass {{X-ray}} Binaries},
  author = {{Mart{\'i}nez-N{\'u}{\~n}ez}, Silvia and Kretschmar, Peter and Bozzo, Enrico and Oskinova, Lidia M. and Puls, Joachim and Sidoli, Lara and Sundqvist, Jon Olof and Blay, Pere and Falanga, Maurizio and F{\"u}rst, Felix and {G{\'i}menez-Garc{\'i}a}, {\'A}ngel and Kreykenbohm, Ingo and K{\"u}hnel, Matthias and Sander, Andreas and Torrej{\'o}n, Jos{\'e} Miguel and Wilms, J{\"o}rn},
  year = 2017,
  month = oct,
  volume = {212},
  number = {1-2},
  eprint = {1701.08618},
  primaryclass = {astro-ph},
  pages = {59--150},
  doi = {10.1007/s11214-017-0340-1},
  archiveprefix = {arXiv},
  journal = {\ssr}
}

@article{matsuoka2009,
  title = {The {{MAXI Mission}} on the {{ISS}}: {{Science}} and {{Instruments}} for {{Monitoring All-Sky X-Ray Images}}},
  author = {Matsuoka, Masaru and Kawasaki, Kazuyoshi and Ueno, Shiro and Tomida, Hiroshi and Kohama, Mitsuhiro and Suzuki, Motoko and Adachi, Yasuki and Ishikawa, Masaki and Mihara, Tatehiro and Sugizaki, Mutsumi and Isobe, Naoki and Nakagawa, Yujin and Tsunemi, Hiroshi and Miyata, Emi and Kawai, Nobuyuki and Kataoka, Jun and Morii, Mikio and Yoshida, Atsumasa and Negoro, Hitoshi and Nakajima, Motoki and Ueda, Yoshihiro and Chujo, Hirotaka and Yamaoka, Kazutaka and Yamazaki, Osamu and Nakahira, Satoshi and You, Tetsuya and Ishiwata, Ryoji and Miyoshi, Sho and Eguchi, Satoshi and Hiroi, Kazuo and Katayama, Haruyoshi and Ebisawa, Ken},
  year = 2009,
  month = oct,
  volume = {61},
  pages = {999},
  doi = {10.1093/pasj/61.5.999},
  langid = {english},
  journal = {\pasj}
}

@article{Merloni12,
  title = {{{eROSITA Science Book}}: {{Mapping}} the {{Structure}} of the {{Energetic Universe}}},
  author = {Merloni, A. and Predehl, P. and Becker, W. and B{\"o}hringer, H. and Boller, T. and Brunner, H. and Brusa, M. and Dennerl, K. and Freyberg, M. and Friedrich, P. and Georgakakis, A. and Haberl, F. and Hasinger, G. and Meidinger, N. and Mohr, J. and Nandra, K. and Rau, A. and Reiprich, T. H. and Robrade, J. and Salvato, M. and Santangelo, A. and Sasaki, M. and Schwope, A. and Wilms, J. and {German eROSITA Consortium}, the},
  year = {2012},
  month = sep,
  publisher = {arXiv},
  journal = {ArXiv e-prints},
  eprint = {1209.3114},
 primaryClass = {astro-ph.HE},
  doi = {10.48550/arXiv.1209.3114}
}

@article{merloni2024,
  title = {The {{SRG}}/{{eROSITA}} All-Sky Survey. {{First X-ray}} Catalogues and Data Release of the Western {{Galactic}} Hemisphere},
  author = {Merloni, A. and Lamer, G. and Liu, T. and {Ramos-Ceja}, M. E. and Brunner, H. and Bulbul, E. and Dennerl, K. and Doroshenko, V. and Freyberg, M. J. and Friedrich, S. and Gatuzz, E. and Georgakakis, A. and Haberl, F. and Igo, Z. and Kreykenbohm, I. and Liu, A. and Maitra, C. and Malyali, A. and Mayer, M. G. F. and Nandra, K. and Predehl, P. and Robrade, J. and Salvato, M. and Sanders, J. S. and Stewart, I. and {Tub{\'i}n-Arenas}, D. and Weber, P. and Wilms, J. and Arcodia, R. and Artis, E. and Aschersleben, J. and Avakyan, A. and Aydar, C. and Bahar, Y. E. and Balzer, F. and Becker, W. and Berger, K. and Boller, T. and Bornemann, W. and Br{\"u}ggen, M. and Brusa, M. and Buchner, J. and Burwitz, V. and Camilloni, F. and Clerc, N. and Comparat, J. and Coutinho, D. and Czesla, S. and Dannhauer, S. M. and Dauner, L. and Dauser, T. and Dietl, J. and Dolag, K. and Dwelly, T. and Egg, K. and Ehl, E. and Freund, S. and Friedrich, P. and Gaida, R. and Garrel, C. and Ghirardini, V. and Gokus, A. and Gr{\"u}nwald, G. and Grandis, S. and Grotova, I. and Gruen, D. and Gueguen, A. and H{\"a}mmerich, S. and Hamaus, N. and Hasinger, G. and Haubner, K. and Homan, D. and Ider Chitham, J. and Joseph, W. M. and Joyce, A. and K{\"o}nig, O. and Kaltenbrunner, D. M. and Khokhriakova, A. and Kink, W. and Kirsch, C. and Kluge, M. and Knies, J. and Krippendorf, S. and Krumpe, M. and Kurpas, J. and Li, P. and Liu, Z. and Locatelli, N. and Lorenz, M. and M{\"u}ller, S. and Magaudda, E. and Mannes, C. and McCall, H. and Meidinger, N. and Michailidis, M. and Migkas, K. and {Mu{\~n}oz-Giraldo}, D. and Musiimenta, B. and {Nguyen-Dang}, N. T. and Ni, Q. and Olechowska, A. and Ota, N. and Pacaud, F. and Pasini, T. and Perinati, E. and Pires, A. M. and Pommranz, C. and Ponti, G. and Poppenhaeger, K. and P{\"u}hlhofer, G. and Rau, A. and Reh, M. and Reiprich, T. H. and Roster, W. and Saeedi, S. and Santangelo, A. and Sasaki, M. and Schmitt, J. and Schneider, P. C. and Schrabback, T. and Schuster, N. and Schwope, A. and Seppi, R. and Serim, M. M. and Shreeram, S. and {Sokolova-Lapa}, E. and Starck, H. and Stelzer, B. and Stierhof, J. and Suleimanov, V. and Tenzer, C. and Traulsen, I. and Tr{\"u}mper, J. and Tsuge, K. and Urrutia, T. and Veronica, A. and Waddell, S. G. H. and Willer, R. and Wolf, J. and Yeung, M. C. H. and Zainab, A. and Zangrandi, F. and Zhang, X. and Zhang, Y. and Zheng, X.},
  year = 2024,
  month = feb,
  volume = {682},
  pages = {A34},
  publisher = {EDP},
  doi = {10.1051/0004-6361/202347165},
  journal = {\aap}
}

@article{mineo2011,
  title = {The Collective {{X-ray}} Luminosity of {{HMXB}} as a {{SFR}} Indicator},
  author = {Mineo, S. and Gilfanov, M. and Sunyaev, R.},
  year = 2011,
  month = may,
  journal = {\an},
  volume = {332},
  pages = {349},
  doi = {10.1002/asna.201011497}
}

@article{mineo2014,
  title = {X-Ray Emission from Star-Forming Galaxies - {{III}}. {{Calibration}} of the {{LX-SFR}} Relation up to Redshift z {$\approx$} 1.3},
  author = {Mineo, S. and Gilfanov, M. and Lehmer, B. D. and Morrison, G. E. and Sunyaev, R.},
  year = 2014,
  month = jan,
  volume = {437},
  pages = {1698--1707},
  publisher = {OUP},
  doi = {10.1093/mnras/stt1999},
  journal = {\mnras}
}

@article{miyasaka2013,
  title = {{{NuSTAR DETECTION OF HARD X-RAY PHASE LAGS FROM THE ACCRETING PULSAR GS}} 0834-430},
  author = {Miyasaka, Hiromasa and Bachetti, Matteo and Harrison, Fiona A. and F{\"u}rst, Felix and Barret, Didier and Bellm, Eric C. and Boggs, Steven E. and Chakrabarty, Deepto and Chenevez, Jerome and Christensen, Finn E. and Craig, William W. and Grefenstette, Brian W. and Hailey, Charles J. and Madsen, Kristin K. and Natalucci, Lorenzo and Pottschmidt, Katja and Stern, Daniel and Tomsick, John A. and Walton, Dominic J. and Wilms, J{\"o}rn and Zhang, William},
  year = 2013,
  month = sep,
  volume = {775},
  number = {1},
  pages = {65},
  publisher = {The American Astronomical Society},
  doi = {10.1088/0004-637X/775/1/65},
  langid = {english},
  journal = {\apj}
}

@article{mushtukov2021,
  title = {Spectrum Formation in {{X-ray}} Pulsars at Very Low Mass Accretion Rate: {{Monte Carlo}} Approach},
  author = {Mushtukov, Alexander A. and Suleimanov, Valery F. and Tsygankov, Sergey S. and Portegies Zwart, Simon},
  year = 2021,
  month = may,
  volume = {503},
  number = {4},
  pages = {5193--5203},
  doi = {10.1093/mnras/stab811},
  langid = {english},
  journal = {\mnras}
}

@article{naze2014,
  title = {X-Ray Emission from Magnetic Massive Stars},
  author = {Naz{\'e}, Y. and P{\'e}tit, V. and Rindbrand, M. and Cohen, D. and Owocki, S. and {ud-Doula}, A. and Wade, G. and Rauw, G.},
  year = 2014,
  month = oct,
  volume = {215},
  number = {1},
  pages = {10},
  publisher = {\apjs},
  doi = {10.1088/0067-0049/215/1/10},
  langid = {english},
  journal = {\apjs}
}

@article{naze2018,
  title = {Hot Stars Observed by {{XMM-Newton}} - {{II}}. {{A}} Survey of {{Oe}} and {{Be}} Stars},
  author = {Naz{\'e}, Ya{\"e}l and Motch, Christian},
  year = 2018,
  month = nov,
  volume = {619},
  pages = {A148},
  publisher = {EDP Sciences},
  doi = {10.1051/0004-6361/201833842},
  langid = {english},
  journal = {\aap}
}

@article{naze2022,
  title = {The {{X-ray}} Emission of {{Be}}+stripped Star Binaries\ding{72}},
  author = {Naz{\'e}, Ya{\"e}l and Rauw, Gregor and Smith, Myron A and Motch, Christian},
  year = 2022,
  month = sep,
  volume = {516},
  number = {3},
  pages = {3366--3380},
  doi = {10.1093/mnras/stac2245},
  langid = {english},
  journal = {\mnras}
}

@article{naze2023,
  title = {{{SRG}}/{{eROSITA}} Survey of {{Be}} Stars},
  author = {Naz{\'e}, Ya{\"e}l and Robrade, Jan},
  year = 2023,
  month = nov,
  volume = {525},
  number = {3},
  pages = {4186--4201},
  doi = {10.1093/mnras/stad2399},
  langid = {english},
  journal = {\mnras}
}

@article{naze2025,
  title = {Another One ({{BH}}+{{OB}} Pair) Bites the Dust},
  author = {Naz{\'e}, Ya{\"e}l and Rauw, Gregor},
  year = 2025,
  month = apr,
  volume = {696},
  pages = {A84},
  publisher = {EDP},
  doi = {10.1051/0004-6361/202453493},
  journal = {\aap}
}

@article{neiner2011,
  title = {{{THE Be STAR SPECTRA}} ({{BeSS}}) {{DATABASE}}},
  author = {Neiner, C. and {de Batz}, B. and Cochard, F. and Floquet, M. and Mekkas, A. and Desnoux, V.},
  year = 2011,
  month = sep,
  volume = {142},
  number = {5},
  pages = {149},
  publisher = {The American Astronomical Society},
  doi = {10.1088/0004-6256/142/5/149},
  langid = {english},
  journal = {\aj}
}

@article{neuhauser2020,
  title = {A Nearby Recent Supernova That Ejected the Runaway Star {$\zeta$} {{Oph}}, the Pulsar {{PSR B1706}}-16, and {{60Fe}} Found on {{Earth}}},
  author = {Neuh{\"a}user, R and Gie{\ss}ler, F and Hambaryan, V V},
  year = 2020,
  month = oct,
  volume = {498},
  number = {1},
  pages = {899--917},
  doi = {10.1093/mnras/stz2629},
  langid = {english},
  journal = {\mnras}
}

@article{neumann2023,
  title = {{{XRBcats}}: {{Galactic High Mass X-ray Binary Catalogue}}\ding{72}},
  author = {Neumann, M. and Avakyan, A. and Doroshenko, V. and Santangelo, A.},
  year = 2023,
  month = sep,
  volume = {677},
  pages = {A134},
  publisher = {EDP},
  doi = {10.1051/0004-6361/202245728},
  journal = {\aap}
}

@article{okazaki2001,
  title = {A Natural Explanation for Periodic {{X-ray}} Outbursts in {{Be}}/{{X-ray}} Binaries},
  author = {Okazaki, A. T. and Negueruela, I.},
  year = 2001,
  month = oct,
  volume = {377},
  number = {1},
  pages = {161--174},
  publisher = {EDP Sciences},
  doi = {10.1051/0004-6361:20011083},
  langid = {english},
  journal = {\aap}
}

@article{perna2006,
  title = {On the {{Spin}}-up/{{Spin}}-down {{Transitions}} in {{Accreting X-Ray Binaries}}},
  author = {Perna, Rosalba and Bozzo, Enrico and Stella, Luigi},
  year = 2006,
  month = mar,
  volume = {639},
  number = {1},
  pages = {363--376},
  doi = {10.1086/499227},
  langid = {english},
  journal = {\apj}
}

@article{persic2004,
  title = {2--10 {{keV}} Luminosity of High-Mass Binaries as a Gauge of Ongoing Star-Formation Rate},
  author = {Persic, M. and Rephaeli, Y. and Braito, V. and Cappi, M. and Ceca, R. Della and Franceschini, A. and Gruber, D. E.},
  year = 2004,
  month = jun,
  volume = {419},
  number = {3},
  pages = {849--862},
  publisher = {EDP Sciences},
  doi = {10.1051/0004-6361:20034500},
  langid = {english},
  journal = {\aap}
}

@article{predehl2021,
  title = {The {{eROSITA X-ray}} Telescope on {{SRG}}},
  author = {Predehl, P. and Andritschke, R. and Arefiev, V. and Babyshkin, V. and Batanov, O. and Becker, W. and B{\"o}hringer, H. and Bogomolov, A. and Boller, T. and Borm, K. and Bornemann, W. and Br{\"a}uninger, H. and Br{\"u}ggen, M. and Brunner, H. and Brusa, M. and Bulbul, E. and Buntov, M. and Burwitz, V. and Burkert, W. and Clerc, N. and Churazov, E. and Coutinho, D. and Dauser, T. and Dennerl, K. and Doroshenko, V. and Eder, J. and Emberger, V. and Eraerds, T. and Finoguenov, A. and Freyberg, M. and Friedrich, P. and Friedrich, S. and F{\"u}rmetz, M. and Georgakakis, A. and Gilfanov, M. and Granato, S. and Grossberger, C. and Gueguen, A. and Gureev, P. and Haberl, F. and H{\"a}lker, O. and Hartner, G. and Hasinger, G. and Huber, H. and Ji, L. and Kienlin, A. V. and Kink, W. and Korotkov, F. and Kreykenbohm, I. and Lamer, G. and Lomakin, I. and Lapshov, I. and Liu, T. and Maitra, C. and Meidinger, N. and Menz, B. and Merloni, A. and Mernik, T. and Mican, B. and Mohr, J. and M{\"u}ller, S. and Nandra, K. and Nazarov, V. and Pacaud, F. and Pavlinsky, M. and Perinati, E. and Pfeffermann, E. and Pietschner, D. and {Ramos-Ceja}, M. E. and Rau, A. and Reiffers, J. and Reiprich, T. H. and Robrade, J. and Salvato, M. and Sanders, J. and Santangelo, A. and Sasaki, M. and Scheuerle, H. and Schmid, C. and Schmitt, J. and Schwope, A. and Shirshakov, A. and Steinmetz, M. and Stewart, I. and Str{\"u}der, L. and Sunyaev, R. and Tenzer, C. and Tiedemann, L. and Tr{\"u}mper, J. and Voron, V. and Weber, P. and Wilms, J. and Yaroshenko, V.},
  year = 2021,
  month = mar,
  volume = {647},
  pages = {A1},
  doi = {10.1051/0004-6361/202039313},
  langid = {english},
  journal = {\aap}
}

@article{ramos-ceja2026,
  title = {The {{SRG}}/{{eROSITA All-Sky Survey DR2}}: {{Cumulative X-ray}} Catalogues from the First Three Surveys and Multi-Wavelength Counterparts in the Western {{Galactic}} Hemisphere},
  author = {{Ramos-Ceja}, M. E. and Lamer, G. and Salvato, M. and Merloni, A. and Sanders, J. S. and Georgakakis, A. and Liu, T. and Bulbul, E. and Buchner, J. and Dennerl, K. and Freyberg, M. J. and Friedrich, S. and Kreykenbohm, I. and Maitra, C. and Nandra, K. and Predehl, P. and Reiprich, T. H. and Robrade, J. and Schwope, A. and Shirley, R. and Stelzer, B. and Stewart, I. and Seppi, R. and Starck, H. and {Tubin-Arenas}, D. and Traulsen, I. and Artis, E. and Aydar, C. and Baldini, P. and Balzer, F. and Becker, W. and Bennedik, M. M. and Bornemann, W. and Brueggen, M. and Brink, J. and Brusa, M. and Burwitz, V. and i Saguer, M. Canal and Clerc, N. and Comparat, J. and Coriat, M. and {Correa-Rodrigues}, J. V. and Czesla, S. and Dauner, L. and Dietl, J. and Ding, Z. and Ducci, L. and Dwelly, T. and Fiorino, L. and Freund, S. and Friedrich, P. and Gaida, R. and Gatuzz, E. and Guida, S. T. and Haemmerich, S. and Haberl, F. and Hartner, G. and {Hernandez-Diaz}, S. and Igo, Z. and Ilic, N. and Kaltenbrunner, D. M. and Khokhriakova, A. and Kink, W. and Kirsch, C. and Kluge, M. and Krippendorf, S. and Krumpe, M. and Kulkarni, S. and Kurpas, J. and Kyritsis, E. and Laktionov, R. and Liu, A. and Lorenz, M. and Malavasi, N. and Mayer, M. G. F. and Meidinger, N. and Mistele, T. and Mueller, S. and {Mu{\~n}oz-Giraldo}, D. and {Nguyen-Dang}, N. T. and Ni, Q. and Ok, S. and Ota, N. and Puehlhofer, G. and Pacaud, F. and Pandya, A. and Perinati, E. and Pommranz, C. and Ponti, G. and Poppenhaeger, K. and Pradeep, K. G. and Rau, A. and Roster, W. and Rukdee, S. and Saeedi, S. and Santangelo, A. and Sasaki, M. and Sheth, S. and Shreeram, S. and Sommer, M. and Srivastava, A. and Suleimanov, V. and Truemper, J. and Vasilas, N. and Veronica, A. and Webb, N. and Weber, P. and Wilms, J. and Yeung, M. C. H. and Zangrandi, F. and Zelmer, S. and Zhang, X. and Zhang, Y. and Zheng, X.},
  year = 2026,
  month = jul,
  number = {arXiv:2607.27772},
  eprint = {2607.27772},
  primaryclass = {astro-ph.HE},
  journal = {submitted to \aap\xspace},
  publisher = {ArXiv},
  doi = {10.48550/arXiv.2607.27772},
  archiveprefix = {arXiv}
}

@article{reig1999,
  title = {Discovery of {{X-ray}} Pulsations in the {{Be}}/{{X-ray}} Binary {{LS}} 992/{{RX J0812}}.4---3114},
  author = {Reig, Pablo and Roche, Paul},
  year = 1999,
  month = jun,
  volume = {306},
  number = {1},
  pages = {95--99},
  doi = {10.1046/j.1365-8711.1999.02463.x},
  journal = {\mnras}
}

@article{reig2011,
  title = {Be/{{X-ray}} Binaries},
  author = {Reig, Pablo},
  year = 2011,
  month = mar,
  volume = {332},
  number = {1},
  eprint = {1101.5036},
  primaryclass = {astro-ph},
  pages = {1--29},
  doi = {10.1007/s10509-010-0575-8},
  archiveprefix = {arXiv},
  journal = {\apss}
}

@article{romano2015,
  title = {Seven Years with the {{Swift Supergiant Fast X-ray Transients}} Project},
  author = {Romano, P.},
  year = 2015,
  month = sep,
  journal = {J. High Energy Astrophys.},
  volume = {7},
  pages = {126--136},
  doi = {10.1016/j.jheap.2015.04.008},
  langid = {english}
}

@article{roucoescorial2017,
  title = {The Low-Luminosity Behaviour of the {{4U}} 0115+63 {{Be}}/{{X-ray}} Transient},
  author = {Rouco Escorial, A. and Bak Nielsen, A. S. and Wijnands, R. and Cavecchi, Y. and Degenaar, N. and Patruno, A.},
  year = 2017,
  month = dec,
  volume = {472},
  pages = {1802--1808},
  publisher = {OUP},
  doi = {10.1093/mnras/stx2111},
  journal = {\mnras}
}

@phdthesis{roucoescorial2019,
title = {Be/{{X-ray}} Transients at Low {{X-ray}} Luminosity},
author = {Rouco Escorial, A.},
year = 2019,
isbn = {978-94-6323-839-7},
langid = {english},
school =  {University of Amsterdam },
addendum = {University of Amsterdam}
}

@article{roucoescorial2020,
  title = {Recurrent Low-Level Luminosity Behaviour after a Giant Outburst in the {{Be}}/{{X-ray}} Transient {{4U}} 0115+63},
  author = {Rouco Escorial, A. and Wijnands, R. and {van den Eijnden}, J. and Patruno, A. and Degenaar, N. and Parikh, A. and Ootes, L. S.},
  year = 2020,
  month = jun,
  volume = {638},
  pages = {A152},
  doi = {10.1051/0004-6361/201936287},
  langid = {english},
  journal = {\aap}
}

@book{shakura_2018,
  title = {Accretion {{Flows}} in {{Astrophysics}}},
  author = {Shakura, Nikolay},
  year = {2018},
  month = {jan},
  series = {Astrophysics and Space Science Library},
  volume = {454},
  publisher = {Springer},
  doi = {10.1007/978-3-319-93009-1}
}

@article{shakura2012,
  title = {Theory of Quasi-Spherical Accretion in {{X-ray}} Pulsars: {{Quasi-spherical}} Accretion},
  author = {Shakura, N. and Postnov, K. and Kochetkova, A. and Hjalmarsdotter, L.},
  year = 2012,
  month = feb,
  volume = {420},
  number = {1},
  pages = {216--236},
  doi = {10.1111/j.1365-2966.2011.20026.x},
  langid = {english},
  journal = {\mnras}
}

@article{shakura2013,
  title = {On the Nature of `off' States in Slowly Rotating Low-Luminosity {{X-ray}} Pulsars},
  author = {Shakura, N. and Postnov, K. and Hjalmarsdotter, L.},
  year = 2013,
  month = jan,
  volume = {428},
  number = {1},
  pages = {670--677},
  doi = {10.1093/mnras/sts062},
  langid = {english},
  journal = {\mnras}
}

@article{shao2014,
  title = {{{ON THE FORMATION OF Be STARS THROUGH BINARY INTERACTION}}},
  author = {Shao, Yong and Li, Xiang-Dong},
  year = 2014,
  month = nov,
  volume = {796},
  number = {1},
  pages = {37},
  publisher = {The American Astronomical Society},
  doi = {10.1088/0004-637X/796/1/37},
  langid = {english},
  journal = {\apj}
}

@article{shtykovskiy2005,
  title = {High-Mass {{X-ray}} Binaries in the {{Small Magellanic Cloud}}: The Luminosity Function},
  author = {Shtykovskiy, P. and Gilfanov, M.},
  year = 2005,
  month = sep,
  volume = {362},
  number = {3},
  pages = {879--890},
  doi = {10.1111/j.1365-2966.2005.09320.x},
  journal = {\mnras}
}

@article{sidoli2008,
  title = {Monitoring {{Supergiant Fast X-Ray Transients}} with {{Swift}}. {{I}}. {{Behavior Outside Outbursts}}},
  author = {Sidoli, L. and Romano, P. and Mangano, V. and Pellizzoni, A. and Kennea, J. A. and Cusumano, G. and Vercellone, S. and Paizis, A. and Burrows, D. N. and Gehrels, N.},
  year = 2008,
  month = nov,
  volume = {687},
  number = {2},
  pages = {1230},
  publisher = {IOP Publishing},
  doi = {10.1086/590077},
  langid = {english},
  journal = {\apj}
}

@article{sidoli2018,
  title = {An {{INTEGRAL}} Overview of {{High-Mass X}}--Ray {{Binaries}}: Classes or Transitions?},
  author = {Sidoli, L and Paizis, A},
  year = 2018,
  month = dec,
  volume = {481},
  number = {2},
  pages = {2779--2803},
  doi = {10.1093/mnras/sty2428},
  langid = {english},
  journal = {\mnras}
}

@article{sidoli2022,
  title = {{{XMM-Newton}} Discovery of Very High Obscuration in the Candidate Supergiant Fast {{X-ray}} Transient {{AX J1714}}.1-3912},
  author = {Sidoli, L. and Sguera, V. and Esposito, P. and Oskinova, L. and Polletta, M.},
  year = 2022,
  month = may,
  volume = {512},
  number = {2},
  pages = {2929--2935},
  doi = {10.1093/mnras/stac691},
  langid = {english},
  journal = {\mnras}
}

@article{sokolova-lapa2021,
  title = {X-Ray Emission from Magnetized Neutron Star Atmospheres at Low Mass Accretion Rates. {{I}}. {{Phase-averaged}} Spectrum},
  author = {{Sokolova-Lapa}, E. and Gornostaev, M. and Wilms, J. and Ballhausen, R. and Falkner, S. and Postnov, K. and Thalhammer, P. and F{\"u}rst, F. and Garc{\'i}a, J. A. and Shakura, N. and Becker, P. A. and Wolff, M. T. and Pottschmidt, K. and H{\"a}rer, L. and Malacaria, C.},
  year = 2021,
  month = jul,
  volume = {651},
  eprint = {2104.06802},
  primaryclass = {astro-ph},
  pages = {A12},
  doi = {10.1051/0004-6361/202040228},
  archiveprefix = {arXiv},
  journal = {\aap}
}

@article{staubert2019a,
  title = {Cyclotron Lines in Highly Magnetized Neutron Stars},
  author = {Staubert, R. and Tr{\"u}mper, J. and Kendziorra, E. and Klochkov, D. and Postnov, K. and Kretschmar, P. and Pottschmidt, K. and Haberl, F. and Rothschild, R. E. and Santangelo, A. and Wilms, J. and Kreykenbohm, I. and F{\"u}rst, F.},
  year = 2019,
  month = feb,
  volume = {622},
  pages = {A61},
  doi = {10.1051/0004-6361/201834479},
  langid = {english},
  journal = {\aap}
}

@article{stella1986,
  title = {Intermittent {{Stellar Wind Acceleration}} and the {{Long-Term Activity}} of {{Population I Binary Systems Containing}} an {{X-Ray Pulsar}}},
  author = {Stella, L. and White, N. E. and Rosner, R.},
  year = 1986,
  month = sep,
  journal = {\apj},
  volume = {308},
  pages = {669},
  doi = {10.1086/164538},
  langid = {english}
}

@article{stierhof2025,
  title = {Don't Torque like That: {{Measuring}} Compact Object Magnetic Fields with Analytic Torque Models},
  author = {Stierhof, J. J. R. and {Sokolova-Lapa}, E. and Berger, K. and Vasilopoulos, G. and Thalhammer, P. and Zalot, N. and Ballhausen, R. and El Mellah, I. and Malacaria, C. and Rothschild, R. E. and Kretschmar, P. and Pottschmidt, K. and Wilms, J.},
  year = 2025,
  month = jun,
  volume = {698},
  pages = {A308},
  doi = {10.1051/0004-6361/202553809},
  langid = {english},
  journal = {\aap}
}

@article{sunyaev2021a,
  title = {{{SRG X-ray}} Orbital Observatory. {{Its}} Telescopes and First Scientific Results},
  author = {Sunyaev, R. and Arefiev, V. and Babyshkin, V. and Bogomolov, A. and Borisov, K. and Buntov, M. and Brunner, H. and Burenin, R. and Churazov, E. and Coutinho, D. and Eder, J. and Eismont, N. and Freyberg, M. and Gilfanov, M. and Gureyev, P. and Hasinger, G. and Khabibullin, I. and Kolmykov, V. and Komovkin, S. and Krivonos, R. and Lapshov, I. and Levin, V. and Lomakin, I. and Lutovinov, A. and Medvedev, P. and Merloni, A. and Mernik, T. and Mikhailov, E. and Molodtsov, V. and Mzhelsky, P. and M{\"u}ller, S. and Nandra, K. and Nazarov, V. and Pavlinsky, M. and Poghodin, A. and Predehl, P. and Robrade, J. and Sazonov, S. and Scheuerle, H. and Shirshakov, A. and Tkachenko, A. and Voron, V.},
  year = 2021,
  month = dec,
  volume = {656},
  pages = {A132},
  doi = {10.1051/0004-6361/202141179},
  langid = {english},
  journal = {\aap}
}

@article{torrejon2001,
  title = {{{BeppoSAX}} Survey of {{Be}}/{{X-ray}} Binary Candidates},
  author = {Torrej{\'o}n, J. M. and Orr, A.},
  year = 2001,
  month = oct,
  volume = {377},
  number = {1},
  pages = {148--155},
  doi = {10.1051/0004-6361:20011070},
  langid = {english},
  journal = {\aap}
}

@article{tsygankov2016,
  title = {Propeller Effect in Two Brightest Transient {{X-ray}} Pulsars: {{4U}} 0115+63 and {{V}} 0332+53},
  author = {Tsygankov, S. S. and Lutovinov, A. A. and Doroshenko, V. and Mushtukov, A. A. and Suleimanov, V. and Poutanen, J.},
  year = 2016,
  month = sep,
  volume = {593},
  pages = {A16},
  doi = {10.1051/0004-6361/201628236},
  langid = {english},
  journal = {\aap}
}

@article{tsygankov2017,
  title = {Stable Accretion from a Cold Disc in Highly Magnetized Neutron Stars},
  author = {Tsygankov, S. S. and Mushtukov, A. A. and Suleimanov, V. F. and Doroshenko, V. and Abolmasov, P. K. and Lutovinov, A. A. and Poutanen, J.},
  year = 2017,
  month = dec,
  volume = {608},
  eprint = {1703.04528},
  primaryclass = {astro-ph},
  pages = {A17},
  doi = {10.1051/0004-6361/201630248},
  archiveprefix = {arXiv},
  journal = {\aap}
}

@article{tsygankov2017a,
  title = {The {{X-ray}} Properties of {{Be}}/{{X-ray}} Pulsars in Quiescence},
  author = {Tsygankov, Sergey S. and Wijnands, Rudy and Lutovinov, Alexander A. and Degenaar, Nathalie and Poutanen, Juri},
  year = 2017,
  month = aug,
  volume = {470},
  number = {1},
  pages = {126--141},
  doi = {10.1093/mnras/stx1255},
  journal = {\mnras}
}

@article{tsygankov2019a,
  title = {Cyclotron Emission, Absorption, and the Two Faces of {{X-ray}} Pulsar {{A}} 0535+262},
  author = {Tsygankov, Sergey S. and Doroshenko, Victor and Mushtukov, Alexander A. and Suleimanov, Valery F. and Lutovinov, Alexander A. and Poutanen, Juri},
  year = 2019,
  month = jul,
  volume = {487},
  number = {1},
  pages = {L30-L34},
  doi = {10.1093/mnrasl/slz079},
  langid = {english},
  journal = {\mnras}
}

@article{voss2010a,
  title = {Swift-{{BAT Survey}} of {{Galactic Sources}}: {{Catalog}} and {{Properties}} of the {{Populations}}},
  author = {Voss, R. and Ajello, M.},
  year = 2010,
  month = oct,
  volume = {721},
  number = {2},
  pages = {1843--1852},
  doi = {10.1088/0004-637X/721/2/1843},
  langid = {english},
  journal = {\apj}
}

@article{walter2007,
  title = {Probing Clumpy Stellar Winds with a Neutron Star},
  author = {Walter, R. and Zurita Heras, J.},
  year = 2007,
  month = dec,
  volume = {476},
  number = {1},
  pages = {335--340},
  doi = {10.1051/0004-6361:20078353},
  langid = {english},
  journal = {\aap}
}

@article{wang2021,
  title = {The {{Detection}} and {{Characterization}} of {{Be}}+{{sdO Binaries}} from {{HST}}/{{STIS FUV Spectroscopy}}},
  author = {Wang, Luqian and Gies, Douglas R. and Peters, Geraldine J. and G{\"o}tberg, Ylva and Chojnowski, S. Drew and Lester, Kathryn V. and Howell, Steve B.},
  year = 2021,
  month = may,
  volume = {161},
  number = {5},
  eprint = {2103.13642},
  primaryclass = {astro-ph},
  pages = {248},
  doi = {10.3847/1538-3881/abf144},
  archiveprefix = {arXiv},
  journal = {\aj}
}

@article{weber2026,
  title = {Ultraluminous {{X-ray}} Sources in the First {{eROSITA}} Survey - {{I}}. {{Candidate}} Catalogs},
  author = {Weber, P. and Roberts, T. P. and H{\"a}mmerich, S. and Kyritsis, E. and Zezas, A. and Mayer, M. G. F. and Zainab, A. and Kreykenbohm, I. and Merloni, A. and Sasaki, M. and Schwope, A. and Middleton, M. and {Basu-Zych}, A. and Hornschemeier, A. and Vulic, N. and Salvato, M. and Webb, N. and Tranin, H. and Schettino, N. and Kirsch, C. and Saeedi, S. and Zangrandi, F. and Lorenz, M. and Dauner, L. and Dauser, T. and Haberl, F. and Maitra, C. and Igo, Z. and Santangelo, A. and Ducci, L. and Wilms, J.},
  year = 2026,
  month = aug,
  volume = {712},
  pages = {A112},
  publisher = {EDP Sciences},
  doi = {10.1051/0004-6361/202659942},
  langid = {english},
  journal = {\aap}
}

@article{wijnands2017,
  title = {Cooling of {{Accretion-Heated Neutron Stars}}},
  author = {Wijnands, Rudy and Degenaar, Nathalie and Page, Dany},
  year = 2017,
  month = sep,
  journal = {\jaa},
  volume = {38},
  number = {3},
  pages = {49},
  doi = {10.1007/s12036-017-9466-5},
  langid = {english}
}

@article{wilson1997,
  title = {A {{Sequence}} of {{Outbursts}} from the {{Transient X-Ray Pulsar GS}} 0834-430},
  author = {Wilson, Colleen A. and Finger, Mark H. and Harmon, B. Alan and Scott, D. Matthew and Wilson, Robert B. and Bildsten, Lars and Chakrabarty, Deepto and Prince, Thomas A.},
  year = 1997,
  month = apr,
  volume = {479},
  number = {1},
  pages = {388--397},
  doi = {10.1086/303841},
  langid = {english},
  journal = {\apj}
}

@article{willingale2013,
  title = {Calibration of {{X}}-Ray Absorption in Our {{Galaxy}}},
  author = {Willingale, R. and Starling, R. L. C. and Beardmore, A. P. and Tanvir, N. R. and O'Brien, P. T.},
  year = 2013,
  month = may,
  volume = {431},
  number = {1},
  eprint = {1303.0843},
  primaryclass = {astro-ph.HE},
  pages = {394--404},
  doi = {10.1093/mnras/stt175},
  archiveprefix = {arXiv},
  journal = {\mnras}
}

@article{xiao2025,
  title = {Propeller Effect in Action: {{Unveiling}} Quenched Accretion in the Transient {{X-ray}} Pulsar {{4U}} 0115+63},
  author = {Xiao, Hua and Tsygankov, Sergey S. and Suleimanov, Valery F. and Mushtukov, Alexander A. and Ji, Long and Poutanen, Juri},
  year = 2025,
  month = oct,
  volume = {702},
  pages = {A216},
  doi = {10.1051/0004-6361/202556527},
  langid = {english},
  journal = {\aap}
}

@incollection{yuan2022,
  title = {The {{Einstein Probe Mission}}},
  booktitle = {Handbook of {{X-ray}} and {{Gamma-ray Astrophysics}}},
  author = {Yuan, Weimin and Zhang, Chen and Chen, Yong and Ling, Zhixing},
  year = 2022,
  pages = {86},
  publisher = {Springer Nature},
  doi = {10.1007/978-981-16-4544-0_151-1},
  isbn = {9789811645440},
  langid = {english}
}

@article{zalot2026,
  title = {A Simple Relation: {{Neutron}} Star Magnetic Field Strength and Spectral Shape at Low Mass Accretion Rates},
  author = {Zalot, Nicolas and {Sokolova-Lapa}, Ekaterina and Zainab, Aafia and Thalhammer, Philipp and Stierhof, Jakob and Berger, Katrin and Pottschmidt, Katja and Ballhausen, Ralf and Malacaria, Christian and Gulbahar, Esin and Wilms, J{\"o}rn},
  year = 2026,
  month = may,
  volume = {709},
  pages = {A182},
  publisher = {EDP Sciences},
  doi = {10.1051/0004-6361/202558692},
  langid = {english},
  journal = {\aap}
}

@article{zorec2016,
  title = {Critical Study of the Distribution of Rotational Velocities of {{Be}} Stars - {{I}}. {{Deconvolution}} Methods, Effects Due to Gravity Darkening, Macroturbulence, and Binarity},
  author = {Zorec, J. and Fr{\'e}mat, Y. and {Domiciano de Souza}, A. and Royer, F. and Cidale, L. and Hubert, A.-M. and Semaan, T. and Martayan, C. and Cochetti, Y. R. and Arias, M. L. and Aidelman, Y. and Stee, P.},
  year = 2016,
  month = nov,
  volume = {595},
  pages = {A132},
  publisher = {EDP Sciences},
  doi = {10.1051/0004-6361/201628760},
  langid = {english},
  journal = {\aap}
}

@article{zorec2017,
  title = {Critical Study of the Distribution of Rotational Velocities of {{Be}} Stars: {{II}}: {{Differential}} Rotation and Some Hidden Effects Interfering with the Interpretation of the {{{\emph{V}}}} Sin {\emph{i}} Parameter},
  author = {Zorec, J. and Fr{\'e}mat, Y. and Domiciano De Souza, A. and Royer, F. and Cidale, L. and Hubert, A.-M. and Semaan, T. and Martayan, C. and Cochetti, Y. R. and Arias, M. L. and Aidelman, Y. and Stee, P.},
  year = 2017,
  month = jun,
  volume = {602},
  pages = {A83},
  doi = {10.1051/0004-6361/201628761},
  langid = {english},
  journal = {\aap}
}

\begin{appendix}
\section{HMXB Catalog and comparison to previous missions}
\label{app:catalog}
The main columns of the updated HMXB catalog used here are \texttt{IDENTIFIER}, which provides the SIMBAD name of the source and the right ascension (RA) and declination (DEC) from \textit{XRBcats} \citep{neumann2023}. The pulse (\texttt{P\_pulse}) and orbital periods (\texttt{P\_orb}) are also retained. The column \texttt{Xray\_Type} of \textit{XRBcats} lists descriptors for each source, including companion type and spectral features typically seen in HMXB spectra, like the presence of cyclotron lines. We compiled information on only the subclass in a separate column called \texttt{TYPE}. The different classes are designated as ``BE'' for BeXRBs, ``SG'' for SgXBs, ``SF'' for SFXTs, ``RSG'' for the rare case of a red supergiant companion, ``MQ'' for microquasars, and ``Gcas'' to indicate the \gamcas nature of a source. In some cases, the system has since been found to host a Be star, with a hot subdwarf companion, and is correspondingly labeled ``sdOB'' to indicate the spectral type of the non-compact companion to the Be star. A source is labeled ``Confused'' if its HMXB characterisation is accepted in literature, but the true nature of the optical companion is still under investigation. The ``Candidate'' label is used if a system has not yet been confirmed as an HMXB, but there is literature categorising it as a candidate of a specific type. We also added a column \texttt{CO} to specify the nature of the compact object, if it is known, with the options being ``NS'' for a neutron star, ``BH'' for a black hole, and ``Unknown'' in the case of inconclusive evidence for either. We added a column \texttt{GAMMA\_BINARY} to indicate whether the system is known to also show $\gamma$-ray emission or not. 

\begin{table}
\caption{HMXBs whose subclass or ancillary information was changed with respect to the \textit{XRBcats}.}
    \renewcommand{\arraystretch}{1.32}
    \renewcommand{\tabcolsep}{2mm}
    \centering
     \begin{tabular}{p{3.2cm}p{2cm}p{2.7cm}}
     \hline
     \hline

    Identifier & Xray\_Type (\textit{XRBcats}) & \texttt{TYPE} \\
     \hline
     1A~1238$-$59 & XP$^{*}$ & Removed \\
     1A~1244$-$60 & XT  & Removed \\      
     AX J1714.1$-$3912 & -- & Candidate (SF) \\
     CCDM~J07474$-$5320A & BE & Removed \\
     1H~0749$-$600 & BE$^{*}$ & Removed \\ 
     HR~4804 & BE & sdOB \\
     HD~96670 & BE, BH & sdOB\\ 
     HD~110432 & BE, Gcas & Gcas\\
     HD~119682 & BE & Gcas\\
     HD~141926 & -- & Removed \\
     HD~249179 & BE & Removed \\
     HESS~J0632$+$057 & GP & BE \\
     IGR~J08262$-$3736 & BE & SG\\
     IGR~J08408$-$4503 & SG, XT & SF \\
     IGR~J11215$-$5952 & SG, XT & SF \\
     IGR~J12341$-$6143 & -- & Candidate (SF) \\
     IGR~J14331$-$6112 & BE & Confused (SG/BE) \\
     IGR~J16328$-$4726 & SG, XT & SF \\
     IGR~J16418$-$4532 & SG, XT & SF \\
     IGR~J16465$-$4507 & SG, XP, XT& SF \\
     IGR~J16479$-$4514 & SG, XT & SF \\
     IGR~J17200$-$3116 & BE & Confused (SG/BE) \\
     MAXI~J0655$-$013 & -- & BE \\
     MAXI~J0709$-$159 & -- & Confused (SF/BE) \\
     MAXI~J0903$-$531 & -- & BE \\
     OAO~1657$-$415 & CL, EB, XP & SG \\
     mu.02~Cru & BE, BH & Gcas \\
     PSR~B1259$-$63 & GP & BE \\
     SGR~0755$-$2933 & -- & BE \\
     XTE J1716$-$389 & -- & Candidate (SG)\\
     XTE~J1739$-$302 & SG, XT & SF \\
     XTE~J1743$-$363 & -- & SF\\
     \hline
     \end{tabular}
     \tablefoot{Sources marked with a * were listed as candidates by \citet{neumann2023}.}
     \label{tab1:reclassified}
\end{table}

 We obtained the labels given in the columns \texttt{CO} and \texttt{TYPE} by parsing the information contained in the \texttt{Xray\_Type} column of \textit{XRBcats}. To populate the \texttt{CO} column, first, the straightforward cases of ``NS'' and ``BH'' were marked correspondingly. In addition, a source was also set to ``NS'' if its descriptors in the original catalog included ``XP'' (X-ray pulsar), ``CL'' (cyclotron line), or ``QPO'' (quasi periodic oscillation), since these are all characteristics of accreting neutron stars. Systems marked as ``MQ'' were also labeled as black holes, as there are no known microquasars with a high mass star and a neutron star. For the \texttt{TYPE} column, a detailed literature search was conducted, to ensure that the information corresponding to each source provided in ``Xray\_Type'' was still up to date. In the course of this, a number of changes were made, all of which are compiled in Table~\ref{tab1:reclassified}. Details on the changes made and the reason for reclassification can be found below.  

\subsection{Reclassification}
We exclude CCDM~J07474$-$5320A and 1H~0749$-$600 are since their HMXB nature is tenuous at best. 

CCDM~J07474$-$5320A was disregarded by \citet{fortin2022} due to a very high parallax, in addition to its initial classification being that of a Be+WD system by \citet{torrejon2001}, corresponding to a very low luminosity that they deemed unrealistic. 

1H~0749$-$600 was already labeled as tentative by \citet{neumann2023} due to \citet{torrejon2001} disregarding its X-ray binary nature, and it is not included by \citet{fortin2023}. The source was, however, included in a sample of $\gamma$-ray binaries by \citep{ducci2023}. 

HD~141926 and HD~249179, were also removed, due to the former being an Herbig Ae/Be star \citep{cody2025} and candidate for hosting a white dwarf companion \citep{torrejon2001}, and the latter being discarded by \citet{fortin2022}. 

HD~110432, and HD~119682 have often been discussed as \gamcas systems \citep{torrejon2001,naze2025}, and $\mu_2$~Cru is included as a Be star in the catalog of isolated Be stars by \citet{naze2023}, but has been proposed to contain a binary companion of unknown nature \citet{neuhauser2020}. We also found it to be detected at a flux and luminosity where only Be stars were found. 
The case of HD~119682 is still disputed \citep{faltova2026}. The four sources above are therefore kept in the catalog as \gamcas systems but not analysed as part of the BeXRB sample. On the other hand, HD~96670 has been reclassified in recent times to not host a black hole companion, but instead as a Be+sdOB system \citep{naze2025}. Not much is known about HR~4804, but it has been included in a search for Be+sdOB \citep{wang2021}. 

The input catalogue does not identify HMXB subtypes beyond the type of the optical companion, so we classified a number of sources to be SFXTs based on literature search. These are as listed in Table~\ref{tab1:reclassified} and can be found in \citet[][and references therein]{bozzo2015}. 

\section{Spectral fitting}
\begin{figure}
    \centering
    \includegraphics[width=0.45\textwidth]{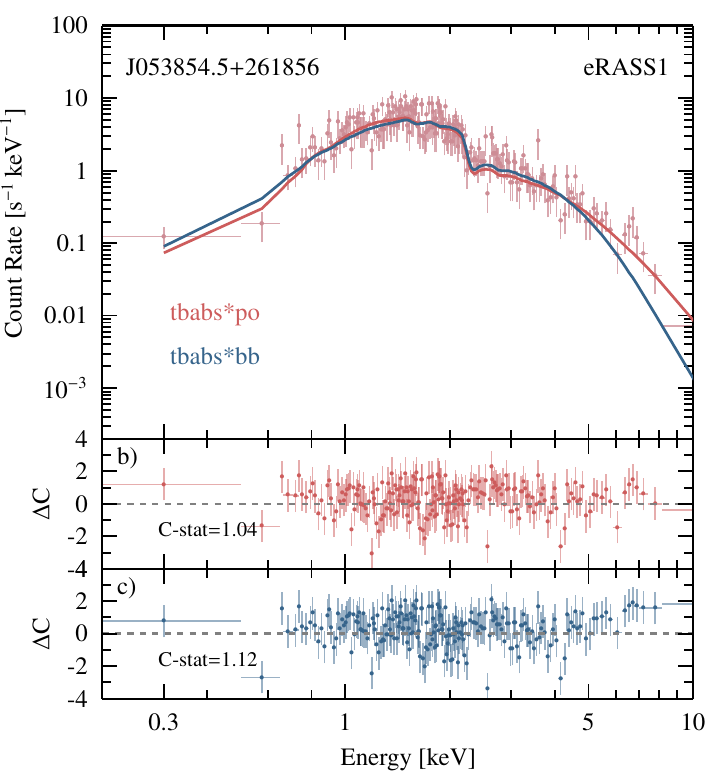}
    \caption{\ero spectrum of J05385.4$+$261856 (1A~0535$+$263) fitted with both an absorbed powerlaw and absorbed black-body model. Panel a) shows the spectral data and the two models. Panel b) and c) contains their respective residuals.}
    \label{fig:eg_spec}
\end{figure}

Figure~\ref{fig:pileup} shows the comparison of the eRASS1 flux distribution before and after pile-up correction. In eRASS1--4, a handful of sources were affected by pile-up. The sources affected were mainly SgXBs -- Cen~X-3, Vela~X-1, Cir~X-1, and 4U~1538$-$52, 4U~1700$-$377, and BeXRBs in Type I outbursts -- A~0538$+$263 and GRO~J1008$-$57, 2S~1417$-$624, and 2S~1553$-$542. 

An example result from fitting the \ero data of the HMXB sample is shown in Fig.~\ref{fig:eg_spec}, from an eRASS1 snapshot of 1A~0535$+$263. 

\begin{figure}
    \centering
    \includegraphics[width=\linewidth]{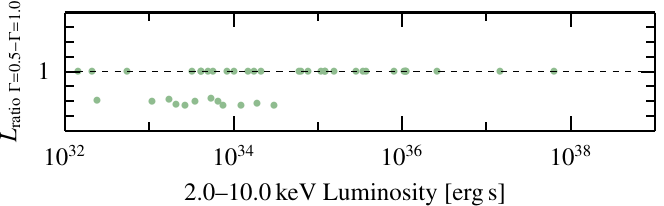}
    \includegraphics[width=\linewidth]{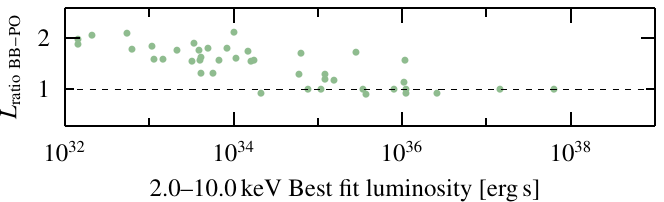}
    \caption{\textbf{Top:} The ratio of the luminosity estimated using an absorbed power law model with $\Gamma=0.5$ to that using $\Gamma=1$, with respect to the $\Gamma=1$ luminosity. \textbf{Bottom:} The ratio between the luminosities computed using a black body model with $kT=1\,\mathrm{keV}$ and a power law model with $\Gamma=1$, with respect to the best fit luminosity used in this work. The luminosity varies at most by a factor ${\sim}2$.}
    \label{fig:lum_comp}
\end{figure}

\begin{figure}
    \centering
    \includegraphics[width=0.45\textwidth]{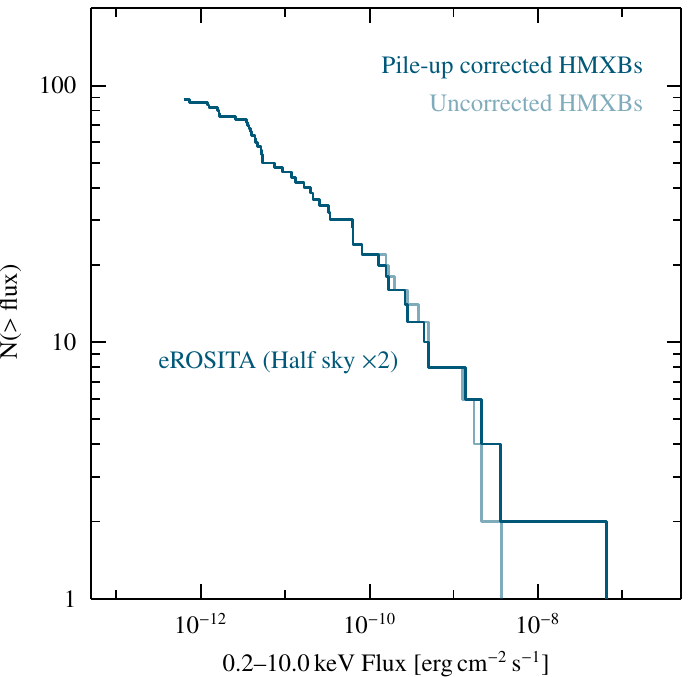}
    \caption{The difference between HMXB fluxes corrected for pile-up compared to no correction performed. This example distribution is shown for eRASS1. The effect is evident at fluxes $\gtrsim10^{-11}\,\mathrm{erg}\,\mathrm{s}^{-1}$}.
    \label{fig:pileup}
\end{figure}

Since the adopted model and model parameters affect the estimated luminosity, especially for those sources where spectral fitting is not feasible, we depict the ratio of the luminosity as obtained using a power law model with $\Gamma=1.0$ and $\Gamma=0.5$ in Fig.~\ref{fig:lum_comp}, as well as its difference to a black body model with a temperature of 0.5\,keV. 

\section{Fits to the \ero \lognl distributions}
\begin{figure*}
    \centering
    \includegraphics[width=0.33\textwidth]{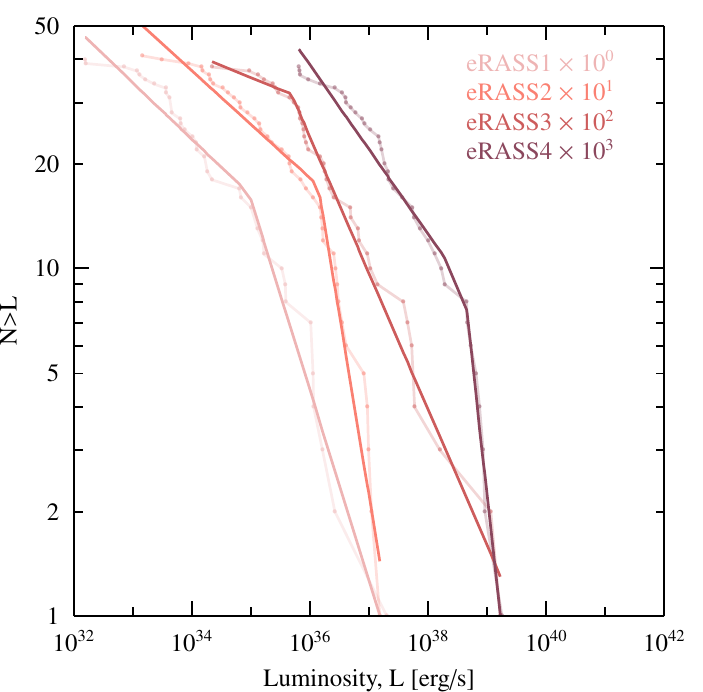}
    \includegraphics[width=0.33\textwidth]{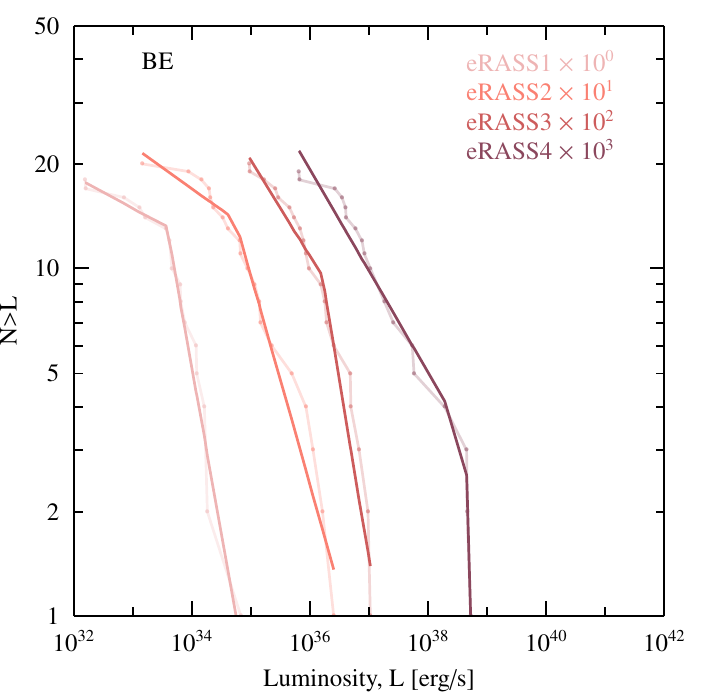}
    \includegraphics[width=0.33\textwidth]{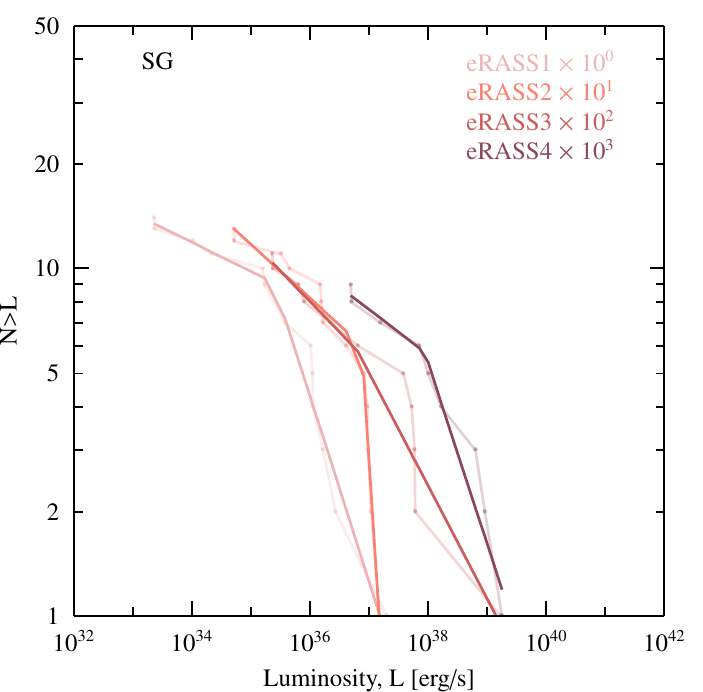}
    \caption{Results of broken power law fits to the eRASS1--4 luminosity distributions of HMXBs (left), BeXRBs (middle) and SgXBs (right). The luminosities in the eRASS2--4 distributions are multiplied by an order of magnitude each, for clarity.}
    \label{fig:fit_pars}
\end{figure*}

\begin{figure*}
    \centering
    \includegraphics[width=0.33\textwidth]{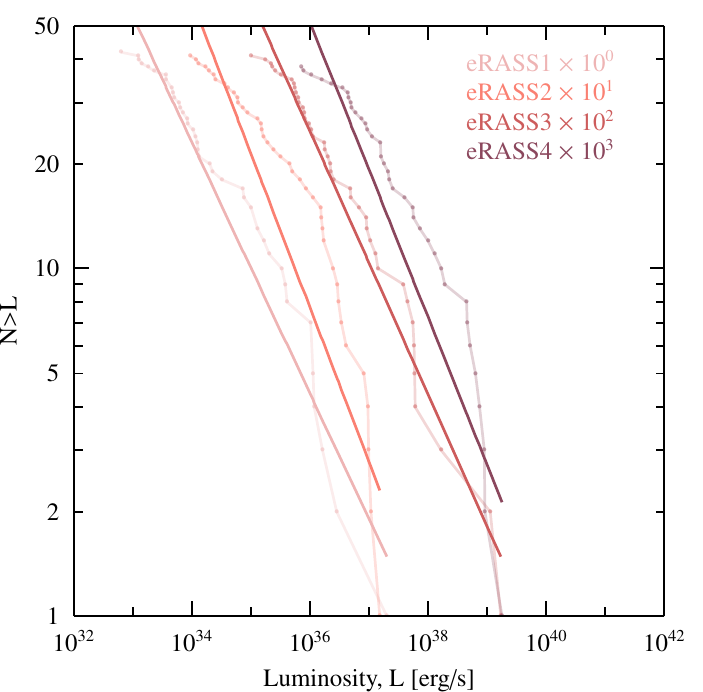}
    \includegraphics[width=0.33\textwidth]{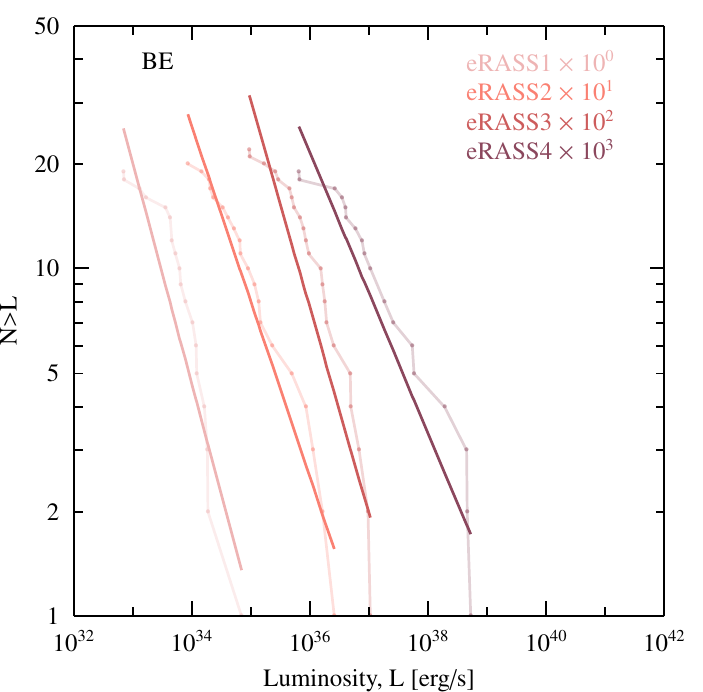}
    \includegraphics[width=0.33\textwidth]{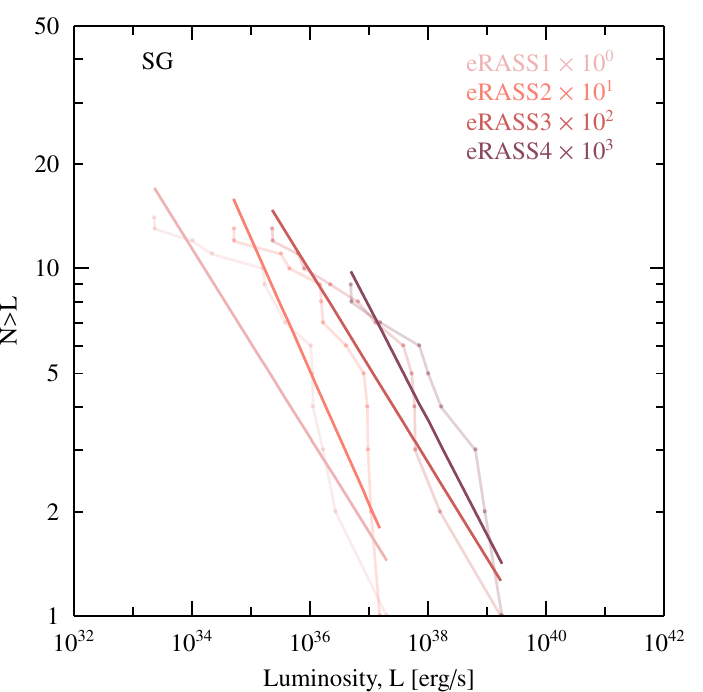}
    \caption{The singular power law fits to the eRASS1--4 luminosity distributions of HMXBs (left), BeXRBs (middle) and SgXBs (right). The luminosity in the eRASS2--4 distributions are multiplied by an order of magnitude each, for clarity.}
    \label{fig:fit_pars_po}
\end{figure*}

We show the fits to the \lognl distributions of the \ero HMXB sample and of the main subclasses in Fig.~\ref{fig:fit_pars}. They each require a broken power law model, as described in the main text. However, there are clear deviations due to intrinsic variability, resulting in the break luminosity varying significantly between individual distributions, often close to either extreme of the covered luminosity regime. Nevertheless, the distributions still require a broken power law, as evidenced by the vast majority of the data being poorly modelled using a singular power law (Fig.~\ref{fig:fit_pars_po}). This is largely also the case for the BeXRBs and SgXBs, but there are instances when the break is unconstrained due to particularly luminous sources, as is seen for the eRASS3 sample of SgXBs. 

\begin{figure*}
    \centering
    \includegraphics[width=\textwidth]{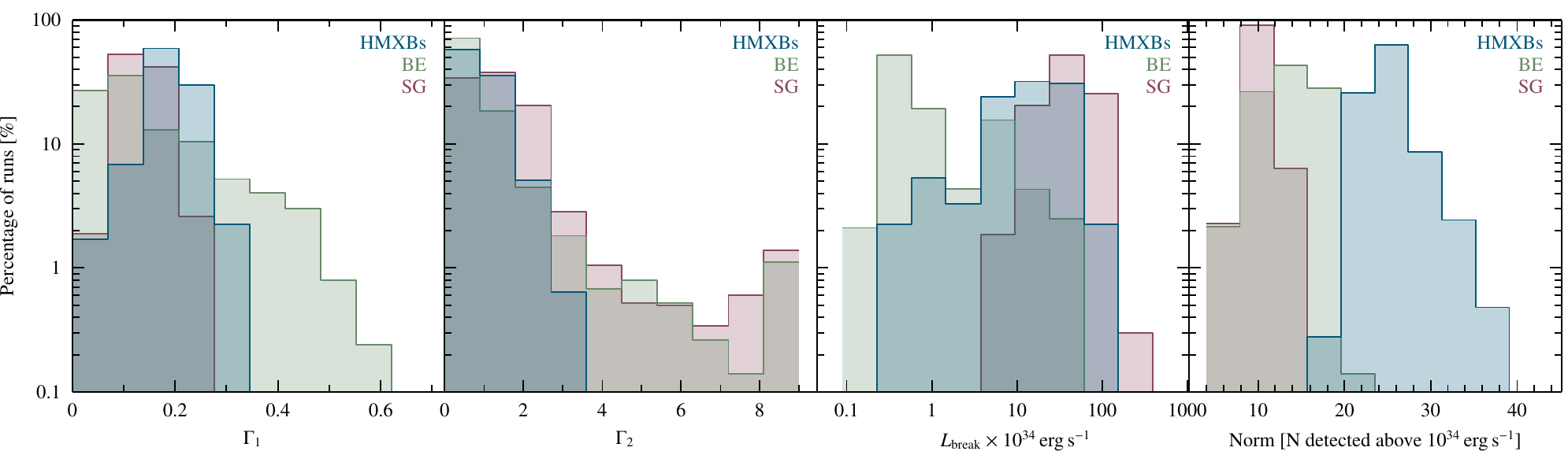}
    \caption{Distribution of fit parameters from the fits conducted to a 1000 simulated \lognl realisations sampled using eRASS1--4 data. \textbf{Left:} The lower luminosity slope is comparable for the overall sample and the subtype distributions. \textbf{Middle:} The higher slope diverges significantly for the BeXRBs, while that of the overall sample follows the SgXB distribution closely. \textbf{Right:} The break luminosity is fairly consistent 99\% of the time, but diverges in the case of outbursts occurring ${\sim}$1\% of the time.}
    \label{fig:lognl_fit_pars}
\end{figure*}

The fit parameters resulting from the fits to the 5000 simulated distributions shown in Fig.~\ref{fig:ero_lum_types} (right panel) are shown in Fig.~\ref{fig:lognl_fit_pars}. The first slope, $\Gamma_1$, of the SgXB distributions follows the trend of the overall HMXB sample, and is constrained around 0.1--0.2, while the BeXRBs show more variance. The three samples show overall agreement for the second slope. There is an additional peak at a higher $\Gamma_2$ for both the SgXBs and BeXRBs (in less than 1\% of the cases), that may indicate distributions where a break is not sufficiently constrained. The distinguishing parameter is the break luminosity, with the BeXRBs and SgXBs peaking at distinctly different cutoffs. This reflects the luminosity regimes that they occupy in the overall luminosity distribution. In a small percentage of cases (5\%), BeXRBs require higher break luminosities, likely corresponding to particularly bright outbursts in those distributions. The norm, which indicates the number of sources detected above $10^{34}\,\mathrm{erg}\,\mathrm{s}^{-1}$, also differs between the two subclasses. The BeXRBs peak around 6--12 sources ($\lesssim30$\% of the sample) above this threshold, while SgXBs predominantly show up above $10^{34}\,\mathrm{erg}\,\mathrm{s}^{-1}$ ($\gtrsim50$\% of the sample). 

\begin{table}
\caption{Kolmogorov-Smirnoff test probability for each fit, which corresponds to the maximum deviation between the fit and the underlying distribution.}
\renewcommand{\arraystretch}{1.32} 
    \renewcommand{\tabcolsep}{2mm}
    \centering
    \begin{tabular}{ccc}
    \hline
    \hline
    Sample & Power law & Broken power law \\
    \hline
    HMXBs & KS-Test & KS-Test \\
    \hline
    eRASS1 & 53.36  & 20.47 \\
    eRASS2 &  50.78 & 21.73 \\
    eRASS3 &  43.90 & 6.66 \\
    eRASS4 &  57.53 & 16.86 \\
    \hline
    SgXBs  & & \\
    \hline
    eRASS1 &  35.51 & 13.56 \\
    eRASS2 &  35.71 & 9.70 \\
    eRASS3 &  20.46 & 15.9 \\
    eRASS4 &  21.40 & 10.6 \\
    \hline
    BeXRBs & & \\
    \hline
    eRASS1 & 38.65  & 9.93 \\
    eRASS2 &  38.91 & 8.86 \\
    eRASS3 &  46.00 & 9.61 \\
    eRASS4 &  41.45 & 20.74 \\
    \end{tabular}
    \label{tab:kstest}
\end{table}

\label{app:fits}

\section{\rxte/ASM and \maxi luminosity distributions}
As discussed in the main body of the paper, we compared the \ero luminosity distributions with \rxte/ASM and \maxi lightcurves. This exercise allows verification of the \ero detections at the high luminosity end and illustrates the limitations of previous instruments. Here, we provide a detailed description of our reproduction of \rxte/ASM \lognl distributions following \citet{grimm2002}, and further details on computing the \maxi luminosity distributions. 
\label{app:rxte}
\subsection{\rxte/ASM}
\begin{figure*}
\centering    
\includegraphics[width=0.45\textwidth]{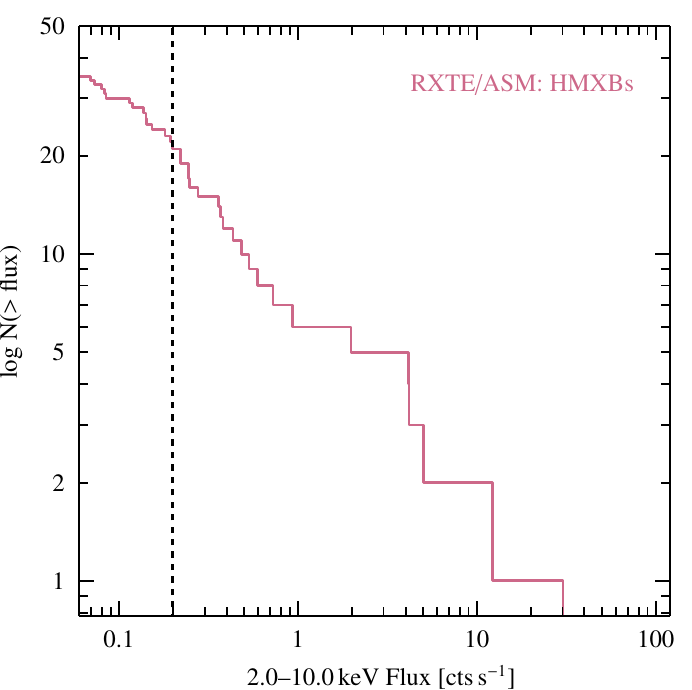}
\includegraphics[width=0.45\textwidth]{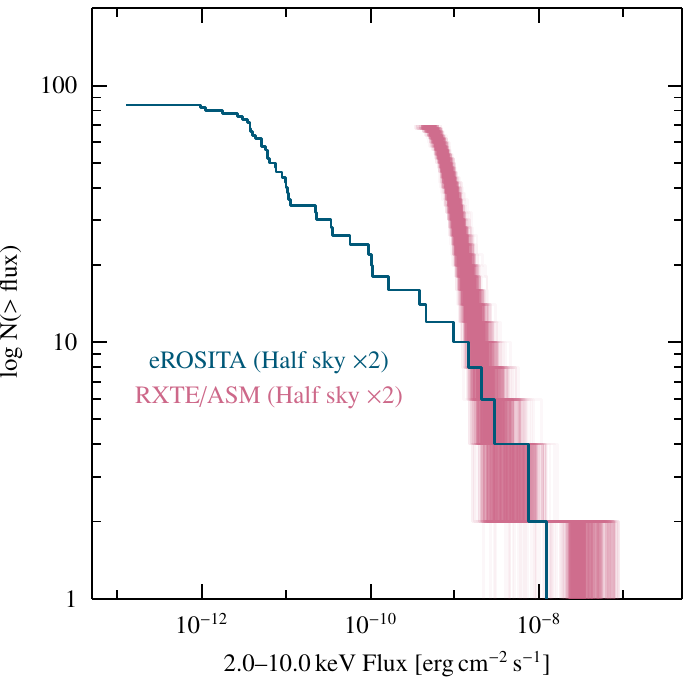}
\caption{\textbf{Left:} The \logns distributions determined using the method followed by \citet{grimm2002}, with the completeness limit indicated by the black dashed line. \textbf{Right:} \logns distributions determined following the method of \citet{grimm2002}, using a brightness cutoff without discarding bins with low signal-to-noise ratio, but taking variability into account.}  
\label{fig:grimm_logns}
\end{figure*}

 \citeauthor{grimm2002} limited the number of sources to those exceeding a flux of 5\,mCrab at any point, corresponding to $0.37\,\mathrm{cts}\,\mathrm{s}^{-1}$ in ASM\footnote{For \rxte/ASM, 1\,Crab corresponds to ${\sim}73.6\,\mathrm{cts}\,\mathrm{s}^{-1}$ \citep{grimm2002}}. We reproduced their \logns (Fig.~5, right, in their manuscript) for comparison with the eROSITA distributions, as shown in Fig.~\ref{fig:grimm_logns} (left). Since the catalogs used by \citet{grimm2002} are not available to us, we obtained the list of sources by matching our current HMXB catalogs with the \rxte/ASM lightcurve database, with a matching radius of $15''$. We then queried the archive for lightcurves of the corresponding matched sources, filtered for sources which reached a count rate of $\geq0.37\,\mathrm{cts}\,\mathrm{s}^{-1}$ at any point since the beginning of \rxte/ASM observations until 2000 April 27 (${\sim}$MJD\,51661), and obtained average rates for the quoted time period, resulting in Fig.~\ref{fig:grimm_logns} (left). We convert \rxte/ASM count rates to flux using the Crab unit, and first reproduced the \rxte/ASM distribution for only the Western hemisphere, introducing a factor 2 to project to the whole sky. 

The brightness cutoff of 5\,mCrab introduces systematic effects. Although it is necessary to account for the sensitivity of the detector, total exclusion of fainter sources from the distribution leads to systematically overestimated fluxes, as does averaging over outbursts of the sources. On removing the brightness cutoff, the distribution appears as shown in Fig.~\ref{fig:grimm_logns} (right), clearly overestimating fluxes at very low count rates. We addressed this by setting a limit on the uncertainty instead, such that the corresponding rate is set to zero if it is below the detection limit of \rxte/ASM, or if its uncertainty is $\gtrsim50$\% of the rate. 

\begin{figure*}
    \centering
    \includegraphics[width=0.45\textwidth]{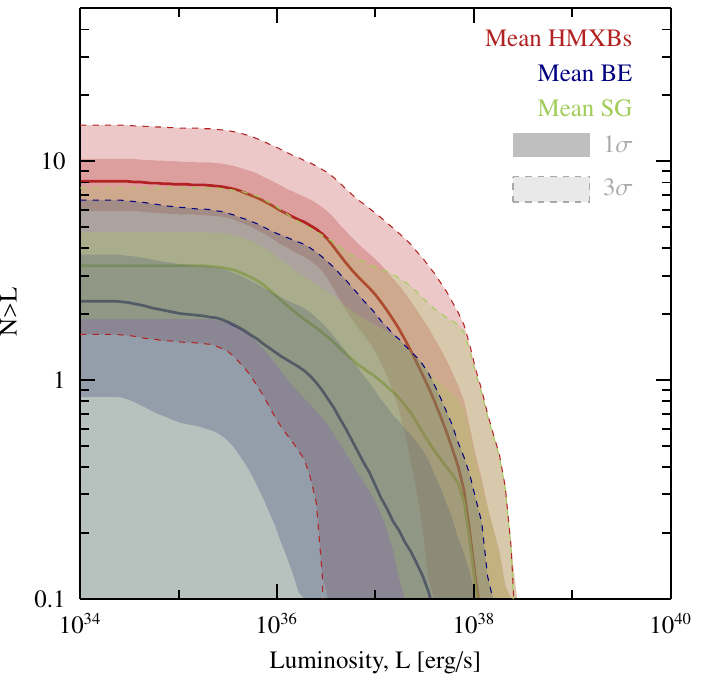}
    \includegraphics[width=0.45\textwidth]{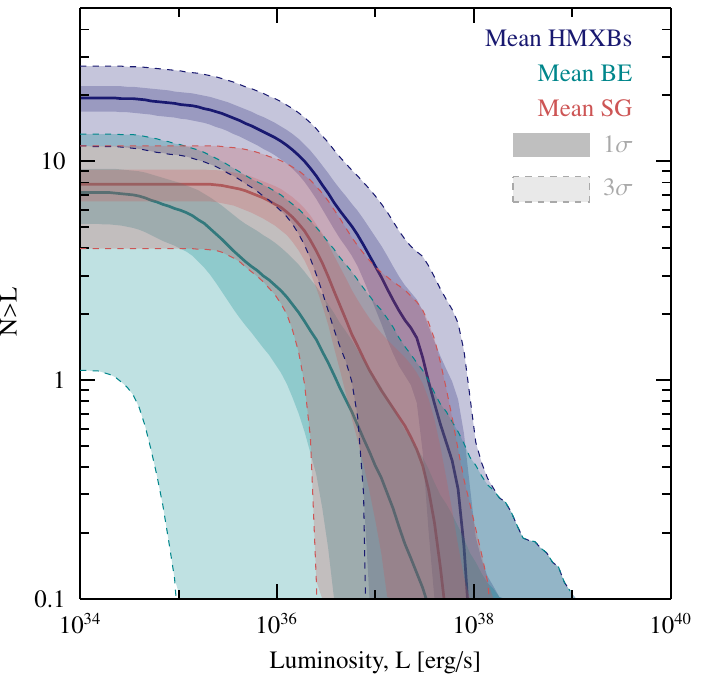}
    \caption{\rxte/ASM (left) and \maxi \lognl (right) distributions split by subclass, with the randomised iteration repeated for the each subclass. The relative dearth of BeXRBs shows how rarely they are detected by \rxte, typically while in outburst. The distinction in behaviour between SgXBs and BeXRBs is much more defined for \maxi.} 
    \label{fig:rxte_lum_types}
\end{figure*}

\subsection{\maxi}
For \maxi, we used the 1\,day binned lightcurves as provided and kept updated by the \maxi team\footnote{\url{https://maxi.riken.jp/top/lc\_ns.html}}, until MJD61213 (2026 June 22nd). The cross-matching is conducted using the same methods as described above. The lightcurves provide count rates in three energy bands, 2--4\,keV, 4--10\,keV, and 10--20\,keV, and we used the cumulative flux contained in 2--10\,keV, normalised  again to the Crab. We conducted the same exercise as for \rxte, randomly drawing fluxes from the lightcurve of each source and building 50000 distributions. Count-rates with uncertainties ${\ge}50$\% of the rate are set to zero and treated as non-detections. 

\subsection{Subclass resolved distributions}
We split both the \rxte/ASM and \maxi distributions into the two main subclasses SgXBs and BeXRBs, and show them in Fig.~\ref{fig:rxte_lum_types} with the mean of the distributions and the $3\sigma$ contour marked. For \rxte/ASM, not even persistent SgXBs are detected all the time, since its sensitivity does not allow for complete coverage below $10^{36}\,\mathrm{erg}\,\mathrm{s}^{-1}$. However, \maxi shows a distinction in the luminosity space occupied by SgXBs and BeXRBs, where SgXBs are predominantly detected above a few $10^{35}\,\mathrm{erg}\,\mathrm{s}^{-1}$. \rxte/ASM only allows for the detection of a 2--3 BeXRBs at any time. Although \maxi as a current monitor fares better with improved coverage, sensitivity, and a more updated source catalog, it detects ${\sim}5$--$10$ BeXRBs on average, for the full sky (in stark comparison to \ero in a singular scan). It is evident that detection thresholds of monitors contributed to BeXRBs being treated as  ``transient'' sources that were mainly active during outbursts.

\begin{figure*}
    \centering
    \includegraphics[width=0.45\textwidth]{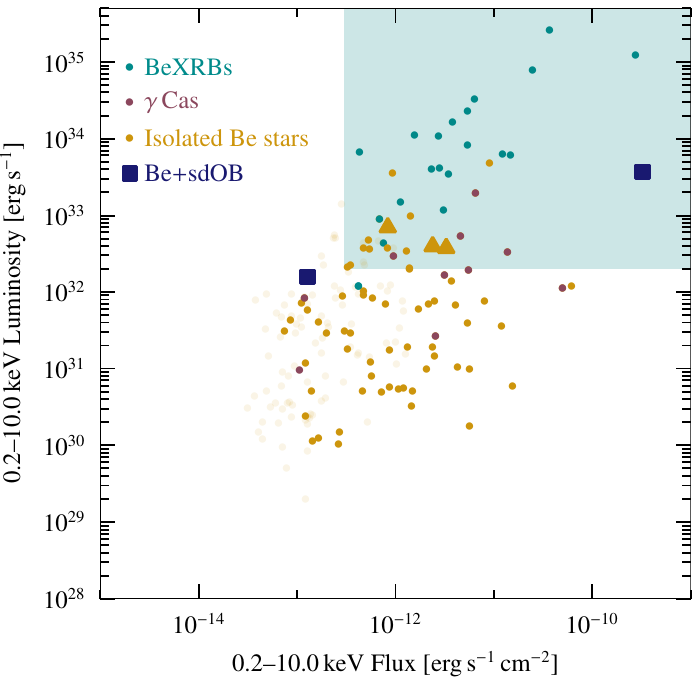}
    \includegraphics[width=0.45\textwidth]{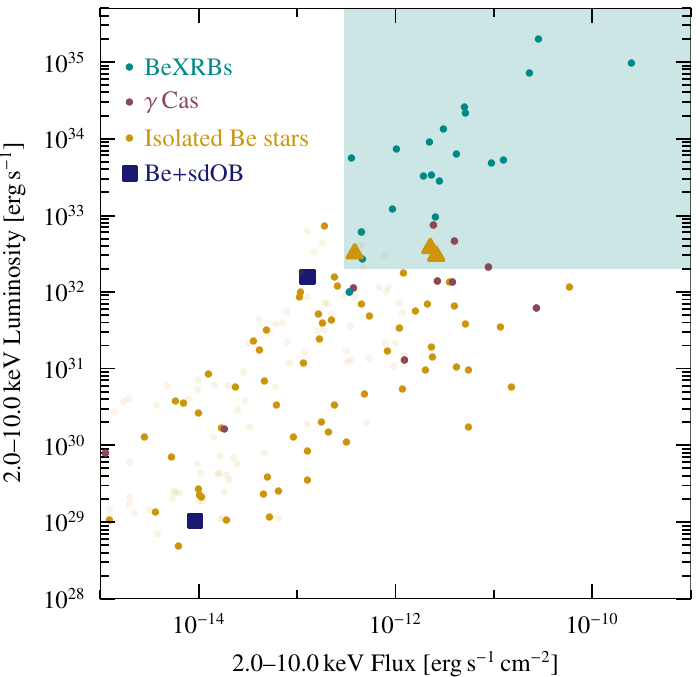}
    \caption{eROSITA fluxes and corresponding luminosities of BeXRBs, isolated Be stars and \gamcas systems, with some especially bright Be stars suspected to be BeXRB candidates marked as triangles. Be+sdOB systems are also depicted in dark blue squares. The lower opacity data points correspond to a \texttt{DET\_LIKE}${\lesssim}20$.}
    \label{fig:flux_lum}
\end{figure*}

\section{BeXRBs as seen by eROSITA}
The properties of BeXRBs in the Western Galactic hemisphere and the information gathered from the \ero survey have been tabulated in Table~\ref{tab:bexrbs_det} and Table~\ref{tab:bexrbs_info}. Table~\ref{tab:bexrbs_info} additionally has a column for hard X-ray detections outside of outburst, in an effort to consolidate ongoing, and support future efforts for low luminosity studies of BeXRBs.  

\begin{table*}[]
    \centering
    \renewcommand{\arraystretch}{1.32}  
    \renewcommand{\tabcolsep}{2mm}
    \caption{BeXRBs in the Western hemisphere as observed by eROSITA, in its four surveys, in descending order of the lowest luminosity at which they were detected. The column eRASS1--4 lists if the source was detected in any of the four eRASSes, while eRASS1:4 lists sources that were detected on combining all four eRASS.}
    \label{tab:bexrbs_det}
    \begin{tabular}{l|l|c|c|c|c|r}
    \hline
    \hline
    SIMBAD Name & eRASS Name & Minimum Flux & Minimum Luminosity & eRASS1--4 & eRASS1:4 & Exposure  \\
    
    & & $\times 10^{-12}\mathrm{erg}\,\mathrm{s}^{-1}\,\mathrm{cm}^{-2}$ & $\times 10^{34}\mathrm{erg}\,\mathrm{s}^{-1}$ & Detection & Detection & s\\
    \hline
    IGR~J13186$-$6257  &   J131825.0$-$625815  &   29.65  &   79.84  &   Yes  &   Yes  &   231.72 \\
LS~1698  &   J103735.3$-$564755  &   39.88  &   12.26  &   Yes  &   Yes  &   253.00 \\
4U~0728$-$25  &   J072853.5$-$260628  &   15.92  &   11.03  &   Yes  &   Yes  &   118.40 \\
SAX~J1324.4$-$6200  &   J132426.6$-$620119  &   3.78  &   10.17  &   Yes  &   Yes  &   229.19 \\
2S~1553$-$542  &   J155748.3$-$542453  &   3.30  &   8.88  &   Yes  &   Yes  &   148.44 \\
2S~1417$-$624  &   J142112.3$-$624155  &   12.92  &   6.44  &   Yes  &   Yes  &   151.35 \\
MAXI~J1409$-$619  &   J140802.7$-$615902  &   2.03  &   5.11  &   No  &   No  &   154.87 \\
IGR~J11435$-$6109  &   J114400.2$-$610736  &   5.00  &   3.52  &   Yes  &   Yes  &   267.55 \\
1A~0535$+$262  &   J053854.5$+$261856  &   53.43  &   1.99  &   Yes  &   Yes  &   94.76 \\
GRO~J1008$-$57  &   J100946.9$-$581735  &   11.93  &   1.77  &   Yes  &   Yes  &   220.24 \\
MXB~0656$-$072  &   J065817.2$-$071235  &   3.05  &   1.20  &   Yes  &   Yes  &   88.03 \\
MAXI~J0903$-$531  &   J090506.8$-$533019  &   0.67  &   0.81  &   Yes  &   Yes  &   214.57 \\
4U~1145$-$619  &   J114800.0$-$621224  &   14.53  &   0.73  &   Yes  &   Yes  &   250.56 \\
GX~304$-$1  &   J130117.0$-$613606  &   16.91  &   0.69  &   Yes  &   Yes  &   262.46 \\
IGR~J10101$-$5654  &   J101011.8$-$565531  &   4.01  &   0.67  &   Yes  &   Yes  &   220.37 \\
1A~1118$-$61  &   J112057.1$-$615500  &   6.04  &   0.60  &   Yes  &   Yes  &   301.11 \\
MAXI~J0655$-$013  &   J065512.4$-$012853  &   3.82  &   0.54  &   Yes  &   Yes  &   85.35 \\
IGR~J06074$+$2205  &   J060726.6$+$220547  &   1.26  &   0.54  &   Yes  &   Yes  &   72.84 \\
Swift~J1626.6$-$5156  &   J162636.5$-$515630  &   0.98  &   0.53  &   Yes  &   Yes  &   139.24 \\
2RXP~J130159.6  &   J130158.7$-$635808  &   1.32  &   0.47  &   Yes  &   Yes  &   262.43 \\
$-$635806 & -- & -- & -- & -- & -- & -- \\
LS~992  &   J081228.3$-$311452  &   0.82  &   0.43  &   Yes  &   Yes  &   129.59 \\
SRGA~J124404.1  &   J124403.7$-$632232  &   0.79  &   0.32  &   Yes  &   Yes  &   262.44 \\
$-$632232 & -- & -- & -- & -- & -- & -- \\
SGR~0755$-$2933  &   J075542.4$-$293353  &   1.74  &   0.23  &   Yes  &   Yes  &   113.62 \\
IGR~J11305$-$6256  &   J113106.8$-$625648  &   0.58  &   0.02  &   Yes  &   Yes  &   303.40 \\
AX~J1700.2$-$4220  &   J170025.2$-$421900  &   0.34  &   0.01  &   Yes  &   Yes  &   119.39 \\
Ginga~0834$-$430  &   J083554.9$-$431117  &   --  &   --  &   No  &   No  &   162.31 \\
IGR~J14059$-$6116  &   J140555.9$-$611629  &   --  &   --  &   No  &   No  &   145.27 \\
IGR~J14488$-$5942  &   J144843.2$-$594213  &   --  &   --  &   No  &   No  &   142.17 \\
XTE~J1543$-$568  &   J154405.1$-$564542  &   --  &   --  &   No  &   No  &   160.08 \\
RX~J1739.4$-$2942  &   J173929.9$-$294209  &   --  &   --  &   No  &   No  &   102.73 \\
\hline
PSR~B1259$-$63  &   J130247.6$-$635008  &   3.01  &   0.17  &   Yes  &   Yes  &   262.47 \\
HESS~J0632$+$057  &   J063259.2$+$054801  &   2.40  &   0.09  &   Yes  &   Yes  &   79.60 \\
PSR~J0635$+$0533  &   J063518.2$+$053306  &   --  &   --  &   No  &   No  &   82.36 \\
\hline
 \end{tabular}
 \tablefoot{The last three sources are not analysed as part of the sample of typical BeXRBs, since two of them are millisecond pulsars with a Be companion star, and HESS J0632+057 is primarily studied in Gamma-rays.}
\end{table*}

\begin{table*}[]
    \centering
    \renewcommand{\arraystretch}{1.32} 
    \renewcommand{\tabcolsep}{2mm}
    \caption{Some important properties of BeXRBs in the Western hemisphere, studied in this work.}
    \label{tab:bexrbs_info}
    \begin{tabular}{l|l|c|c|c|c|c}
    \hline
    \hline
    SIMBAD Name & eRASS Name & Distance & $P_\mathrm{pulse}$ & $P_\mathrm{orb}$ & CRSF & Detection $>10\,\mathrm{keV}$  \\
    & & kpc & s & d & keV & \\
    \hline 
1A~0535$+$262  &  J053854.5$+$261856  &  1.77  &  103.47  &  110.60  &  50 &  \textit{NuSTAR} \\ 
IGR~J06074$+$2205  &  J060726.6$+$220547  &  5.98  &  373.20  &  --  &  -- &  \textit{NuSTAR*} \\ 
MAXI~J0655$-$013  &  J065512.4$-$012853  &  3.44  &  1129.10  &  27.90  &  -- &  \textit{NuSTAR} \\ 
MXB~0656$-$072  &  J065817.2$-$071235  &  5.73  &  160.70  &  101.00  &  -- &  \textit{NuSTAR} \\ 
4U~0728$-$25  &  J072853.5$-$260628  &  7.61  &  103.14  &  34.50  &  -- & \textit{NuSTAR*} \\ 
SGR~0755$-$2933  &  J075542.4$-$293353  &  3.29  &  308.00  &  59.52  &  -- &  \textit{NuSTAR} \\ 
LS~992  &  J081228.3$-$311452  &  6.66  &  31.91  &  81.30  &  -- &  \textit{NuSTAR*} \\
Ginga~0834$-$430  &  J083554.9$-$431117  &  4.50  &  12.30  &  105.80  &  -- & \textit{NuSTAR*} \\
MAXI~J0903$-$531  &  J090506.8$-$533019  &  10.00  &  14.05  &  57.00  &  -- &  \\
GRO~J1008$-$57  &  J100946.9$-$581735  &  3.52  &  93.50  &  249.50  &  78 &  \textit{NuSTAR} \\ 
IGR~J10101$-$5654  &  J101011.8$-$565531  &  3.74  &  --  &  --  &  -- &  \textit{NuSTAR*} \\ 
LS~1698  &  J103735.3$-$564755  &  5.07  &  862.00  &  60.90  &  -- &   \textit{NuSTAR*} \\ 
1A~1118$-$61  &  J112057.1$-$615500  &  2.88  &  405.00  &  24.00  &  55 &   \textit{NuSTAR*} \\ 
IGR~J11305$-$6256  &  J113106.8$-$625648  &  1.54  &  --  &  120.83  &  -- & \textit{NuSTAR*} \\ 
IGR~J11435$-$6109  &  J114400.2$-$610736  &  7.67  &  161.80  &  52.46  &  -- & \\
4U~1145$-$619  &  J114800.0$-$621224  &  2.05  &  292.40  &  187.50  &  -- &  \textit{NuSTAR*} \\ 
SRGA~J124404.1  &  J124403.7$-$632232  &  5.85  &  538.70  &  138.00  &  -- & \textit{NuSTAR} \\ 
$-$632232 & & & & & \\
GX~304$-$1 &  J130117.0$-$613606  &  1.85  &  272.00  &  133.00  &  54 & \textit{NuSTAR} \\ 
2RXP~J130159.6$-$635806  &  J130158.7$-$635808  &  5.43  &  643.00  &  --  &  -- &  \textit{NuSTAR} \\ 
IGR~J13186$-$6257  &  J131825.0$-$625815  &  --  &  --  &  19.99  &  -- &  \\ 
SAX~J1324.4$-$6200  &  J132426.6$-$620119  &  --  &  172.84  &  1.13  &  -- &  \\ 
IGR~J14059$-$6116  &  J140555.9$-$611629  &  --  &  --  &  13.71  &  -- & \\
MAXI~J1409$-$619  &  J140802.7$-$615902  &  14.50  &  503.00  &  14.70  &  44, 73, 128 &  \textit{NuSTAR} \\ 
2S~1417$-$624  &  J142112.3$-$624155  &  6.45  &  17.60  &  42.12  &  -- & \textit{NuSTAR*} \\ 
IGR~J14488$-$5942  &  J144843.2$-$594213  &  --  &  33.42  &  49.63  &  -- & \\
XTE~J1543$-$568  &  J154405.1$-$564542  &  5.92  &  27.12  &  75.56  &  -- &  \textit{NuSTAR*} \\
2S~1553$-$542  &  J155748.3$-$542453  &  --  &  9.28  &  30.60  &  23--27 &  \\
Swift~J1626.6$-$5156  &  J162636.5$-$515630  &  10.00  &  15.36  &  132.90  &  10--18 & \\
AX~J1700.2$-$4220  &  J170025.2$-$421900  &  1.50  &  54.20  &  44.12  &  -- & \textit{NuSTAR*} \\
\hline
PSR~B1259$-$63  &  J130247.6$-$635008  &  2.17  &  0.05  &  1236.72  &  -- &  \\ 
HESS~J0632$+$057  &  J063259.2$+$054801  &  1.75  &  --  &  316.80  &  -- & \\
PSR~J0635$+$0533  &  J063518.2$+$053306  &  6.29  &  0.03  &  11.20  &  -- &  \\ 
\hline
 \end{tabular}
 \tablefoot{*Sources part of a NuSTAR Cycle 12 campaign to build a volume limited hard X-ray sample of BeXRBs within 7.5\,kpc. The other \textit{NuSTAR} data are part of published work \citep[][and references therein]{zalot2026}. For the CRSF column, the values are taken from \citet{staubert2019a}. Distances are based on \citep{bailer-jones2021b} where available, or taken from \citep{neumann2023}.}
\end{table*}

\section{Final Catalog}
We provide a catalogue along with this paper as auxiliary material. A representative list of the information contained in the catalogue is summarised in Table.~\ref{tab:catalog}, using eRASS1 as an example. The columns pertaining to subsequent eRASSes have the corresponding suffix of \texttt{e2}--\texttt{e4}. In addition, for each fitted parameter, we provide the upper and lower 90\% confidence intervals labeled with a \texttt{hi} and \texttt{lo} suffix added to the column name corresponding to a parameter, for example, \texttt{Gamma\_e1\_lo} and \texttt{Gamma\_e1\_hi}, list the uncertainties on the photon index of the power law fit. In the case of upper limits, only the flux columns are populated, and no uncertainties are provided. 

\begin{table*}[]
\caption{A representative list of fields from the final catalog, using eRASS1 as an example.}
\renewcommand{\arraystretch}{1.32} %% increase table row spacing
    \renewcommand{\tabcolsep}{2mm}
    \centering
    \begin{tabular}{l|l}
    \hline
    \hline
     Field    & Description \\
    \hline
    \texttt{SRC\_NAME} & eRASS name of the source \\
    \texttt{e1\_RA}    & Right Ascension from eRASS1 \\
    \texttt{e1\_DEC}  &  Declination from eRASS1 \\
    \texttt{SIMBAD\_NAME} & Source name  \\
    \texttt{CAT\_RA}   & Right Ascension from \citet{neumann2023}\\ 
    \texttt{CAT\_DEC}  & Declination from  \citet{neumann2023} \\
    \texttt{Gaia\_ID} & Gaia DR3 ID from \citet{collaboration2023a}\\
    \texttt{DISTANCE} & Distance used* \\         
    \texttt{BJ\_rpgeo} & Photogeometric distance \citet{bailer-jones2021b} \\
    \texttt{TYPE} & The sub-type of the source, as described in Section~\ref{app:catalog} \\
    \texttt{CO} & Type of compact object\\
    \texttt{GAMMA\_BINARY} & Presence of Gamma-ray emission \citep{neumann2023} \\
    \texttt{P\_SPIN} & Spin period of the neutron star \citep{neumann2023} \\
    \texttt{P\_ORB} & Orbital period of the system \citep{neumann2023} \\
    \texttt{RXTE\_LC} & Availability of RXTE/ASM monitoring data \\
    \texttt{MAXI\_LC} & Availability of MAXI monitoring data\\
    \texttt{BAT\_LC} &  Availability of \swiftbat monitoring data \\
    \texttt{DET\_LIKE\_e1} & Detection likelihood from eRASS1B catalogue \citep{merloni2024}\\
    \texttt{Soft\_Flux\_e1} & Fitted flux in the 0.2--2.0\,keV energy range [$\mathrm{erg}\,\mathrm{s}^{-1}\,\mathrm{cm}^{-2}$]\\
    \texttt{Hard\_Flux\_e1} & Fitted flux in the 2.0--10.0\,keV energy range [$\mathrm{erg}\,\mathrm{s}^{-1}\,\mathrm{cm}^{-2}$]\\
    \texttt{Total\_Flux\_e1} & Fitted flux in the 0.2--10.0\,keV energy range [$\mathrm{erg}\,\mathrm{s}^{-1}\,\mathrm{cm}^{-2}$]\\
    \texttt{N\_H} & Fitted or fixed [$N_{H}\times10^{22}\,\mathrm{cm}^{-2}$]\\
    \texttt{Gamma\_e1} & Fitted photon index for the power-law (PL) model\\
    \texttt{PL\_norm\_e1} & Normalisation for the power law fit\\
    \texttt{kT\_e1} & Temperature for the black body model [keV]\\
    \texttt{BB\_norm\_e1} &  Normalisation for the black body model \\
    \texttt{C-Stat\_BB} & C-stat fit statistic for the BB model\\
    \texttt{C-Stat\_PL} & C-stat fit statistic for the PL model\\
    \texttt{PILE\_UP} & eRASSes which are affected by pile-up \\
    \end{tabular}
    \label{tab:catalog}
    \tablefoot{*In the absence of a reliable \citet{bailer-jones2021b} distance, the distance from \textit{XRBcat} is used. Each fitted parameter has two additional columns for the uncertainties. The same format is used for eRASS2--4.}
\end{table*}

\end{appendix}

\end{document}